\documentclass[twocolumn]{aastex631}

\usepackage{gensymb}

\usepackage{savesym}
\savesymbol{tablenum}
\usepackage{siunitx}
\restoresymbol{SIX}{tablenum}

\usepackage{hyperref}
\usepackage{nth}

\usepackage{enumitem}
\usepackage{amsmath,amssymb}
\usepackage{xcolor}

\begin{document}

\title{Self-Consistent JWST Census of Star Formation and AGN activity at $z=5.5-13.5$}

\correspondingauthor{Jordan C. J. D'Silva}
\email{jordan.dsilva@research.uwa.edu.au}
\author[0000-0002-9816-1931]{Jordan C. J. D'Silva}
\affiliation{International Centre for Radio Astronomy Research (ICRAR) and the
International Space Centre (ISC), The University of Western Australia, M468,
35 Stirling Highway, Crawley, WA 6009, Australia}
\affiliation{ARC Centre of Excellence for All Sky Astrophysics in 3 Dimensions (ASTRO 3D), Australia}

\author[0000-0001-9491-7327]{Simon P. Driver}
\affiliation{International Centre for Radio Astronomy Research (ICRAR) and the
International Space Centre (ISC), The University of Western Australia, M468,
35 Stirling Highway, Crawley, WA 6009, Australia}

\author[0000-0003-3021-8564]{Claudia D. P. Lagos}
\affiliation{International Centre for Radio Astronomy Research (ICRAR) and the
International Space Centre (ISC), The University of Western Australia, M468,
35 Stirling Highway, Crawley, WA 6009, Australia}
\affiliation{ARC Centre of Excellence for All Sky Astrophysics in 3 Dimensions (ASTRO 3D), Australia}

\author[0000-0003-0429-3579]{Aaron S. G. Robotham}
\affiliation{International Centre for Radio Astronomy Research (ICRAR) and the
International Space Centre (ISC), The University of Western Australia, M468,
35 Stirling Highway, Crawley, WA 6009, Australia}


\author[0000-0003-4875-6272]{Nathan J. Adams}
\affiliation{Jodrell Bank Centre for Astrophysics, Alan Turing Building,
University of Manchester, Oxford Road, Manchester M13 9PL, UK}

\author[0000-0003-1949-7638]{Christopher J. Conselice} 
\affiliation{Jodrell Bank Centre for Astrophysics, Alan Turing Building,
University of Manchester, Oxford Road, Manchester M13 9PL, UK}

\author[0000-0003-1625-8009]{Brenda Frye} 
\affiliation{Department of Astronomy/Steward Observatory, University of Arizona, 933 N Cherry Ave,
Tucson, AZ, 85721-0009, USA}

\author[0000-0001-6145-5090]{Nimish P. Hathi}
\affiliation{Space Telescope Science Institute, 3700 San Martin Drive, Baltimore, MD 21218, USA}

\author[0000-0002-4130-636X]{Thomas Harvey} 
\affiliation{Jodrell Bank Centre for Astrophysics, Alan Turing Building,
University of Manchester, Oxford Road, Manchester M13 9PL, UK}

\author[0000-0002-6610-2048]{Anton M. Koekemoer} 
\affiliation{Space Telescope Science Institute,
3700 San Martin Drive, Baltimore, MD 21218, USA}

\author[0000-0002-6150-833X]{Rafael {Ortiz~III}} 
\affiliation{School of Earth and Space Exploration, Arizona State University,
Tempe, AZ 85287-1404, USA}

\author[0000-0003-4223-7324]{Massimo Ricotti}
\affiliation{Department of Astronomy, University of Maryland, College Park, 20742, USA}

\author[0000-0002-5404-1372]{Clayton Robertson}
\affiliation{Department of Physics and Astronomy, University of Louisville, Natural Science Building 102, Louisville, KY 40292, USA}

\author[0000-0001-7016-5220]{Michael~J.~Rutkowski}
\affiliation{Minnesota State University-Mankato, Dept. of Physics \& Astronomy, Trafton Science Center North 141, Mankato, MN, 56001 USA}
\email{michael.rutkowski@mnsu.edu}

\author[0000-0001-6564-0517]{Ross M. Silver}
\affiliation{Astrophysics Science Division, NASA Goddard Space Flight Center, Greenbelt, MD 20771, USA}

\author[0000-0003-3903-6935]{Stephen M. Wilkins}
\affiliation{Astronomy Centre, University of Sussex, Falmer, Brighton BN1 9QH, UK}

\author[0000-0001-9262-9997]{Christopher N. A. Willmer} 
\affiliation{Steward Observatory, University of Arizona,
933 N Cherry Ave, Tucson, AZ, 85721-0009, USA}

\author[0000-0001-8156-6281]{Rogier A. Windhorst}
\affiliation{School of Earth and Space Exploration, Arizona State University,
Tempe, AZ 85287-1404, USA}

\author[0000-0003-3329-1337]{Seth H. Cohen} 
\affiliation{School of Earth and Space Exploration, Arizona State University,
Tempe, AZ 85287-1404, USA}

\author[0000-0003-1268-5230]{Rolf A. Jansen} 
\affiliation{School of Earth and Space Exploration, Arizona State University,
Tempe, AZ 85287-1404, USA}

\author[0000-0002-7265-7920]{Jake Summers} 
\affiliation{School of Earth and Space Exploration, Arizona State University,
Tempe, AZ 85287-1404, USA}

\author[0000-0001-7410-7669]{Dan Coe} 
\affiliation{Space Telescope Science Institute, 3700 San Martin Drive, Baltimore, MD 21218, USA}
\affiliation{Association of Universities for Research in Astronomy (AURA) for the European Space Agency (ESA), STScI, Baltimore, MD 21218, USA}
\affiliation{Center for Astrophysical Sciences, Department of Physics and Astronomy, The Johns Hopkins University, 3400 N Charles St. Baltimore, MD 21218, USA}

\author[0000-0001-9440-8872]{Norman A. Grogin} 
\affiliation{Space Telescope Science Institute,
3700 San Martin Drive, Baltimore, MD 21218, USA}

\author[0000-0001-6434-7845]{Madeline A. Marshall} 
\affiliation{Los Alamos National Laboratory, Los Alamos, NM 87545, USA}

\author[0000-0001-6342-9662]{Mario Nonino} 
\affiliation{INAF-Osservatorio Astronomico di Trieste, Via Bazzoni 2, 34124
Trieste, Italy} 

\author[0000-0003-3382-5941]{Nor Pirzkal} 
\affiliation{Space Telescope Science Institute,
3700 San Martin Drive, Baltimore, MD 21218, USA}

\author[0000-0003-0894-1588]{Russell E. Ryan, Jr.} 
\affiliation{Space Telescope Science Institute,
3700 San Martin Drive, Baltimore, MD 21218, USA}

\author[0000-0001-7592-7714]{Haojing Yan} 
\affiliation{Department of Physics and Astronomy, University of Missouri,
Columbia, MO 65211, USA}



\begin{abstract}
The cosmic star formation history (CSFH) and cosmic active galactic nuclei (AGN) luminosity history (CAGNH) are self consistently measured at $z = 5.5-13.5$. This is achieved by analyzing galaxies detected by the James Webb Space Telescope from $\approx 400 \, \mathrm{arcmin^{2}}$ fields from the PEARLS, CEERS, NGDEEP, JADES and PRIMER surveys. In particular, the combination of spectral energy distribution fitting codes, EAZY and \textsc{ProSpect}, are employed to estimate the photometric redshifts and astrophysical quantities of $3751$ distant galaxies, from which we compute the stellar mass, star formation rate and AGN luminosity distribution functions in four redshift bins. Integrating the distribution functions, we find that the CSFH rises by $\approx 1$~dex over $z = 13.5 - 5.5$ and the CAGNH rises by $\approx 1$~dex over $z = 10.5 - 5.5$. We connect our results of the CSFH and CAGNH at $z=13.5-5.5$ to that from $z= 5-0$ to determine the summary of $\gtrsim 13$ Gyr of star formation and AGN activity, from the very onset of galaxy formation to the present day.
\end{abstract}

\keywords{}


\section{Introduction} \label{sec:intro}
Two of the most important physical processes occurring in galaxies are the assembly of stellar mass and the growth of supermassive black holes (SMBHs) powering active galactic nuclei (AGN). The prominence of these two processes is inferred by their relative contributions to the spectral energy distributions (SEDs) of galaxies at virtually all wavelengths from the X-rays through to the radio \citep[e.g.,][]{conroyModelingPanchromaticSpectral2013a,padovaniActiveGalacticNuclei2017a}. Because both AGN and star formation emit across the full SED, inferences on the physics of star formation rely on appropriately accounting for the AGN component and vice-versa.

In previous studies, this accounting has often been accomplished by separating galaxies into either star forming or AGN with diagnostic parameters such as optical emission lines \citep[e.g., the BPT diagram,][]{baldwinClassificationParametersEmissionline1981b} and X-ray emission \citep[e.g.,][]{brandtCosmicXraySurveys2015}. Especially at high redshift, this strategy does not work well. Common emission line ratios cannot significantly separate AGN from star-forming galaxies in the low metallicity terrain of the $z\gtrsim 5$ Universe \citep{maiolinoJADESDiversePopulation2023a,ublerGANIFSMassiveBlack2023a}. In fact, identification of AGN via optical emission lines is challenging due to the host galaxy emission, even in low mass galaxies at low redshift \citep[e.g.,][]{trumpBIASESOPTICALLINERATIO2015}, similar to the conditions expected in the early Universe. Finally, X-ray detections of confirmed broad-line $z\gtrsim 5$ AGN are sparse, suggesting either significant X-ray absorption or naturally X-ray faint AGN \citep[e.g.,][]{yangCEERSKeyPaper2023,anannaXRayViewLittle2024,yueStackingXrayObservations2024,maiolinoJWSTMeetsChandra2025a}.

Vast multiwavelength surveys have afforded the ability to simultaneously treat star formation and AGN when interpreting the emission of galaxies. Using the multiwavelength Galaxy and Mass Assembly survey \citep[GAMA,][]{driverGalaxyMassAssembly2011a,driverGalaxyMassAssembly2022a} and Deep Extragalactic VIsible Legacy Survey \citep[DEVILS,][]{daviesDeepExtragalacticVIsible2018,daviesDeepExtragalacticVIsible2021a}, \citet{thorneDEVILSCosmicEvolution2022} demonstrated that the SED fitting software \textsc{ProSpect} \citep{robothamProSpectGeneratingSpectral2020} accurately disentangles the relative flux contributions of star formation and AGN from the far ultraviolet (FUV) to far infrared (FIR) SED up to $z\approx 5$.

Treating these two critical mechanisms as if they are independent prevents us from better reconciling the union of star formation and AGN activity with the physics of galaxy formation. This is especially true given that feedback from AGN is thought to be an important element that regulates star formation and the baryon cycle in galaxies in the local Universe \citep[e.g.,][]{katsianisEvolvingMassdependentSsSFRM2019,daviesDeepExtragalacticVIsible2022,wrightBaryonCycleModern2024}. The interplay of these processes may lead to a tight relationship between the properties of the SMBH and the surrounding geography of galaxies, such as velocity dispersion of the gas and the mass of the bulge \citep{magorrianDemographyMassiveDark1998,gebhardtRelationshipNuclearBlack2000a,merloniSynthesisModelAGN2008,kormendyCoevolutionNotSupermassive2013}. This connection between star formation and the AGN activity produces a coevolving cosmic star formation history (CSFH) and cosmic AGN bolometric luminosity history (CAGNH) that has persisted for at least $11$~Gyr up to the present day \citep{kauffmannUnifiedModelEvolution2000,madaudickinson2014,dsilvaGAMADEVILSCosmic2023}, illustrative of the broad connection between stellar mass assembly and the growth of SMBHs. To further explore these quantities at $z\gtrsim 5$, where much of the current interest lies, we must use facilities that can probe the high-redshift Universe such as the James Webb Space Telescope \citep[JWST,][]{gardnerJamesWebbSpace2006a}.  

Quantifying how efficiently galaxies form stars and grow their SMBHs at $z\gtrsim 5$ is critical to uncovering the details of the epoch of reionisation (EOR) and how the EOR sets the initial conditions for all subsequent galaxy formation and evolution. Before JWST, observations of the UV luminosity function down to $M_{UV} = -13$ at $z\approx2-9$ \citep{bouwensUVLuminosityFunctions2015a, oeschDearth10Galaxies2018, stefanonGalaxyStellarMass2021, bouwensNewDeterminationsUV2021} favoured a paradigm where star formation was the dominant source of ionising photons. However, recent JWST results have renewed interest in the AGN contribution to reionisation, with AGN now claiming a higher ionising emissivity, up to $\approx 50$\% of what star-forming galaxies contribute, \citep{harikaneJWSTNIRSpecFirst2023b} compared to pre-JWST results \citep[$\approx 10$\%,][]{matsuokaSubaruHighzExploration2018a}.

Indeed, a striking finding from JWST after roughly 1000 days in space is its reported evidence that AGN activity is widespread at $z\gtrsim 5$, as inferred from detections of broad H$\alpha$ line components \citep[i.e., $\gtrsim 1000 \, \mathrm{km \, s^{-1}}$,][]{schneiderAreWeSurprised2023b,mattheeLittleRedDots2024,kocevskiHiddenLittleMonsters2023a,juodzbalisEPOCHSVIIDiscovery2023,harikaneJWSTNIRSpecFirst2023b,larsonCEERSDiscoveryAccreting2023a}. AGN activity as it appears in little red dots \citep[LRDs,][]{mattheeLittleRedDots2024,kocevskiRiseFaintRed2024a} are especially prevalent, indicating $\approx 1$~dex more SMBH accretion per unit comoving volume at $z\approx 4-5$ than we would have expected from simply extrapolating the X-ray luminosity function \citep{akinsCOSMOSWebOverabundancePhysical2024,yangCEERSKeyPaper2023}. At the same time, JWST results indicate rapid stellar mass assembly at $z\gtrsim 5$ \citep[e.g.][]{harikaneComprehensiveStudyGalaxies2023,harikanePureSpectroscopicConstraints2024,adamsEPOCHSIIUltraviolet2024,donnanJWSTPRIMERNew2024,behrooziAverageStarFormation2013a,harikaneGOLDRUSHIVLuminosity2022,lovellExtremeValueStatistics2023}, producing $\approx 10^{11} \, \mathrm{M_{\odot}}$ galaxies \citep[e.g.,][]{labbePopulationRedCandidate2023,glazebrookMassiveGalaxyThat2024} only a few 100 Myr after the Big Bang. It should be noted however, that the presence of an AGN that is not accounted for could skew the recovered $\mathrm{M_{\star}}$ to higher masses \citep[e.g.,][]{thorneDeepExtragalacticVIsible2022,dsilvaStarFormationAGN2023}.

While broad, $\gtrsim 1000 \, \mathrm{km \, s^{-1}}$, H$\alpha$ components as detected by NIRSpec seem to be effective in detecting AGN activity at $z\approx 5$ \citep[e.g.,][]{harikaneJWSTNIRSpecFirst2023b,larsonCEERSDiscoveryAccreting2023a,maiolinoJADESDiversePopulation2023a}, it is unfeasible to obtain high signal to noise spectra for a representative sample of galaxies. As such, in this work we leverage broadband, $\approx 0.2 \to 4.4 \, \mu\mathrm{m}$ SEDs of thousands of $z\gtrsim 5$ galaxies as observed by both the Hubble Space Telescope (HST) and JWST. We follow the approach of \citet{thorneDeepExtragalacticVIsible2021,thorneDEVILSCosmicEvolution2022} and use \textsc{ProSpect} to obtain the astrophysical parameters, $\mathrm{M_{\star}}$, SFR and bolometric $\mathrm{L_{AGN}}$, from the SEDs of galaxies between $z=5.5-13.5$. Using those quantities we calculate the stellar mass distribution function (SMF), the star formation rate distribution function (SFRF) and the AGN luminosity distribution function (AGNLF). Stellar mass assembly and the growth of SMBHs are explored from the perspectives of the cosmic stellar mass history (CSMH), CSFH and CAGNH, expanding on the analysis of \citet{dsilvaGAMADEVILSCosmic2023} who presented measurements of the same quantities from $z\approx 0-5$. Hence, we measure the cosmic densities of star formation and the growth of SMBHs in a self-consistent manner from the EOR through to the local Universe.

In Section~\ref{sec:data}, we describe the JWST imaging data and our novel processing methods. In Section~\ref{sec:multiband}, we describe our multiband photometry and extraction pipeline. In Section~\ref{sec:sedfitting}, we detail our SED fitting methods and our $z\geq 5.5$ sample. In Section~\ref{sec:results}, we present our key results on the CSFH and CAGNH. In Section~\ref{sec:caveats} we discuss caveats and future work. Finally, In Section~\ref{sec:summary}, we present a summary and concluding remarks. We use standard concordance $\Lambda$CDM cosmology with $H_{0} = 70 \, \mathrm{km \, s^{-1} \, Mpc^{-1}}$, $\Omega_{\Lambda} = 0.7$ and $\Omega_{M} = 0.3$. We use the AB magnitude system \citet{okeSecondaryStandardStars1983} and the \citet{chabrierGalacticStellarSubstellar2003} initial mass function (IMF). Throughout this text, we denote the uniform distribution with boundaries $a$ and $b$ as $U(a,b)$ and the normal distribution with mean, $\mu$, and variance, $\sigma^{2}$, as $\mathcal{N}(\mu, \sigma^{2})$.
 
\section{Imaging data} \label{sec:data}
In this paper we used both proprietary and publicly available HST and JWST data. All of our JWST fields were processed with the novel package: \textsc{JumProPe}, which stands for the JWST UWA Multiwavelength ProTools Processing Endeavour\footnote{\url{https://github.com/JordanDSilva/JUMPROPE}} \citep{jordandsilva_2025_15086450}. While \textsc{JumProPe} and its many processing steps have been introduced previously \citep{dsilvaStarFormationAGN2023,windhorstJWSTPEARLSPrime2023,robothamDynamicWispRemoval2023}, we here provide a complete description of the processing package for reference. 

\subsection{Query and calibration}
\verb|UNCAL| files, the rawest form of JWST data with which the typical scientist interacts, were downloaded from the Mikulski Archive for Space Telescopes (MAST). \textsc{JumProPe} includes this functionality by means of the \verb|astroquery| module, enabling automatic querying of MAST.

After downloading this data, the STScI developed JWST Calibration Pipeline was used to process \verb|UNCAL| files to \verb|CAL|. It is this step where, for example, cosmic ray impacts are flagged, sky dark and flat fields are removed and the World Coordinate System (WCS) is applied. In this step, the user must specify the JWST Calibration Reference Data System (CRDS) denomination (e.g., the reference file pmap). Our reductions used five different CRDS pmap and calibration pipeline version combinations because of the frequency of data ingest, but at a minimum used pmap=1100 and pipeline version \texttt{1.11.2}. 

\begin{figure}
    \centering
    \includegraphics[width = \columnwidth]{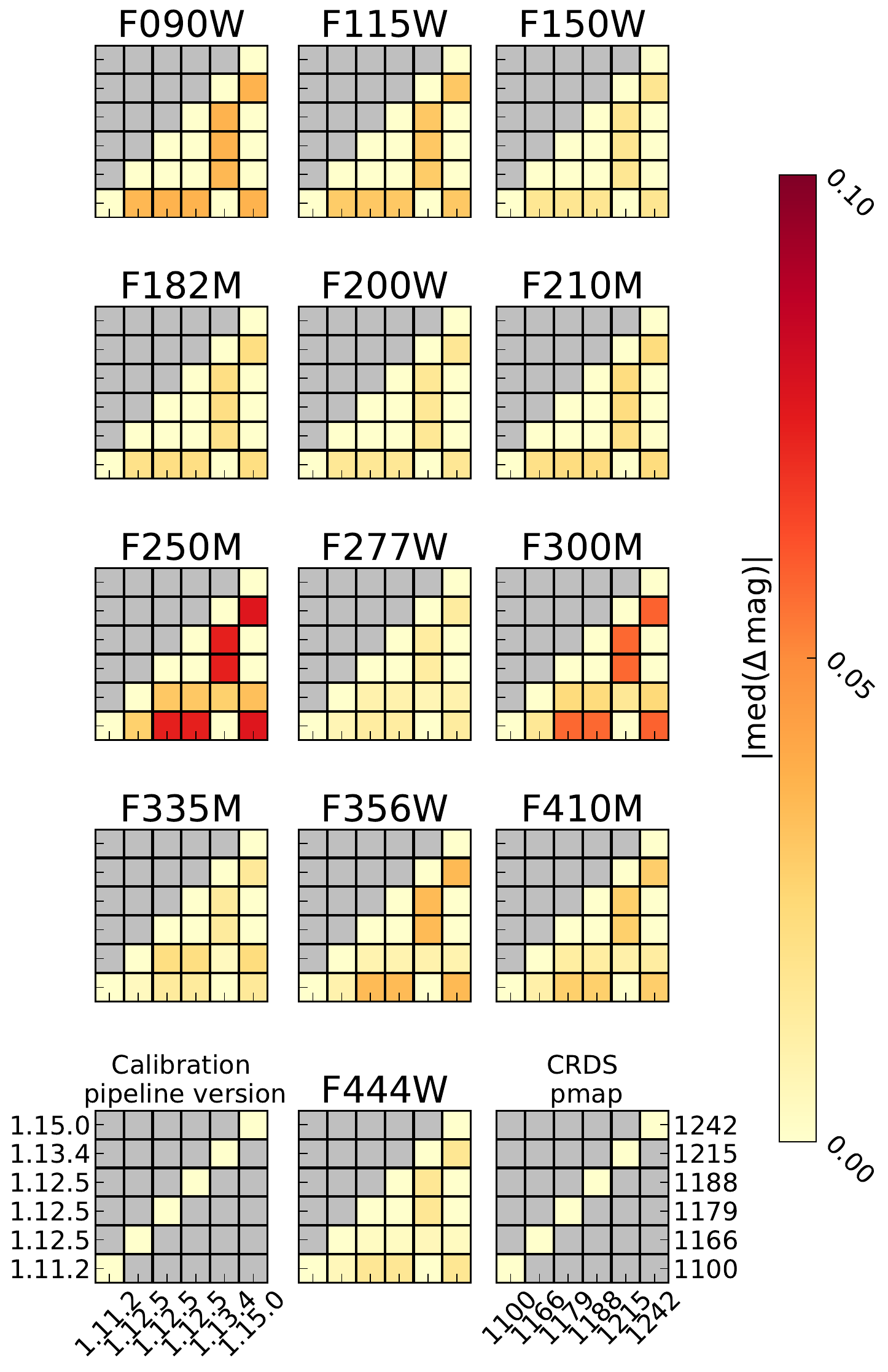}
    \caption{Corner plots of the absolute median magnitude differences of each pixel between calibration pipeline versions and CRDS pmaps per filter. This was computed for the \texttt{CAL} returned from the calibration pipeline.}
    \label{fig:calver_pmap}
\end{figure}

Figure~\ref{fig:calver_pmap} shows the absolute median magnitude difference per pixel between each permutation of the calibration pipeline version and CRDS pmap per filter. These runs were performed on data from programs 1210 and 3215 in GOODS-South over all modules (A and B) and sub-modules ($1\to 4$ for the short wavelength channel) of NIRCam. For the data used in this work, the most out-of-date combination is calibration pipeline version$=1.11.2$, pmap$=1100$ and the most up-to-date is calibration pipeline version$=1.13.4$, pmap$=1215$. We also show the results for calibration pipeline version$=1.15.0$, pmap$=1242$ which was the most modern combination as of July 2024. The largest difference is $\approx 0.1$~mag in the F250M, but overall the differences are mostly negligible and we used the original pmaps applied to each data set. To reflect systematic zero-point differences between the filters as shown in Figure~\ref{fig:calver_pmap}, we assumed a $10\%$ uncertainty floor on all photometeric measurements that were used for SED fitting.

\subsection{Additional artefact removal}
Having produced \verb|CAL| files, the JWST Calibration Pipeline can then make mosaics using \textsc{AstroDrizzle}. However, at this stage, we intercept the data and perform three crucial processing steps not present in the default pipeline to significantly improve the quality of the data. 

\subsubsection{$1/f$ removal}
Because the NIRCam detectors export data along read-out lines, variations in the amplifiers can cause banding structures, known as $1/f$ noise, in the images that can compromise the photometry. To correct this, we employed \textsc{ProFound} \citep{robothamProFoundSourceExtraction2018} to compute clipped column and row-wise medians on the individual detectors, producing a spatial map of the $1/f$ pattern that was then subtracted from the images. Because the presence of real astronomical sources may bias the row and column medians, the $1/f$ correction has three levels of aggressiveness depending on how crowded the image is. As we produced mostly blank fields, we used the strongest $1/f$ removal. Weaker versions of the $1/f$ algorithm are most appropriate for images of galaxy clusters \citep[e.g.,][]{fryeJWSTPEARLSView2023} or in nearby objects with large projected angular sizes that fill the frame \citep[e.g.,][]{keelJWSTPEARLSDust2023a} where they would otherwise be oversubtracted. 

\subsubsection{Wisp removal}
Wisps are diffuse artefacts in NIRCam images caused by stray light from bright stars, outside the field of view. Often, the wisps are confined to fixed locations on the detectors. However, slight variations in the wisp positions due to the exact optical path of the stray photons limits the effectiveness of subtracting pre-determined wisp templates (though this situation will likely improve as JWST continues to build on its existing dataset). Fortunately, the geometry of the NIRCam detector layout means that the long-wavelength channels are virtually unaffected by the wisps. As such, a special algorithm was developed to remove the wisps \citep{robothamDynamicWispRemoval2023} that we employed on the data in this work. In essence, the unaffected long wavelength channel is used to isolate the wisp. The algorithm first produces a deep mosaic of the longest wavelength image that overlaps with the affected short channels to encapsulate most, if not all, genuine astronomical sources. The long channel is then appropriately scaled to match the pixel values of the short channel filters as closely as possible. Because the average SED of astronomical objects ensures that real sources will most likely be detected in at least the long, red filters, the residual when the long is subtracted from the short will be the wisp pattern. The wisp pattern can then be subtracted from the affected image. We note that the presence of wisps affects the $1/f$ removal as the pattern can bias the column and row-wise subtraction, meaning that the wisp removal must be performed before the $1/f$ removal for the short wavelength channels. 

\subsubsection{Sky removal}
Finally, we performed sky subtraction. Again, we used \textsc{ProFound} to detect sources and compute sky maps from the largely artefact-free \verb|CAL| files. The sky maps were then stacked to compute super skies for each module and filter combination, allowing us to constrain global detector properties. For the short wavelength channel, each module of NIRCam is made of a further four sub-modules, and so we also subtracted a global pedestal value to ensure smooth pixel distributions over the entire breadth of the module. The sky subtraction in detail is represented as 
\begin{multline}
        (SCI - SKY) = SCI - \\ (M \times SKYSUPER + B + P),
\end{multline}
where $SCI$ is the science extension, $SKY$ is the sky model, $M$ and $B$ are the coefficients of a linear fit to $SKYSUPER$, the super sky, and $P$ is the pedestal value. In detail, each pixel in the $SCI$ frame is fit with a linear function with the corresponding pixel in the $SKYSUPER$ frame, and this fit is subtracted from $SCI$ to effectively remove the sky background. This sky model is tested against a simple \textsc{ProFound} run, where if the latter option provides a better sky solution, as per the global $\chi^{2}$ for a normally distributed sky, then the simplified \textsc{ProFound} sky is instead subtracted.

This pipeline of $1/f$ removal, wisp correction and sky subtraction is used externally by the PEARLS collaboration, and further details of the pipeline steps can be found in \citet{windhorstJWSTPEARLSPrime2023,adamsEPOCHSIIUltraviolet2024,conseliceEPOCHSDiscoveryStar2024}. 

\subsection{Mosaicking with \textsc{ProPane}}
Having produced flat and artefact-free \verb|CAL| files, the next step is to turn them into mosaics using the image stacking software \textsc{ProPane} \citep{robotham_propane_2024}. The \textsc{ProFound} run performed earlier to calculate the sky statistics preserves the RMS maps of the sky that are used for inverse-variance weighted stacking ($\mathrm{inVar = skyRMS^{-2}}$). \textsc{ProPane} also computes a median stack of the input frames that is more resilient to outlier pixels compared to the inverse-variance stack. Outlier pixels in the inverse-variance stack, which are mostly cosmic rays that survived the initial jump detection in the Calibration Pipeline, are filled in with the equivalent pixels from the median stack. This way, we leveraged the superior depth of the inverse-variance stack while using the median stack to patch most of the remaining cosmetic artefacts. \textsc{ProPane} also propagates the mosaic of the inverse-variance image that we used for source detection. 

We also provided a reference catalogue, obtained from previous reductions of the JWST fields and observations with ground-based telescopes, to ensure astrometric accuracy. We note, that this pipeline is well suited to adapt source catalogues from forthcoming wide-field surveys with instruments like EUCLID and LSST. 

\subsection{Field selection}
In this work we used $\approx 400$~amin$^2$ of JWST NIRCam imaging drawn from the Prime Extragalactic Areas of Reionization and Lensing Science (PEARLS, PIs: R. Windhorst \& H. Hammel, PIDs: 1176 \& 2738) survey \citep{windhorstJWSTPEARLSPrime2023}, the JWST Advanced Deep Extragalactic Survey \citep[JADES, PIDs: 1180 PI: D. Eisenstein, 1210 PI: N. Luetzgendorf \& 3215, PI: D. Eisenstein \& R. Maiolino, ][]{eisensteinOverviewJWSTAdvanced2023a,riekeJADESInitialData2023}, the Next Generation Deep Extragalactic Exploratory Public (NGDEEP, PID: 2079, PIs: S. Finkelstein, C. Papovich, N. Pirzkal) survey, \citep{bagleyNextGenerationDeep2024}, the Cosmic Evolution Early Release Science \citep[CEERS, PID: 1345, PI: S. Finkelstein,][]{bagleyCEERSEpochNIRCam2023a,finkelsteinCEERSKeyPaper2023a} and, finally, the Public Release IMaging for Extragalactic Research (PRIMER, PID:1837, PI: J. Dunlop) survey. Key information for each field per survey is presented in Table~\ref{tab:survey_limits}.

We required that each field contained at least 8 NIRCam filters: F090W; F115W; F150W; F200W; F277W; F356W; F410M; F444W; to enable robust SED fitting. This is true for all of our fields except NGDEEP and CEERS where the F090W filter is missing but is made up for with the availability of HST ACS F814W imaging, covering a similar pivot wavelength. The JADES imaging enjoys additional medium band imaging that we to use in our SED fitting. 

Because PEARLS also images cluster environments, we only use the off-cluster, parallel imaging of ACT-CL J0102$-$4915 (El Gordo), MACS J0416.1$-$2403 (MACS0416), MACS J1149.5$+$2223 (MACS1149), and PLCK G165.7$+$67.0 (G165). PHz G191.24$+$62.04 (G191) is also a PEARLS-Clusters target, however it is a Planck-selected proto-cluster at $z\approx 2$ and so we used the entire footprint. All of these fields were selected for study because they are largely blank fields, ensuring that our galaxy measurements are not overly biased by dense clusters, intra-cluster light or the effects of strong gravitational lensing. 

\section{Source detection and multiband measurements} \label{sec:multiband}
After the creation of image mosaics, source detection and multiband photometric measurements were pursued. This pipeline is also implemented in \textsc{JumProPe} as described here.

\subsection{Star masking}
The hexagonal mirror segments of JWST produce the unmistakable 6-pointed point spread function (PSF) pattern when observing stars. The presence of diffraction spikes introduces contaminants during source detection. This is especially significant when folding in observations with the HST, where the flux from the JWST diffraction spikes is missing, leading to SEDs that can resemble $z\gtrsim 5$ drop-out galaxies. Often, the strategy to remove diffraction spikes either involves filtering spurious detections in catalogue space or manually masking the affected areas of the region before source detection. 

\textsc{JumProPe} employs a novel method to mask bright stars before source detection. First, The GAIA source catalogue was used to isolate the positions of bright stars in each of the stacked images. Only objects in the GAIA catalogue with $\mathtt{classprob\_dsc\_combmod\_star} > 0.98$ $\wedge$ $\mathtt{phot\_bp\_rp\_excess\_factor} < 2.5 $ were masked to avoid erroneously masking distant, point-source quasars. 

\begin{figure}[h!]
    \centering
    \includegraphics[width=\columnwidth]{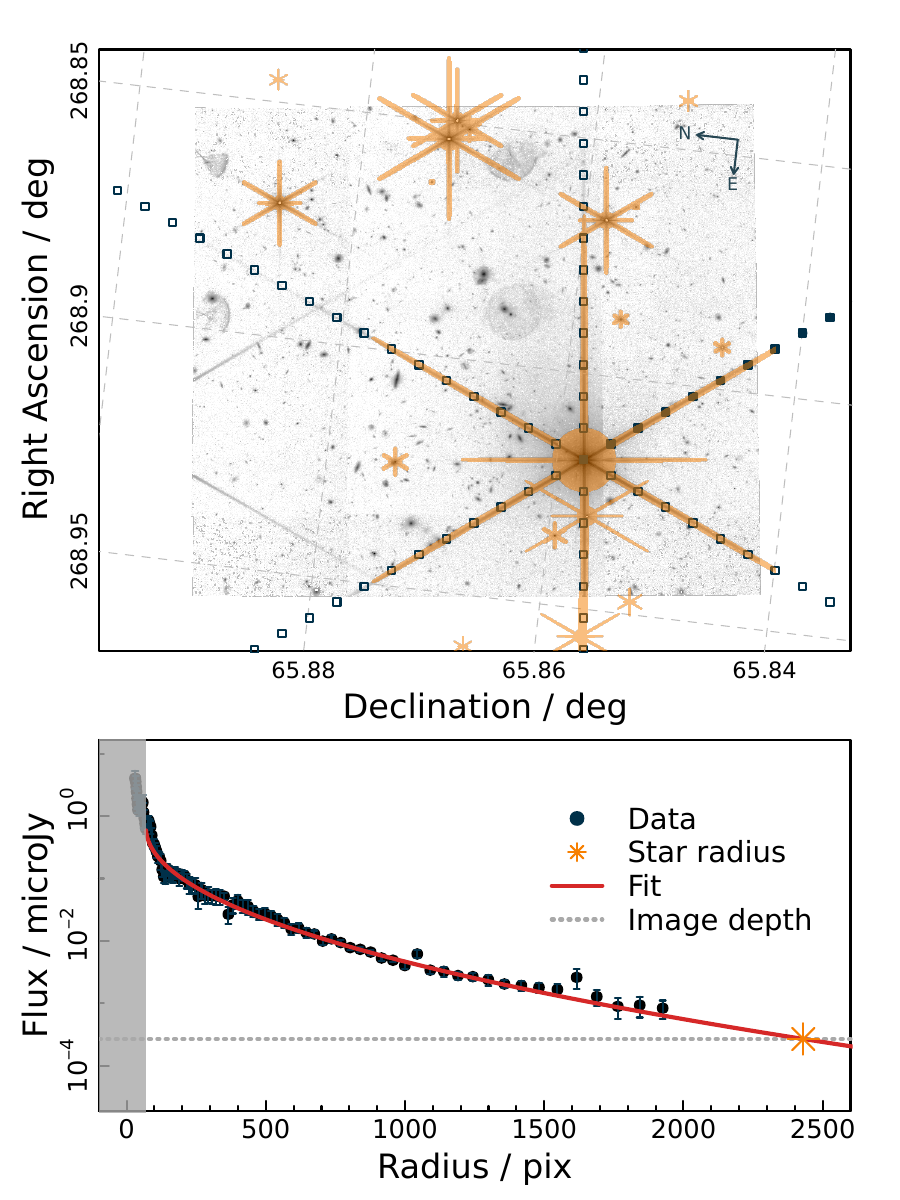}
    \caption{\textit{Top:} Example GAIA star mask of the field containing 2MASS J17554042+6551277 in the F277W filter. The dark blue square points demonstrate the 6 radial rays extended out from the star centre. The orange shapes show the extent of the masks pasted on each GAIA star. \textit{Bottom:} The radial flux profile of 2MASS J17554042+6551277 (the brightest star in the field). The solid red line is the 1D Sersic profile fit. The vertical grey band is the inner pixels that are masked due to being saturated. The horizontal dashed grey line is the approximate local image depth and the orange pointed star is the radius at which the profile descends below the depth.}
    \label{fig:star_mask}
\end{figure}

To find the extent of the diffraction spike, 6 rays were extended out in the image matrix from the centre of the star at angles of $30\degree, 90\degree, 150\degree, 210\degree, 270\degree, 330\degree$ and the radial flux profiles were calculated. This allowed us to fit a 1D S\'ersic profile to the diffraction spike, which we assumed was best represented by the ray with the maximum amount of flux, and note the radius from the star's centre at which the flux descends below the local skyRMS value. With the central positions and the radii of each GAIA star, a simplistic PSF mask of ellipses was transplanted onto a blank matrix of the same dimensions as the image and saved. One ellipse (circle) was used for the central core of the star, three ellipses were used for the largest spikes and one final ellipse was used for the final horizontal diffraction feature caused by the support structures of the primary mirror. This process was performed on the stacks of each 10 digit visit ID, where the orientation is natively the position angle of the telescope at the time of observation, ensuring that the orientations of the PSF patterns on the image are all the same. Any residual flux is captured by enabling \textsc{ProFound} to dilate the star masks until the curve-of-growth criterion of $\lesssim 1$\% change is satisfied. An example of this star masking algorithm is presented in Figure~\ref{fig:star_mask}, showing the field of 2MASS J17554042+6551277. 

In some cases, the star mask does not fully capture all of the diffraction spike or the stars are too far out of the field of view. The residual diffraction spike artefacts can be removed in catalog space using improbable colours and proximity to heavily deblended groups of segments (we elaborate further on artefact masking in Section~\ref{subsect:artefacts}). To use these masks on the large mosaics, we simply warped the input masks onto the same astrometric grid. 

\subsection{Photometry with \textsc{ProFound}}
We used \textsc{ProFound} for source detection, optimizing for the faint objects in the images. Only sources that were $\gtrsim 1.5$ times the skyRMS and more than 7 pixels in extent were detected. Four iterations of segment dilations were also enabled to capture the majority of the flux of the objects. Segmentation dilations are essentially curve-of-growth calculations to iteratively grow the initial segments until the fractional change of the added flux is $\lesssim 1\%$. Source detection was performed on an inverse-variance-weighted stack of the F277W, F356W and F444W filters. In total, 283,314 objects across all fields were detected with \textsc{ProFound}.

The segmentation map was then passed onto each of the images to extract multiband photometry with \textsc{ProFound}. An additional phase of dilation was allowed for each image to capture the remaining flux in each band. We measured photometry in both the undilated and dilated segments that were produced by the source detection. Undilated photometry allows for high S/N colours, most appropriate for photometric redshifts, while dilated photometry captures more flux and is thus more appropriate for astrophysical quantities like $\mathrm{M_{\star}}$. The dilated segments conform to the morphology of the galaxies as they appear in the images meaning that we do not need to employ aperture or PSF corrections to obtain an accurate measurement. The flux uncertainties produced by \textsc{ProFound} do not account for covariance between pixels that are induced during the mosaicking phase. Blank apertures of various sizes were placed on the images to determine the distribution of sky pixels, capturing the covariance, and the \textsc{ProFound} flux uncertainties were appropriately scaled to account for this additional source of noise. 

\begin{figure}[h!]
    \centering
    \includegraphics[width = \linewidth]{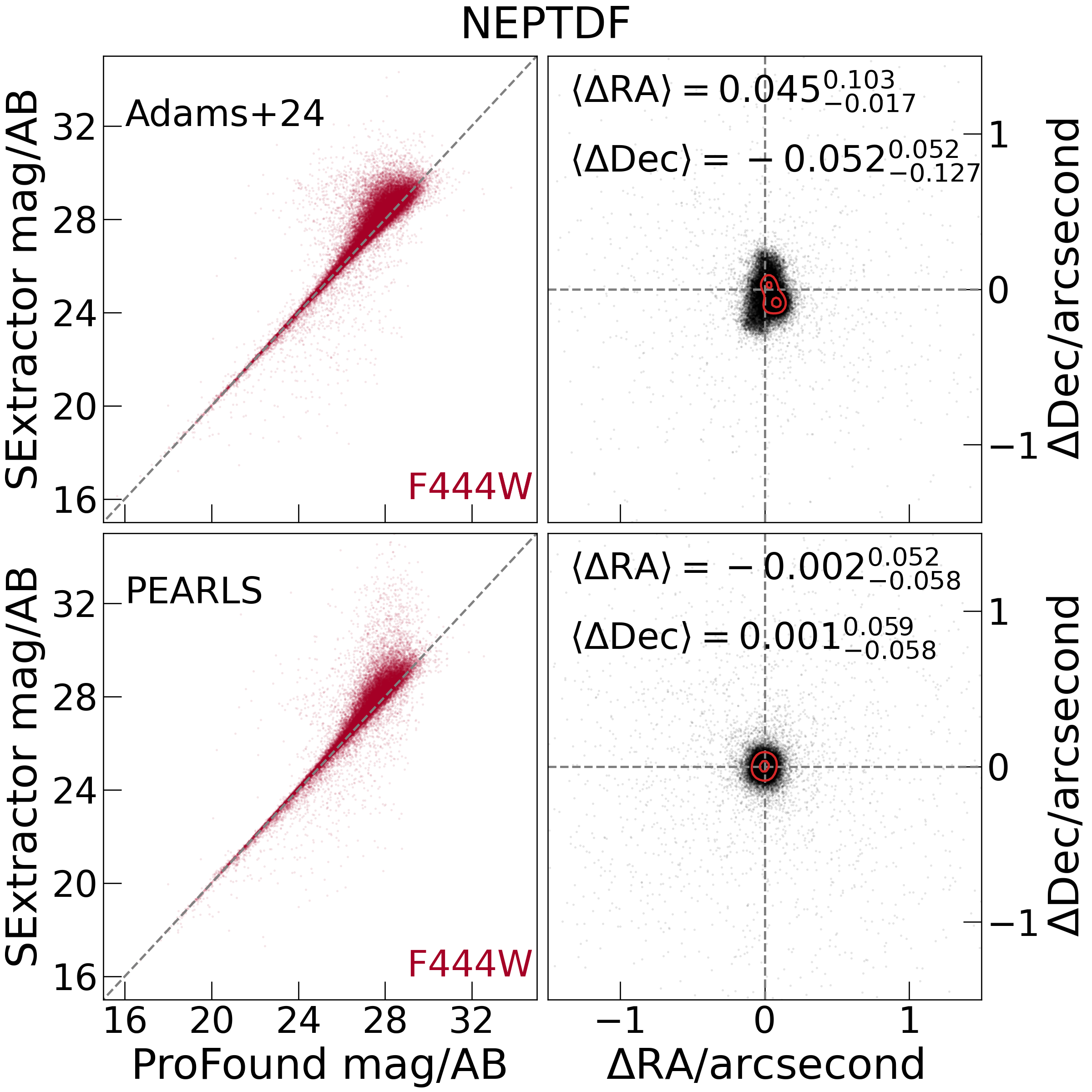}
    \caption{\textit{Left: } F444W \textsc{ProFound} photometry against \textsc{SourceExtractor} from the EPOCHS data set \citep[\textit{top},][]{conseliceEPOCHSDiscoveryStar2024,adamsEPOCHSIIUltraviolet2024} and the PEARLS team (\textit{bottom}). \textit{Right: } Astrometry difference between EPOCHS and PEARLS. The red contours show the $1\sigma$ and $2\sigma$ levels. Median and dispersions of $\Delta$RA and $\Delta$Dec are also quoted.}
    \label{fig:nep_compare}
\end{figure}

To find the detection sensitivities of each of our fields, we placed \SI{0.15}{\arcsecond} apertures onto blank areas of the $\mathrm{F277W+F356W+F444W}$ detection images and computed the $5\sigma$ limiting magnitudes from the blank sky pixel distribution. 

\begin{table*}[t!]
    \centering
    \begin{tabular}{c c c c c c}
        Field & $5\sigma$ depth/mag & Total area & Effective area/arcmin$^{2}$ & Right Ascension & Declination\\
        \hline
        \hline
         PEARLS El Gordo Parallel & 29.36 & 5.01 & 4.90 & 15.74 & -49.21\\ 
         PEARLS G165 Parallel  & 29.09 & 5.01 & 4.92 & 171.76 & 42.47\\ 
         PEARLS MACS0416 Parallel  & 29.51 & 14.32 & 13.84 & 64.07 & -24.07\\ 
         PEARLS MACS1149 Parallel  & 29.31 & 5.026 & 4.99 & 177.40 & 22.35\\ 
         PEARLS G191 & 29.27 & 10.03 & 9.80 & 161.16 & 33.83\\ 
         PEARLS NEPTDF & 29.58 & 65.10 & 60.95 & 260.70 & 65.82\\ 
         PEARLS WFC3-ERS &  29.37 & 10.03 & 9.78 & 53.18 & -27.70\\ 
         JADES GOODS-S & 30.46 & 26.77 & 26.23 &  53.16 & -27.79\\ 
         JADES Origin GOODS-S & 30.43 & 9.33 & 9.05 & 53.06 & -27.87\\ 
         NGDEEP GOODS-S & 29.81 & 10.90 & 10.27 & 53.25 & -27.84\\ 
         CEERS Extended Groth Strip& 29.06 & 102.43 & 99.50& 214.91 & 52.87\\ 
         PRIMER COSMOS & 28.93 & 145.47 & 138.38 & 150.13 & 2.33 \\ \hline
    \end{tabular}
    \caption{$5\sigma$ magnitude depths in \SI{0.15}{\arcsecond} apertures for the $\mathrm{F277W+F356W+F444W}$ detect band} and area in arcmin$^{2}$ covered by our fields. The effective area is the area remaining after masking. The right ascension and declination are the J2000 central coordinates of the fields in degrees.
    \label{tab:survey_limits}
\end{table*}
 Table~\ref{tab:survey_limits} shows the $5\sigma$ limiting magnitudes in the $\mathrm{F277W+F356W+F444W}$ detection image for our fields and the area covered. 
 
Figure~\ref{fig:nep_compare} shows a comparison between our \textsc{ProFound} photometry in the NEPTDF and \textsc{SourceExtractor} photometry from the EPOCHS data set \citep{conseliceEPOCHSDiscoveryStar2024,adamsEPOCHSIIUltraviolet2024} and the PEARLS collaboration (Cohen, S priv. comm.). The comparison data sets used \textsc{AstroDrizzle} to produce the mosaics, as opposed to \textsc{ProPane}. A key result is the propensity for \textsc{ProFound} to obtain $\approx 10$\% more flux, especially at mag~$\gtrsim26$, compared to \textsc{SourceExtractor} on account of its segment dilation logic. Though only the F444W photometry is shown in this figure, similar trends are exhibited in all other filters. We also see that astrometric differences are $\lesssim \SI{1}{\arcsecond}$, with the median differences close to zero and low dispersion, as quoted in Figure~\ref{fig:nep_compare}. Overall, the systematic differences between \textsc{AstroDrizzle}/\textsc{SourceExtractor} photometry and ours are minimal and as expected.

\subsection{Folding in HST observations} \label{subsect:hst}
HST observations were included to further sample the $\lambda < 0.9\, \mu$m part of the observed galaxy SEDs. This was possible for all fields except the parallel fields of G165, G191, and El Gordo. The vast majority of HST observations, such as from the Hubble Frontier Fields \citep{lotzHFF2017}, targeted the main cluster in those cases. 

For PRIMER, CEERS, NGDEEP and JADES we used ACS/WFC F606W and F814W data from the CANDELS HST program \citep{koekemoerCANDELS2011,groginCANDELS2011}, warping the images to the JWST astrometric grid and using the magnitude zero-points as reported by the CANDELS repository\footnote{\url{https://archive.STScI.edu/hlsp/candels}}. For the NEPTDF we used ACS/WFC F435W and F606W data from HST programs GO 15278 and GO 16252/16793 \citep{obrienTREASUREHUNTTransientsVariability2024}. For the WFC3-ERS field we used WFC3/UVIS F225W, F275W, F336W data from the HST Early Release Science programs 11359 \citep{windhorstHubbleSpaceTelescope2011a}. These data were obtained from the PEARLS collaboration and we used their reported magnitude zero-points. 

The HST data were generated using \textsc{AstroDrizzle} as opposed to \textsc{ProPane}, but the difference between the two codes is negligible \citep{robotham_propane_2024}.

\subsection{Artefact flagging}\label{subsect:artefacts}
While the \textsc{ProFound} detection was tuned to minimize the incidence of false-positive sources, we imposed further selections in the catalogue to reject as many resilient artefacts as possible. 

Hot pixels caused by cosmic ray impacts were rejected if all three of these criteria were satisfied: the effective radii of sources were below the approximate angular resolution of NIRCam at $2.7\, \mu$m ($\approx0.11''$), more than a quarter of the total flux was in the central pixel and $S/N>5$ of the whole source. The motivation for this is that energetic cosmic rays strike the detector without passing through the optics and deposit their energy into basically a single pixel of the CCD.

Stars, not in the GAIA catalogue, were flagged if the sources inhabited the stellar locus in the plane of detected magnitude versus effective radius and in the plane of the F150W versus F277W-F444W colour. In both cases the stellar locus was identified by fitting kernel density estimations in bins of magnitude and identifying the minimum between the loci of galaxies and stars in this plane of F277W-F444W-detect magnitude. A similar strategy, using these planes to discern stars from genuine galaxies, was used by \citet{bellstedtGalaxyMassAssembly2020a,windhorstJWSTPEARLSPrime2023} and will be used for galaxy number counts as part of the SKYSURF project (Tompkins in prep.).

Residual diffraction spikes were flagged if sources inhabited the stellar locus in the plane of the F150W versus F277W-F444W colour and were near to bright groups. \textsc{ProFound} provides segmentation information for neighbouring segments, referred to as groups. We matched diffraction spikes with stellar colours that are within $5\%$ of the total-light radius of groups of at least $2$ touching segments and brighter than $22$ magnitudes in the detect band, in a similar manner as \citet{bellstedtGalaxyMassAssembly2020a}. The motivation for this selection is that the diffraction spikes that survive the initial star masking will be highly fragmented because of the substructure in the NIRCam PSF and exhibit stellar colours. 

Only sources with ${\mathtt{edge\_excess}}\leq 1$ and ${\mathtt{edge\_frac}} > 0.5$, which are measurements reported by \textsc{ProFound}, were included in the subsequent analysis of this paper. \texttt{Edge\_excess} is defined as the ratio of segment edge pixels to the expected number given the elliptical geometry measurements of the galaxy, and a value $\geq1$ is likely indicative of compromised photometry. \texttt{Edge\_frac} is the fraction of segment edge pixels that are touching the sky, and so we required at least $50$\% of pixels to be touching the sky to limit sources with many boundary neighbours and/or over-fragmented.   

\subsection{Milky Way extinction correction}
In addition to the \textit{in-situ} dust screen of the extragalactic sources, dust in the Milky Way is an additional source of attenuation. To obtain rest-frame fluxes of the sources, Milky Way attenuation must therefore be corrected. 

$E(B-V)$ values from the \citet{adePlanckIntermediateResults2013} all-sky map of microwave dust haze were combined with the \citet{fitzpatrickAnalysisShapesInterstellar2007a} extinction curve and $R_{V}=3.1$ to compute the attenuation value: $A(\lambda) > 0$. The observed magnitudes, and likewise fluxes, were then corrected as 
\begin{equation}
    \mathrm{mag_{emit}}(\lambda) = \mathrm{mag_{observed}}(\lambda) - A(\lambda).
\end{equation}

\section{SED fitting} \label{sec:sedfitting}
The SED is a record of the astrophysical processes that have occurred. Using SED fitting, we extracted the photometric redshift ($z_{\mathrm{phot}}$), $\mathrm{M_{\star}}$, SFR averaged over $10$~Myr and $\mathrm{L_{AGN}}$. A two phased approach of first fitting $z_{\mathrm{phot}}$ and then $\mathrm{M_{\star}}$, SFR and $\mathrm{L_{AGN}}$ was employed to reduce degeneracies of the astrophysical parameters with distance. 

\subsection{Photometric redshifts with EAZY} \label{sec:eazy}
The $z_{\mathrm{phot}}$ was initially fitted with the widely used \textsc{EAZY} package \citep{brammerEAZYFastPublic2008}. No prior was assumed and absorption from the IGM was enabled. The undilated colour photometry was used in this initial phase of $z_{\mathrm{phot}}$ fitting.

We used the standard template set included in \textsc{EAZY}, derived from the Flexible Stellar Population Synthesis (FSPS) code \citep{conroyPropagationUncertaintiesStellar2010}, supplemented with updated templates from \citet{larsonSpectralTemplatesOptimal2023}. The updated templates include model SEDs with bluer UV colours that were found to be more representative for simulated $z\gtrsim7$ galaxies than the standard \textsc{EAZY} templates. EAZY makes an SED model of linear combinations of templates and uses least-squares fitting to minimise the residual of that model and the data. As such, the degrees of freedom, $dof$, for EAZY models is the difference between the number of data points and the number of template combinations. 

With a limited number of filters, spectral features, such as the Lyman break at $\lambda \approx 0.1216 \, \mu$m and the Balmer break at $\lambda \approx 0.3645\, \mu$m in the rest-frame, are sparsely sampled meaning that confusion can arise in the fitting and lead to a distribution of possible $z_{\mathrm{phot}}$ solutions. As such, striving to discern genuine $z \gtrsim 5$ galaxies, we used the following selection:

\begin{itemize}
    \item $\int^{25}_{5} P(z_{\mathrm{phot}}) dz \geq 0.8$ to ensure a robust probability of the high redshift solution. 
    
    \item $S/N_{\mathrm{Ly\alpha}} < 2\sigma$, where $S/N_{\mathrm{Ly\alpha}}$ is the combined signal-to-noise for fluxes blueward of the Lyman break to ensure significant absorption of Lyman continuum flux as expected given the $z_{\mathrm{phot}}$. 
    
    \item EAZY$_{hiz}$ must be better fit than EAZY$_{lowz}$ as per the $\chi^{2}$ test at the 2$\sigma$ level. 
\end{itemize}

For the last item of the selection, EAZY$_{hiz}$ refers to one EAZY fit where $z_{\mathrm{phot}} \in [0.01, 24.99]$ and EAZY$_{lowz}$ is where $z_{\mathrm{phot}} \in [0.01, 4.5]$. The use of two EAZY runs in this manner allowed us to better visualise the low $z_{\mathrm{phot}}$ solution. The critical value of $z_{\mathrm{phot}}=4.5$ was chosen as a buffer range to compute $\mathrm{M_{\star}}$, SFR and $\mathrm{L_{AGN}}$ distributions at $z_{\mathrm{phot}} \geq 5.5$. 

We used the $\chi^{2}$ distribution to omit poor fits to the data and to discern between EAZY runs. We accepted $z_{\mathrm{phot}} \geq 4.5$ provided EAZY$_{hiz}$ better represented the data compared to EAZY$_{lowz}$ with $2\sigma$ significance. The choice of $2\sigma$ corresponds to a $95\%$ confidence in the redshift, to balance adequate number statistics and fidelity. The $2\sigma$ level has been used previously in high-redshift studies \citep[e.g.,][]{bowlerLackEvolutionVery2020a,finkelsteinCensusBright85112022,adamsEPOCHSIIUltraviolet2024}.

If this test showed that both runs represented the data similarly, we considered the $dof = N_{\mathrm{templates}} - N_{\mathrm{fluxes}}$ of the two runs and used the $\chi^{2}$ distribution to test whether the model with higher $dof$ provided a justifiably better fit than the model with lower $dof$. In this case, we accepted $z_{\mathrm{phot}} \geq 4.5$ provided that the $dof$ for EAZY$_{hiz}$ was greater than that of EAZY$_{lowz}$ and that we could justify the use of the more complex model with $2\sigma$ significance. When both runs had the same $dof$, then there is no statistical difference and we did not subsume that source into our $z_{\mathrm{phot}} \geq 4.5$ sample. The number of templates used for each EAZY$_{hiz}$ and EAZY$_{lowz}$ run may be different because some templates out of the total 18 may have zero weight and therefore not participate in the fit.

\begin{figure}[h!]
    \centering
    \includegraphics[width = \columnwidth]{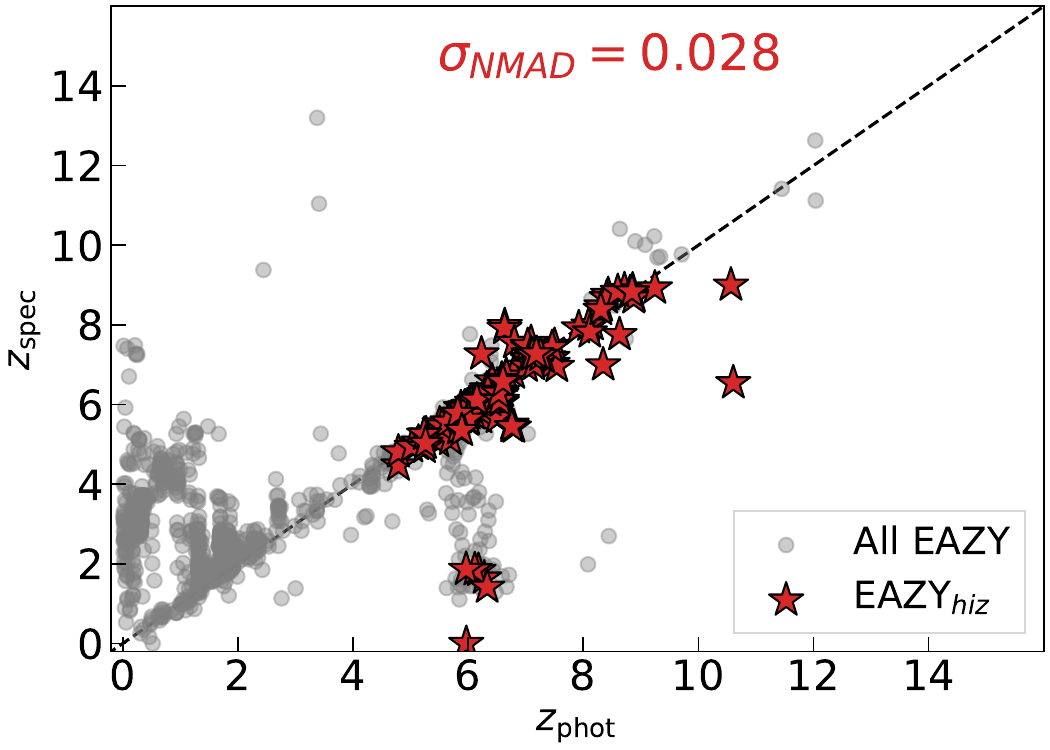}
    \caption{$z_{\mathrm{spec}}$ against $z_{\mathrm{phot}}$ from the CEERS Extended Groth Strip and JADES GOODS-South programs. Grey points are all EAZY$_{hiz}$ $z_{\mathrm{phot}}$. Red points are only the sources that survived the $z_{\mathrm{phot}} > 4.5$ selection. The dashed, grey line is the equality line.}
    \label{fig:specz_photoz}
\end{figure}

Figure~\ref{fig:specz_photoz} shows some of the $z_{\mathrm{phot}}$ against confirmed $z_{\mathrm{spec}}$ in the CEERS and JADES Goods-South fields, as determined by performing a coordinate match with an accuracy of \SI{1}{\arcsecond}. CEERS and JADES are the two greatest repositories of spectroscopic redshifts to date \citep{arrabalharoSpectroscopicConfirmationCEERS2023a,nakajimaJWSTCensusMassMetallicity2023a,bunkerJADESNIRSpecInitial2023}. Overall, we exhibit reasonable agreement against the spectroscopic results. The $z_{\mathrm{phot}}$ outliers, in grey, are most likely caused by a combination of confusion of the Lyman and Balmer breaks as per the EAZY templates or potentially AGN dominated SEDs not accounted for in the EAZY templates, complicating the redshift extraction. The agreement further improves when only the sources that pass the aforementioned $z_{\mathrm{phot}} \geq 4.5$ selection are included, as evidenced by the low $\sigma_{\mathrm{NMAD}} = 0.028$. The agreement is quantified using the normalised-median-absolute-deviation:
\begin{equation}
    \sigma_{\mathrm{NMAD}} = 1.48 \times \mathrm{med}\left( \left|\frac{\Delta z - \mathrm{med}(\Delta z)}{1 + z_{\mathrm{spec}}} \right| \right),
\end{equation}
where $\Delta z = z_{\mathrm{phot}} - z_{\mathrm{spec}}$. In total, 4585 $z_{\mathrm{phot}} > 4.5$ candidates were obtained with this selection.

Interestingly, a few $z_{\mathrm{spec}}\gtrsim 10$ sources did not survive our photometric selection, meaning that we could not, with at least $2\sigma$ confidence, rule out those sources as lower redshift interlopers, as per their $z_{\mathrm{phot}}$, with this strategy. In most of these cases, both the $z_{\mathrm{phot}}$ and $z_{\mathrm{spec}}$ are informed by the Lyman break, and since this is only a single spectral feature we could not significantly separate the EAZY$_{lowz}$ and EAZY$_{hiz}$ solutions. Where possible, we subsumed the $z_{\mathrm{spec}}$ into our sample and adopted a normal $P(z)$ centred on $z_{\mathrm{spec}}$ and $\sigma = 0.028$ for the sake of SED fitting with \textsc{ProSpect}. Note, only $\approx 5$\% of the entire sample have secure $z_{\mathrm{spec}}$ and are thus not expected to introduce a significant bias in the results. 

As a further assessment of the photometric redshift performance, we computed mock photometry with \textsc{ProSpect} in the wide-band NIRCam filters. The mock photometry were generated for 10000 synthetic galaxies with SED parameters drawn from the parameter distributions described in \citet[][see their Tab. 1]{thorneDEVILSCosmicEvolution2022}, and the redshifts were drawn from a uniform distribution between $z=0-13.5$. We also drew a further 10000 samples where we included an AGN component, for a total of 20000. The EAZY templates that we used do not accommodate dominant AGN, and so this test with artificial photometry allowed us to investigate whether the EAZY templates introduced any bias for AGN dominated galaxies.

\begin{figure*}
    \centering
    \includegraphics[width = \textwidth]{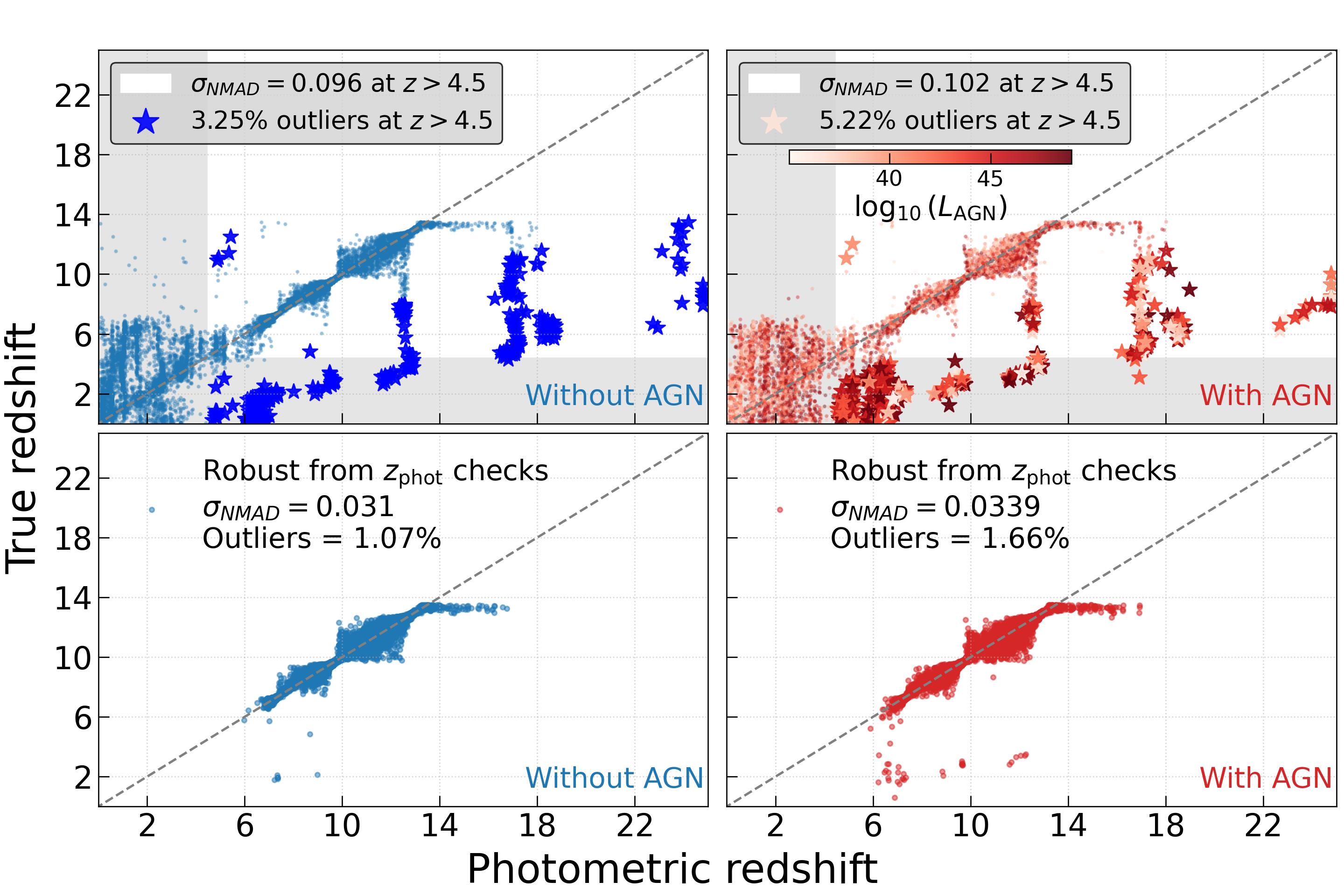}
    \caption{EAZY photometric redshift performance for 20000 synthetic photometry samples generated with \textsc{ProSpect}. \textit{Top left:} 10000 samples generated without an AGN component. The blue stars are the outliers obtained with Equation~\ref{eq:zphot_outliers}. The grey shaded regions are visual aids to de-emphasise the relevant parameter space. The grey line is the 1:1 relation. \textit{Top right:} Same as the top left panel but for the 10000 samples generated with an AGN component. In this panel, the points are colour coded by the $\mathrm{L_{AGN}}$. \textit{Bottom:} Same as the two top panels but for only for the sample that survived our  selection criteria to obtain a robust sample.}
    \label{fig:zphot_mock}
\end{figure*}

Figure~\ref{fig:zphot_mock} shows the results of our mock photometry test. We calculated the catastrophic outliers at $z_{\mathrm{phot}} > 4.5$ \citep{brammerEAZYFastPublic2008} as
\begin{equation} \label{eq:zphot_outliers}
    \mathrm{Outliers} = \frac{\Delta z}{1+z_{\mathrm{true}}} > 5\sigma_{\mathrm{{NMAD}}},
\end{equation}
finding that the absence of AGN templates in EAZY introduces $\approx 2\%$ more outliers. Comparing the mock photometry test with the $z_{\mathrm{spec}}$, we confirm that the locus of outliers about $z_{\mathrm{phot}} \approx 6$ are most likely galaxies with a dominant AGN component that are confused as dusty galaxies by the EAZY templates. One might expect that the robust sample that we obtained would be biased to low $\mathrm{L_{AGN}}$ because our set up of EAZY ostensibly struggles to recover the true $z_{\mathrm{phot}}$ for completely AGN dominated SEDs on account of template incompleteness. Fortunately, at $z>4.5$ EAZY can still recover the $z_{\mathrm{phot}}$ for AGN dominated SEDs because of the Lyman break due to neutral hydrogen absorption that is present even when there is a minimal host galaxy component. Indeed, comparing the $\mathrm{L_{AGN}}$ distribution of the 10000 mock galaxies with an included AGN to the final robust sample, we found that the distributions could not be statistically distinguished with the Kolmogorov-Smirnov test indicating a p-value, $p_{val}=0.385$.

Using the same selection criteria to obtain a robust sample, we find that $\sigma_{\mathrm{NMAD}} \approx 0.03$ for both samples with and without an AGN, and the outlier fraction is $\lesssim 1.6\%$. Hence, this exercise demonstrates the validity of our robust selection.

\subsection{\textsc{ProSpect}}\label{sec:prospect}
Having generated $P(z)$'s for our sample, we then used the SED fitting software \textsc{ProSpect} to determine the astrophysical parameters of interest: $\mathrm{M_{\star}}$, SFR, $\mathrm{L_{AGN}}$. We ran \textsc{ProSpect} on our sample of $z_{\mathrm{phot}} \geq 4.5$ candidates, using the total fluxes. A brief description of \textsc{ProSpect} is given here:

\begin{enumerate}
    \item Star formation history (SFH). \textsc{ProSpect} models the SFHs of galaxies parametrically with flexible, four-parameter models, which are essentially skewed normal distributions \citep[see equations 1-5 in][]{dsilvaGAMADEVILSCosmic2023}. While this cannot reproduce stochasticity in the true SFH, it has been shown that the SFH parametrisation in \textsc{ProSpect} reproduces the average behaviour of galaxy SFHs \citep{robothamProSpectGeneratingSpectral2020}. As such, the true power of \textsc{ProSpect} is derived from its application to large samples as presented here. The SFH is then combined with the \citet{chabrierGalacticStellarSubstellar2003} IMF and \citet{bruzualStellarPopulationSynthesis2003} stellar spectral libraries to test SEDs against the data in the generative model. For reference, \citet{bellstedtProGenyIIImpact2024} showed that, in general, the SFR and $\mathrm{M_{\star}}$ of $z\approx 0$ galaxies only marginally change (within $\sim 0.3$~dex) when adopting the \citet{kroupaInitialMassFunction2002} or \citet{laceyUnifiedMultiwavelengthModel2016} IMFs. From the high redshift perspective, \citet{harveyEPOCHSIVSED2024} showed that adopting a top-heavy IMF, which may be more characteristic of the conditions of the early Universe compared to the \citet{chabrierGalacticStellarSubstellar2003}, could reduce the $\mathrm{M_{\star}}$ of $z\approx 12$ galaxies by $\approx 0.5$~dex.
    \item Metallicity history (ZH). A key aspect of \textsc{ProSpect} is its flexible metallicity evolution linearly tied to the SFH. The motivation for this is to adequately capture the physics of chemical enrichment as a result of stellar evolution. \citet{bellstedtGalaxyMassAssembly2020} found that the metallicity evolution was crucial to reproduce the expected peak of CSFH at $z\approx1.5-2$. 
    \item AGN. \textsc{ProSpect} employs the \citet{fritz06agnmodel} templates, further expanded by \citet{feltreSmoothClumpyDust2012a}, to model the AGN component. Here, the SED of the primary source is assumed to be a combination of power laws as 
    \begin{equation}
            L(\lambda) \propto
                \begin{cases} 
                  \lambda^{1} & 10 \leq \lambda \, [\text{\AA}] \leq 500 \\
                  \lambda^{-0.2} & 500 < \lambda \, [\text{\AA}] \leq 1250 \\
                  \lambda^{-1.5} & 1250 < \lambda \, [\text{\AA}] \leq 10^{4} \\
                  \lambda^{-4} & \lambda \, [\text{\AA}] > 10^{4}. 
               \end{cases}
        \end{equation} The torus is assumed to be a flared disk of a range of dust grain sizes and with a smooth spatial distribution. A free parameter of the \citet{fritz06agnmodel} model is the optical depth of the torus and the viewing angle, which is needed to explain how the primary source is attenuated. Of use in this work is the bolometric luminosity of the AGN, $\mathrm{L_{AGN}}$, which is a free parameter in the \citet{fritz06agnmodel} model. The flexibility of this model encompasses both type-I and type-II AGN.
    \item Attenuation and reemission. \textsc{ProSpect} uses the \citet{charlotSimpleModelAbsorption2000} model to describe attenuation by dust. Here, stellar light can be attenuated by both the interstellar dust screen spread throughout the galaxy and the dust enshrouded birth clouds of the stellar nurseries. Physically, the attenuated light is then reemited in the IR and beyond, and this is prescribed in \textsc{ProSpect} with the \citet{daleTwoparameterModelInfrared2014} remission model. Note, however, that we cannot constrain the dust emission due to our photometry not probing wavelengths longer than the rest-frame optical at $z\gtrsim5$. We therefore fixed the parameters $\alpha_{SF}=1$ and $\alpha_{BC}=3$, essentially fixing the dust temperature and the wavelength of the peak of the dust emission. 
    \item Photometric redshift. Because \textsc{ProSpect} is a fully generative SED model, directly fitting $z_{\mathrm{phot}}$ with \textsc{ProSpect} mitigates template incompleteness that may be present in EAZY, allowing us to further refine the $z_{\mathrm{phot}}$. We modified \textsc{ProSpect} to make it more appropriate for photometric redshift fitting, essentially tying the maximum age of stars to the age of the Universe at the test $z_{\mathrm{phot}}$. The original design of \textsc{ProSpect} was to ingest $z_{\mathrm{spec}}$, where this would not have been an issue. The $P(z)$ of each source obtained from EAZY was used as a prior on $z_{\mathrm{phot}}$, ensuring that the desired astrophysical parameters have realistic uncertainties \citep[e.g.,][]{acquavivaSimultaneousEstimationPhotometric2015}. The IGM absorption is implemented as a function of redshift as a cumulative normal distribution\footnote{The R function is \texttt{pnorm(z,3.8,1.2)}} with mean$=3.8$ and standard deviation$=1.2$, such that the IGM is totally ionised by $z\approx6$ and all flux blue-ward of the Lyman break is absorbed and reemited as emission lines. 
\end{enumerate}
In conjunction with \textsc{ProSpect}, the optimisation software \textsc{Highlander} was used to sample the multimodal posterior distributions. \textsc{Highlander} switches between a genetic algorithm and Markov-Chain-Monte-Carlo (MCMC) to traverse local extrema and identify the global maximum \citep[e.g.,][]{thorneDeepExtragalacticVIsible2021}. This allowed us to refine the $z_{\mathrm{phot}}$ even in the presence of the EAZY $P(z)$ because, as we further elaborate on in Section~\ref{subsec:final_sample}, the \textsc{ProSpect} $z_{\mathrm{phot}}$ is not necessarily at the maximum likelihood EAZY $z_{\mathrm{phot}}$. Afforded by its Bayesian implementation, the uncertainty on $z_{\mathrm{phot}}$ is propagated into the uncertainties on the astrophysical quantities.

Correctly disentangling of the light contributions from stars and from AGN is essential when we wish to study these two processes in unison. As such, we performed two \textsc{ProSpect} runs, with and without an AGN component. Throughout the text, the two runs are referred to as $\mathrm{Pro_{Stellar+AGN}}$ and $\mathrm{Pro_{Stellar}}$. Other than the inclusion of the AGN model, the set up between the two runs was identical. 

Because $\mathrm{Pro_{Stellar+AGN}}$ introduces more free parameters, we used the Deviance Information Criterion (DIC) to judge whether the extra complexity justifiably improves the SED fit. The DIC is a measure of the quality of the fitted model, similar to the Akaike Information Criterion (AIC), but uses more information about the shape of the entire posterior distribution.
\begin{equation}
    \mathrm{DIC} = \left< D(\theta) \right> - p_{D},
\end{equation} where $\left< D(\theta) \right> = \left< -2\log(p(y|\theta)) \right>$ is the log-posterior averaged over the posterior samples and $p_{D}=\frac{1}{2} \mathrm{var}(D(\theta))$ is the effective number of parameters. The DIC penalises models with needlessly more parameters than needed to reflect the characteristics of the data. We preferred the $\mathrm{Pro_{Stellar+AGN}}$ fit provided that the difference in DIC between the two models,
\begin{equation}
    \label{eq:dic}
    \Delta \mathrm{DIC} = \mathrm{DIC_{Stellar}} - \mathrm{DIC_{Stellar+AGN}},
\end{equation} was greater than 2. The median $\Delta \mathrm{DIC}$ was $\approx 0.3$. We experimented with different $\Delta \mathrm{DIC}$ thresholds between the models of $0,5,10$, finding that the differences in the final results of the CSMH and CSFH did not change significantly outside of their error bars. As expected, for $\Delta \mathrm{DIC}$ of $5,10$ we had lower proportions of AGN, with no preferred $\mathrm{Pro_{Stellar+AGN}}$ at $z\geq 9.5$ for the most extreme $\Delta \mathrm{DIC}=10$. In any case, as explained below, we can only obtain lower and upper limits on the $\mathrm{L_{AGN}}$ densities and CAGHN with the data on hand.

In cases where the $\mathrm{Pro_{Stellar}}$ was preferred either that means that the galaxies do not host an AGN component or that the AGN component could not be detected with the data at hand and so, we considered two extremes of the $\mathrm{L_{AGN}}$ to effectively bound the possible AGN contribution. The lower limit ($\mathrm{L_{AGN}^{lolim}}$) was set such that $\mathrm{L_{AGN}} = 0$. The upper limit ($\mathrm{L_{AGN}^{uplim}}$) was set by finding the $\mathrm{L_{AGN}}$ for an unobscured AGN template from the \citet{fritz06agnmodel} models that contributed no more that $20\%$ of the flux in the summed F277W$+$F356W$+$F444W $\mathrm{Pro_{Stellar}}$ SED model fluxes. We note that SED contributions from AGN above $10\%$ usually identified with significant AGN \citep[e.g.,][]{thorneDeepExtragalacticVIsible2022}; we doubled the threshold in this work because we also assumed a $10\%$ uncertainty floor on the flux measurements that went into the SED fitting. This meant that we could constrain a realistic bound of possible AGN contribution to the galaxy SEDs where the true AGN contribution must lie somewhere in between. We may miss AGN activity with the data at hand for those galaxies that preferred the $\mathrm{Pro_{Stellar}}$ fit because we do not probe redder wavelengths to uncover a faint AGN component. 

\begin{figure*}[t!]
    \centering
    \includegraphics[width=\textwidth]{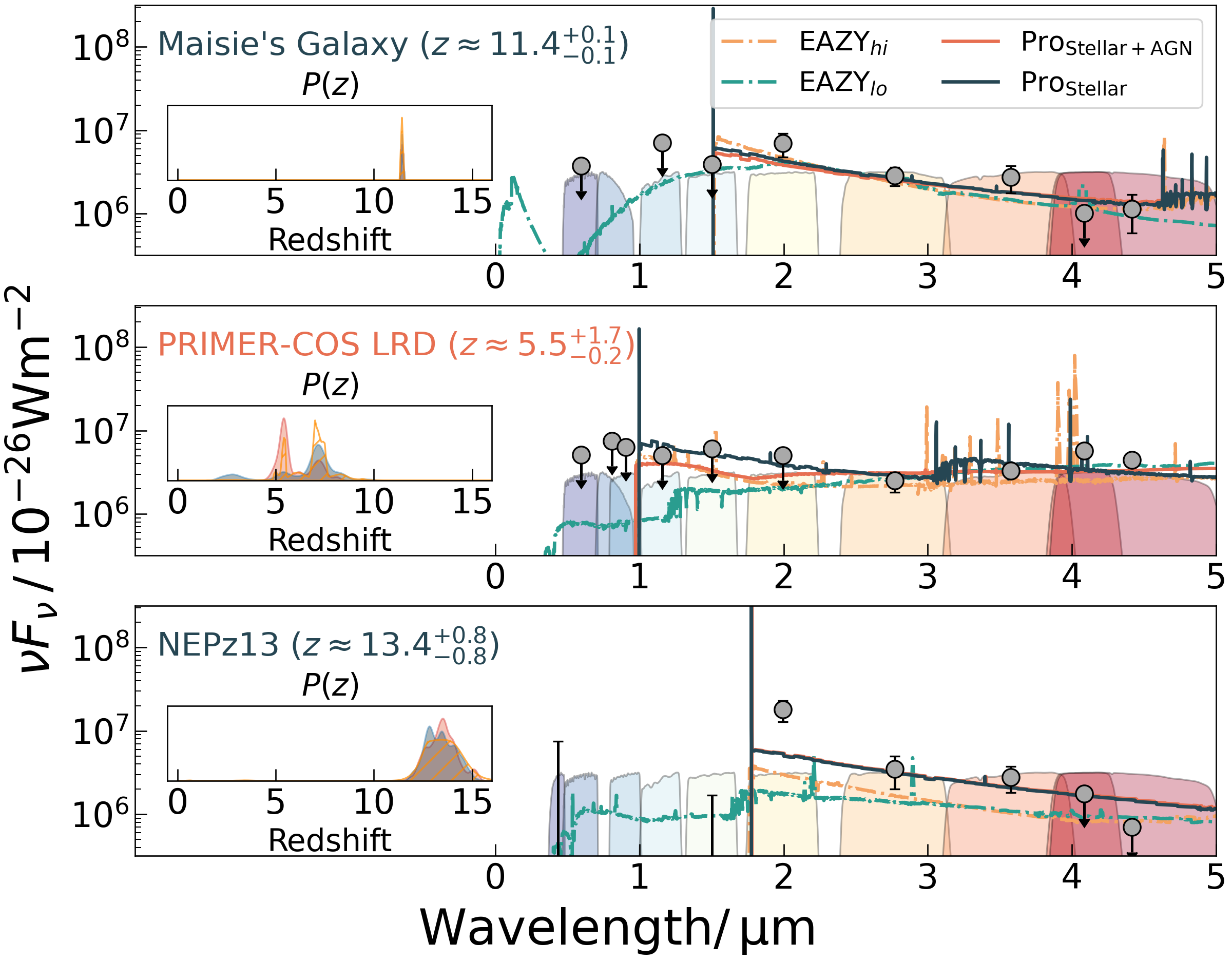}
    \caption{Select SEDs of noteworthy sources. In all panels, the grey points with error bars are our photometry. $2\sigma$ upper limits are noted with arrows. The dot-dashed orange line is the $\mathrm{EAZY_{hiz}}$ SED fit at the median of the posterior distribution (and close to the peak of the $P(z)$ as such), while the dot-dashed green line is the $\mathrm{EAZY_{lowz}}$ SED fit. The solid orange line is the $\mathrm{Pro_{Stellar+AGN}}$ SED fit while the solid blue line is the $\mathrm{Pro_{Stellar}}$ SED fit. The HST and JWST filter transmission curves are shown with rainbow colours. Each panel is inset with the $P(z)$ from $\mathrm{EAZY_{hiz}}$, $\mathrm{Pro_{Stellar+AGN}}$ and $\mathrm{Pro_{Stellar}}$, with a similar colour scheme as the SED lines. \textit{Top: } SED of Maisie's Galaxy at $z_{\mathrm{spec}} = 11.4$ \citep{finkelsteinLongTimeAgo2022}. \textit{Middle: } SED of one of the LRDs in the COSMOS field \citep{kocevskiRiseFaintRed2024a}. \textit{Bottom: } SED of a $z_{\mathrm{phot}} \approx 13$ galaxy from the NEP-TDF. In every panel, we denote with colour which \textsc{ProSpect} fit was preferred and the $z_{\mathrm{phot}}$.}
    \label{fig:select_SEDs}
\end{figure*}

Figure~\ref{fig:select_SEDs} shows SED fits to a compilation of noteworthy sources including Maisie's Galaxy at $z_{\mathrm{spec}} = 11.4$ \citep{finkelsteinLongTimeAgo2022}, a LRD in the COSMOS field \citep{kocevskiRiseFaintRed2024a} and a $z_{\mathrm{phot}} \approx 13$ galaxy from the NEP-TDF. 

\subsection{Vetting brown dwarf contaminants}
\label{subsect:browndwarf}
Absorption features of brown dwarf stars in the Milky Way can mimic similar spectral features of $z\geq3$ galaxies hosting AGN \citep{langeroodiLittleRedDots2023}. We employed the Sonora Bobcat models \citep{marleySonoraBrownDwarf2021a} to test whether the SEDs were better fit by galaxy or brown dwarf SEDs. 

We interpolated the solar metallicity models covered by the evolutionary tracks, encompassing effective temperature, $\mathrm{T_{eff} = 200-2000K}$ and surface gravity $\mathrm{\log(g / cm \, s^{-2}) = 3-5.2504}$. We also included a normalisation parameter to accommodate distance. We propagated the template spectra through both the same HST-JWST filters used for \textsc{ProSpect} and fitted the brown dwarf models in a similar manner with \textsc{Highlander}. We preferred brown dwarf models provided that $\Delta \mathrm{DIC}$ between the preferred \textsc{ProSpect} model was $>2$ and provided that the effective radius in the detection image was $\leq$ \SI{0.08}{\arcsecond}, which is approximately the full-width-half-maximum of the PSF in the F277W filter. The motivation by this additional morphological constraint is to restrict our brown dwarf search to only unresolved point-like sources. Out of the 4585 candidates, 21 of those were identified as brown dwarf stars. As our fields are mostly pointed out of the Milky Way disk, we expect that brown dwarf contamination in our final sample is negligible if not null after filtering these 21 objects.

\subsection{Final sample} \label{subsec:final_sample}
Interestingly, $1132$ out of the $4585$ galaxies with $z_{\mathrm{phot}} \geq 4.5$ as identified by EAZY ended up with lower $z_{\mathrm{phot}}$ when fitted with \textsc{ProSpect}, even when passing in the EAZY $P(z)$. We refitted those $1132$ mismatched galaxies but expanded the number of iterations of both the genetic algorithm and the MCMC steps by a factor of three, in case the mismatch in the $z_{\mathrm{phot}}$ was due to the optimiser settling on a local maximum of the posterior distribution instead of the desired global maximum. We found that $365$ out of those $1132$ then had $z_{\mathrm{phot}} \geq 4.5$, leaving us with a sample of $3751$ robust high redshift galaxies.

A potential reason for this $z_{\mathrm{phot}}$ mismatch may be template incompleteness as the generative model in \textsc{ProSpect} allows for more flexibility in the model SEDs to compare against the data in contrast to the 18 SED templates used in EAZY. The 18 fixed templates used for our EAZY fitting, 12 from the default FSPS templates \citep{conroyPropagationUncertaintiesStellar2010} and 6 from the expanded set from \citet{larsonSpectralTemplatesOptimal2023}, which were added to better encapsulate the SEDs of $z>6-7$ galaxies, appear to overvalue the $z_{\mathrm{phot}} > 4.5$ in some cases, while the flexible \textsc{ProSpect} instead settles on the lower redshift solution. This suggests that the 18 EAZY templates struggle to accommodate the diversity of galaxy SED shapes at $z<4.5$ compared to the fully generative \textsc{ProSpect}.

\begin{figure}
    \centering
    \includegraphics[width=\columnwidth]{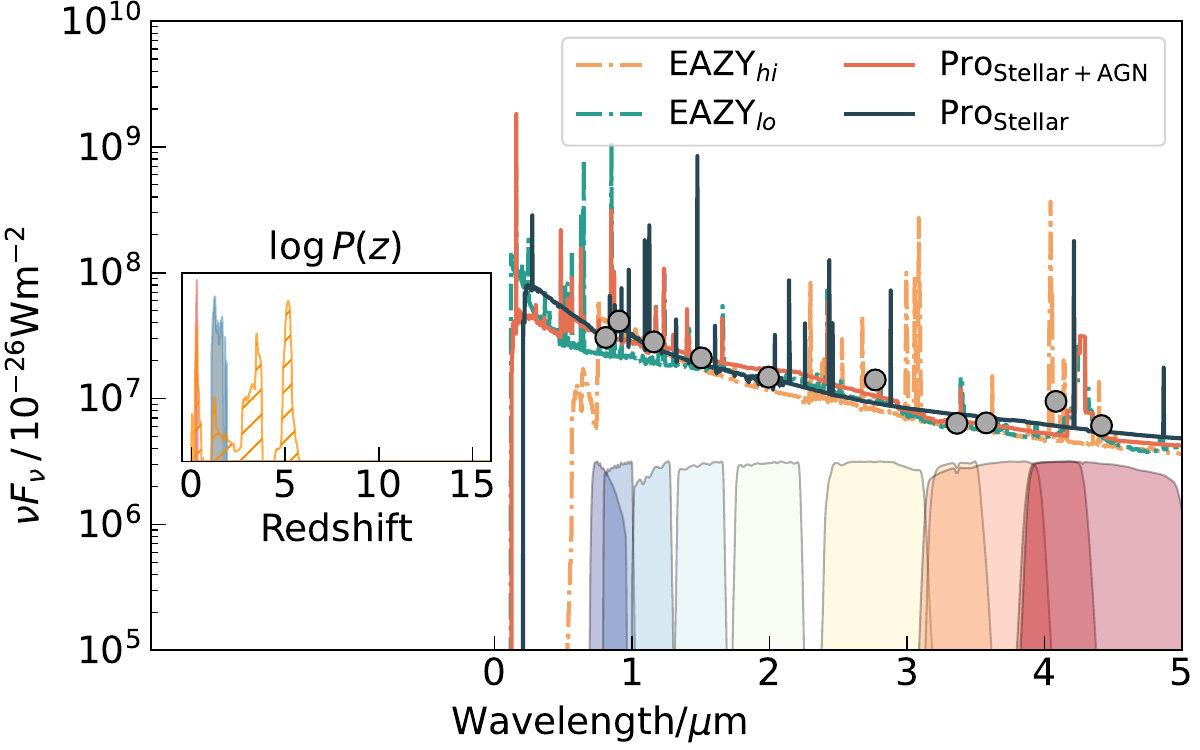}
    \caption{SED of galaxy 101396 from the JADES field in our photometric catalogue. The grey points with error bars show the photometry. The dot-dashed orange line is the $\mathrm{EAZY_{hiz}}$ SED fit while the dot-dashed green line is the $\mathrm{EAZY_{lowz}}$ SED fit. The solid orange line is the $\mathrm{Pro_{Stellar+AGN}}$ SED fit while the solid blue line is the $\mathrm{Pro_{Stellar}}$ SED fit. The HST and JWST filter transmission curves are shown with rainbow colours. The inset panel is the $\log (P(z))$ from $\mathrm{EAZY_{hiz}}$, $\mathrm{Pro_{Stellar+AGN}}$ and $\mathrm{Pro_{Stellar}}$, with a similar colour scheme as the SED lines.}
    \label{fig:galaxy_sed1}
\end{figure}

Figure~\ref{fig:galaxy_sed1} shows an example SED of a galaxy in the JADES field where the $z_{\mathrm{phot}}$ determined from \textsc{ProSpect} did not agree with the initial $z_{\mathrm{phot}} \geq 4.5$ determination from EAZY. Comparing the $P(z)$'s, it can be seen that the increased flexibility with \textsc{ProSpect} prefers the local maxima from the EAZY $P(z)$, in contrast to the global maximum at $z_{\mathrm{zphot}} \approx 5.0$. In fact, out of the $4585$ EAZY candidates, $\approx 20\%$ of those ended up with $z_{\mathrm{phot}} < 4.5$. Part of initial EAZY robust-sample selection required that at least $80\%$ of the probability mass of the $P(z)$ be $z_{\mathrm{phot}} > 4.5$, suggesting some consistency with the fraction of $z_{\mathrm{phot}}$ mismatches we found between SED fitting codes.

In the subsequent analyses of this paper, we consider the robust sample of $3751$, being the galaxies whose  $z_{\mathrm{phot}} > 4.5$ survived the two phased filtration through EAZY and \textsc{ProSpect}. 

\begin{figure}[h!]
    \centering
    \includegraphics[width=\columnwidth]{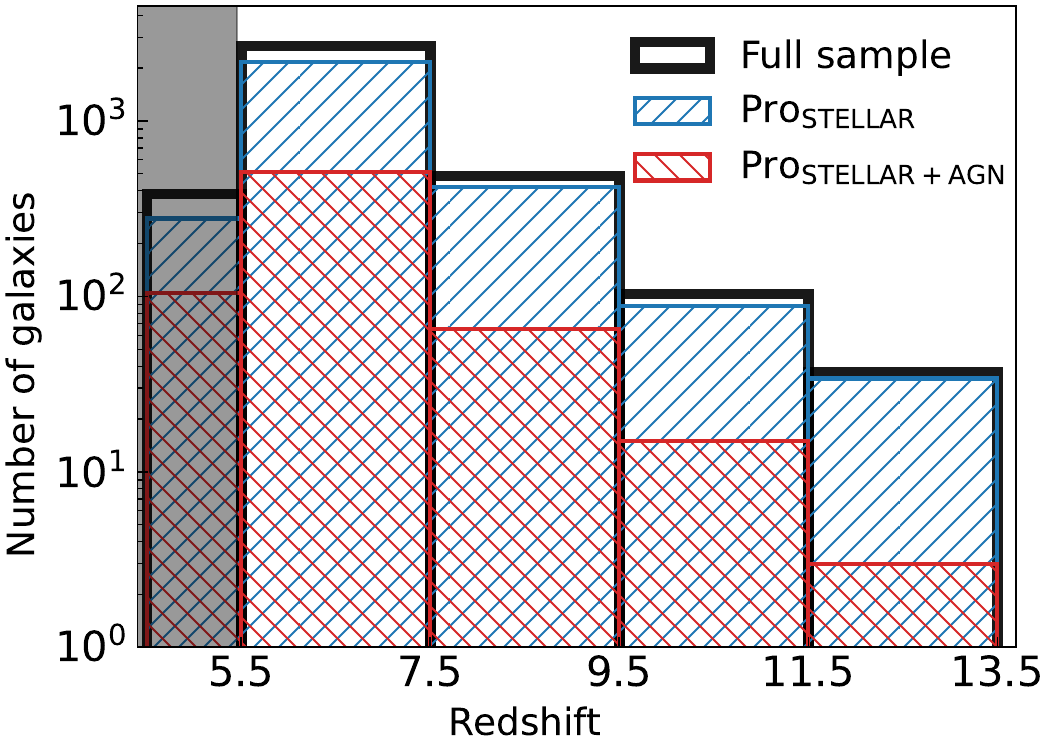}
    \caption{Redshift distribution of $3751$ galaxies. The black histogram is the distribution for the entire sample. The blue hatched histogram is the distribution for the subset of the sample better fit with $\mathrm{Pro_{Stellar}}$, while the red hatched histogram is the same for $\mathrm{Pro_{Stellar+AGN}}$. The grey band is the redshift buffer range of $z_{\mathrm{phot}} = 4.5-5.5$ used when deciding between $\mathrm{EAZY_{lowz}}$ and $\mathrm{EAZY_{hiz}}$.}
    \label{fig:z_distribution}
\end{figure}
Figure~\ref{fig:z_distribution} shows the redshift distribution of the final sample. 

\section{Results} \label{sec:results}
With the final sample of redshifts and astrophysical quantities, we explored the interface of stellar mass assembly and SMBH growth in four bins of redshift from $z=5.5$ to $z=13.5$. The upper edge of $z=13.5$ was chosen because extreme emission line galaxies at $z\approx 4-5$ can mimic the colours of $z\gtrsim 13.5$ galaxies, introducing a pathological source of systematic uncertainty in constraining the redshift from photometry alone \citep{zavalaDustyStarburstsMasquerading2023a,arrabalharoConfirmationRefutationVery2023}. 

\subsection{Effective volumes}
Due to sensitivity limits of astronomical instruments, inherently faint galaxies will be missed from sample selections. This causes an unphysical turn-over in the histogram toward low luminosities, known as the Malmquist bias. 

We computed magnitude completeness functions to account for the implicit selection imposed by our \textsc{ProFound} source finding parameters and the sensitivity limits of the data. In each of our fields a representative $12.25$ amin$^2$ area was chosen where source injection-recovery simulations were performed \citep[e.g.,][]{leethochawalitQuantitativeAssessmentCompleteness2022a}. The magnitude-completeness was calculated as 
\begin{equation}
    C(\mathrm{mag}_{i}) = \frac{N(\mathrm{mag_{rec}}_{,i}) - N(\mathrm{FP}_{i})}{N(\mathrm{mag_{in}}_{,i})},
\end{equation}
where $C(\mathrm{mag}_{i})$ is the completeness value in the $i^{th}$ magnitude bin, $N(\mathrm{mag_{rec}}_{,i})$ is the number of recovered magnitudes, $N(\mathrm{mag_{in}}_{,i})$ is the number of injected magnitudes and $N(\mathrm{FP}_{i})$ is the number of false-positives. $N(\mathrm{FP}_{i})$ was calculated by performing a coordinate match of the recovered sources against the injected ones and finding those that were unmatched. We used the 2D profile fitting and generation code \textsc{ProFit} \citep{robothamProFitBayesianProfile2017} to generate 200 fake sources. The redshift distribution was assumed to be uniform  and the distributions for effective radius, axial ratio and S\'ersic index were adapted from observations of $z\gtrsim 3$ galaxies observed by JWST \citep[e.g.,][]{kartaltepeCEERSKeyPaper2023a,ferreiraJWSTHubbleSequence2023}. The exact same \textsc{ProFound} set up was then employed to recover the injected sources. This process of source injection and recovery was repeated 50 times to obtain a Monte-Carlo uncertainty on the completeness values, and the completeness curve was fitted with the logistic function \citep[e.g.,][]{leethochawalitQuantitativeAssessmentCompleteness2022a}. 

\begin{figure}[h!]
    \centering
    \includegraphics[width=\columnwidth]{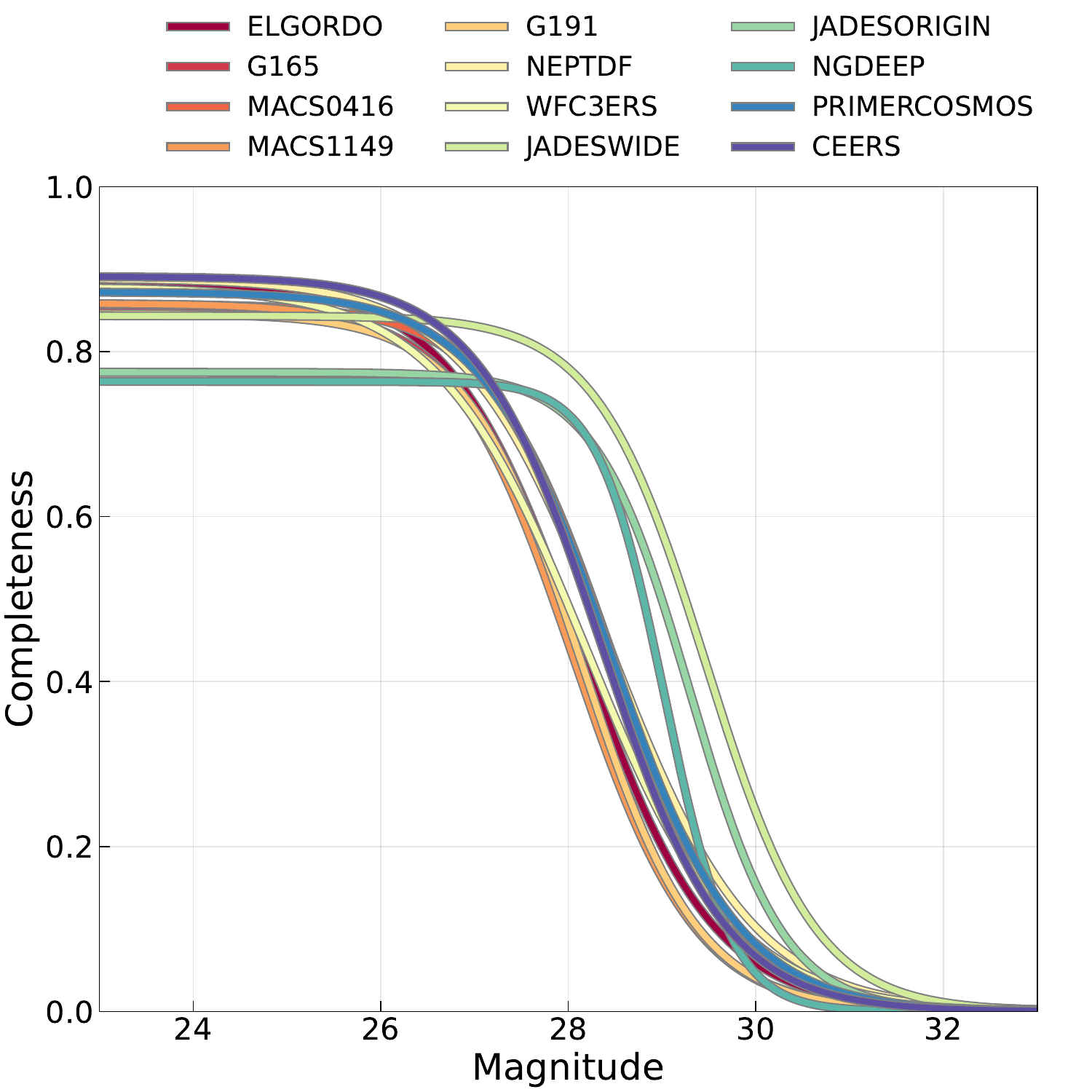}
    \caption{Magnitude completeness for all fields used in this work, as described in the figure legend. Magnitudes are obtained from the combined F277W, F356W and F444W used in the detection image.}
    \label{fig:magnitude_completeness}
\end{figure}
Figure~\ref{fig:magnitude_completeness} shows the magnitude-completeness fitted logistic functions for each of our fields. We hit a maximum of $\approx 80$\% completeness toward bright magnitudes as those bright galaxies are fragmented in the source detection process. The reader is reminded that the source detection pipeline was optimised to extract the faint sources, where overdeblending is not an issue. In most fields we are $\approx 50$\% complete up to $28$~mag, except for JADES and NGDEEP that both enjoy superior depth with $\approx 50$\% completeness up to $29$~mag. 

The effective volume was then calculated for each galaxy of observed magnitude, $m$, in the final sample as 
\begin{equation}
    V_{\mathrm{eff}}(m) = \int^{z_{max}}_{z_{min}} \Omega_{\mathrm{tot}}\frac{dV}{dzd\Omega}C(m) dz,
\end{equation}
where $C(m)$ is the magnitude completeness function, $\Omega_{\mathrm{tot}}$ is our total effective survey area, $\frac{dV}{dzd\Omega}$ is the differential comoving volume element and $z_{max}/z_{min}$ are the upper/lower limits of the redshift bins.

\subsection{Stellar mass, SFR, AGN luminosity completeness}
The completeness limit of $\mathrm{M_{\star}}$, SFR, $\mathrm{L_{AGN}}$ is necessary when constructing their distribution functions as galaxies exhibit a range of mass-to-light ratios that can introduce biases into the distribution \citep{weigelStellarMassFunctions2016}. For each galaxy, we computed 

\begin{multline}
    \log_{10}(\mathrm{M_{lim}}_{,i}) =\\ \log_{10}(\mathrm{M_{\star}}_{,i})+0.4\times (\mathrm{mag}_{i} - \mathrm{mag_{\lim}})
\end{multline}

where $\mathrm{M_{lim}}_{,i}$ is the limiting mass for which we could detect the $i^{th}$ galaxy given its observed $\mathrm{M_{\star}}_{,i}$ and $\mathrm{mag}_{i}$ as per the survey $\mathrm{mag_{lim}}$, presented in Table~\ref{tab:survey_limits}. Then, in each bin of redshift, we calculated the \nth{95} percentile of the limiting masses and for galaxies with greater than $50$\% magnitude completeness, allowing us to compute a mass completeness limit as a function of redshift. For consistency, a similar calculation was performed for the SFR and $\mathrm{L_{AGN}}$. 

\subsection{Distribution functions} \label{sec:distributions}
The census of $\mathrm{M_{\star}}$, SFR, $\mathrm{L_{AGN}}$ was obtained by computing essentially the histogram of those quantities per unit comoving volume. In each redshift bin and for each quantity, we fitted a double power law function to the distributions:
\begin{equation}
    \label{eq:dpl}
    \phi(x) = \frac{\phi^{*}}{(x/x^{*})^{-\alpha} + (x/x^{*})^{-\beta}},
\end{equation}
where $\phi^{*}$ is a normalization term, $\alpha$ is the slope of the function for $x < x^{*}$, and $\beta$ is the slope for $x > x^{*}$. 

Priors were also used to guide the fit, and the same priors used in every redshift bin. 
\begin{table*}
    \centering
    \begin{tabular}{c c c c c}
         Quantity & $\log_{10}(\Phi^{*})$ & $\alpha$ & $\beta$ & $\log_{10}(x^{*})$ \\
        \hline
        \hline
        $\mathrm{M_{\star}}$ & $\mathcal{N}(-2.0, 5.0^{2})$ & $\mathcal{N}(-0.5, 0.2^{2})$ & $\mathcal{N}(-2.0, 3.0^{2})$ & $\mathcal{N}(8.5, 1.0^{2})$  \\
        \hline
        $\mathrm{SFR}$ & $\mathcal{N}(-2.0, 5.0^{2})$ & $\mathcal{N}(-0.5, 0.2^{2})$ & $\mathcal{N}(-2.0, 3.0^{2})$ & $\mathcal{N}(1.0, 1.0^{2})$  \\
        \hline
        $\mathrm{L_{AGN}^{uplim}}$ & $\mathcal{N}(-4.0, 5.0^{2})$ & $\mathcal{N}(-0.5, 0.2^{2})$ & $\mathcal{N}(-2.0, 1.0^{2})$ & $\mathcal{N}(45.0, 1.0^{2})$ \\ 
        \hline
    \end{tabular}
    \caption{Prior distributions used for fitting the double power law functions to the SMF, SFRF and AGNLF. We denote the normal distribution with mean, $\mu$, and variance, $\sigma^{2}$, as $\mathcal{N}(\mu, \sigma^{2})$.} $\phi^{*}$ is a normalization term, $\alpha$ is the slope of the function for $x < x^{*}$, and $\beta$ is the slope for $x > x^{*}$.
    \label{tab:priors}
\end{table*}
A tight prior on the faint-end slope, $\alpha$, was employed for the SMF and SFRF, being informed by results from the $z\lesssim5$ Universe \citep[e.g.,][]{thorneDeepExtragalacticVIsible2021}. 

Solutions where $\alpha \leq -1$ produce divergent measurements for the CSMH/CSFH, under the assumption that the distributions extend monotonically as a power law toward low $\mathrm{M_{\star}}$ and SFR (see Equation~\ref{eq:rho_integral}). This situation would be avoided if the SMF/SFRF flattens or turns over at low values. With this data, we do not probe faint enough galaxies to constrain this behaviour. However, HST observations of faint galaxies magnified by lensing clusters show that the UV luminosity function at $z\gtrsim 6$ seems to extend as a power law even down to $\mathrm{M_{UV}} \approx -13\to -15$ \citep{atekExtremeFaintEnd2018,bouwens29GalaxiesMagnified2022}. With respect to the $\mathrm{M_{\star}}$ and $\mathrm{SFR}$, solutions where $\alpha \geq 0$ are suggestive of incompleteness in the domain that we probe with this data. 

The double power law was fitted using the Bayesian fitting software, \textsc{dftools} \citep{obreschkowEddingtonDemonInferring2018}. \textsc{dftools} employs bin-free fitting and correct treatment of the Eddington bias as a result of the uncertainties on the quantities and the shape of the distribution functions. Instead of treating the astrophysical quantities as point-estimates, they are treated as Gaussian probability distributions with the mode given by the measured value and width given by the uncertainty range from the inferred posterior distributions \citep[for further details, the reader is directed to Eqs. 6-10 in][]{obreschkowEddingtonDemonInferring2018}.

A comparison of the double power law and the Schechter function \citep{schechterAnalyticExpressionLuminosity1976a} showed that the double power law was the preferred model to describe the SMF, SFRF and AGNLF as per the posterior odds reported by \textsc{dftools}. In general, for all distributions there is excess number density toward large values that could not be adequately encapsulated by the exponential decline of the Schechter function beyond the characteristic knee. 

To compute the AGNLF we used the $\mathrm{L_{AGN}}$ of the $\mathrm{Pro_{Stellar+AGN}}$ preferred fits as is, whilst we used the $\mathrm{L_{AGN}^{uplim}}$ from the $\mathrm{Pro_{Stellar}}$ preferred SED fits, as described in Section~\ref{sec:sedfitting}. The AGNLF fit that we here report is thus an upper limit. We could not perform an equivalent fit for only the $\mathrm{Pro_{Stellar+AGN}}$ preferred SED fits due to the dearth, of data points, especially in the highest redshift bin. For those, we instead show the binned quantities. Note, that this situation will improve with increased survey area, to better sample bright $\mathrm{L_{AGN}}$, and redder bands to better constrain fainter $\mathrm{L_{AGN}}$ from the SED fitting. The upper and lower limits converge at the bright end of the AGNLF implying direct detections in that luminosity regime. We only fit for $\mathrm{L_{AGN}} < 10^{46} \, \mathrm{erg \, s^{-1}}$ because many of those objects may have unphysical $\mathrm{L_{AGN}}$ (we elaborate on this in Section~\ref{sect:sed_model}). 

\begin{figure*}[h]
    \centering
    \includegraphics[width=\textwidth]{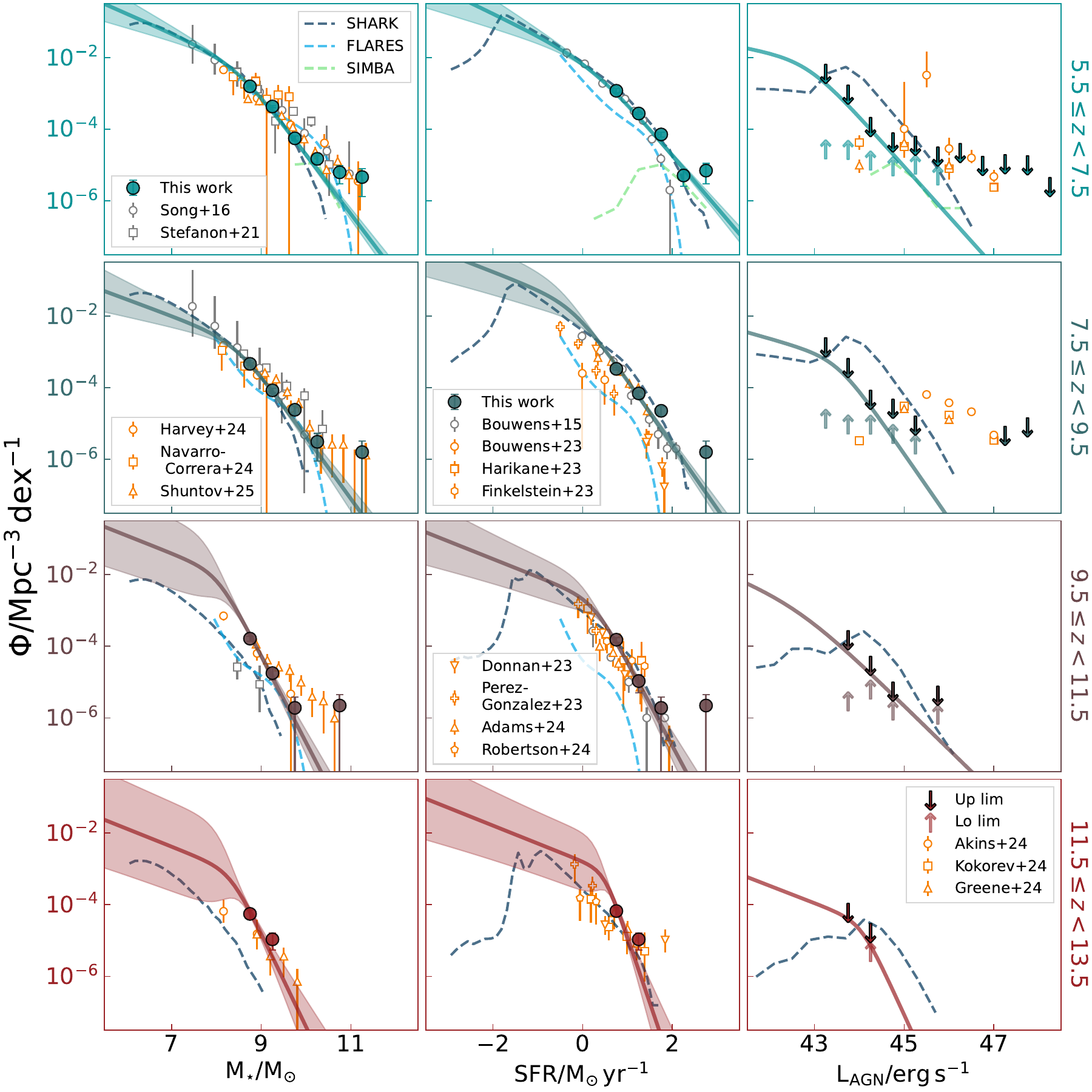}
    \caption{Number density distributions of $\mathrm{M_{\star}}$ (\textit{left column}), $\mathrm{SFR}$ (\textit{middle column}) and $\mathrm{L_{AGN}}$ (\textit{right column}), computed at $z=5.5-7.5$ (\textit{top row}), $z=7.5-9.5$, $z=9.5-11.5$ and $z=11.5-13.5$ (\textit{bottom row}). In each panel, the double power law function fits and $1\sigma$ uncertainty ranges are shown as solid lines and shaded regions, with the colour transitioning from blue to red at $z = 5.5 \to 13.5$. Filled points and $1\sigma$ error bars show the binned $1/V_{\mathrm{eff}}$ distributions (some of the error bars are smaller than the points). For the AGNLF, the downward facing arrows show the upper limit $\mathrm{L_{AGN}}$ and the upward facing arrows show the equivalent lower limit, explained in Section~\ref{sec:prospect}. We only fitted the $\mathrm{L_{AGN}}$ upper limits and restricted the fit to $\mathrm{L_{AGN}} < 10^{46} \, \mathrm{erg \, s^{-1}}$. Comparisons to the literature were obtained from \citet{songEvolutionGalaxyStellar2016,stefanonGalaxyStellarMass2021,harveyEPOCHSIVSED2024,navarro-carreraConstraintsFaintEnd2023,shuntovCOSMOSWebStellarMass2025} for the SMF, \citet{bouwensUVLuminosityFunctions2015a,bouwensEvolutionUVLF2023a,harikaneComprehensiveStudyGalaxies2023, finkelsteinCEERSKeyPaper2023a, donnanEvolutionGalaxyUV2023,perez-gonzalezLife30Probing2023a,adamsEPOCHSIIUltraviolet2024,robertsonEarliestGalaxiesJADES2024} for the SFRF and \citet{akinsCOSMOSWebOverabundancePhysical2024,kokorevCensusPhotometricallySelected2024,greeneUNCOVERSpectroscopyConfirms2024} for the AGNLF. The SFRF literature comparisons are UV luminosity number density measurements converted to SFR as $\mathrm{SFR = 0.63 \times 1.28 \times 10^{-28} \times L_{UV}}$, where $\mathrm{L_{UV}}$ is the UV luminosity at \SI{1500}{\angstrom} in units of $\mathrm{erg \, s^{-1} \, Hz^{-1}}$ and the factor of $0.63$ is the relevant conversion of a Salpeter to Chabrier IMF. Results from both hydrodynamical \citep[e.g., \textsc{simba}, \textsc{Flares},][]{daveSIMBACosmologicalSimulations2019,lovellFirstLightReionisation2020,vijayanFirstLightReionization2021a} and semi-analytic \citep[e.g., \textsc{Shark},][]{lagosSharkIntroducingOpen2018,lagosQuenchingMassiveGalaxies2024b} models are shown as dashed lines.
    }
    \label{fig:all_distributions}
\end{figure*}

Figure~\ref{fig:all_distributions} shows the fitted distributions. For comparison, we also calculated the $1/V_\mathrm{eff}$ distributions in bins of $0.5$~dex. Overall, our results exhibit broad agreement in comparison to previous measurements in the literature. We wish to highlight that these astrophysical quantities have been inferred simultaneously with \textsc{ProSpect}, and that we have endeavoured to obtain a representative census of $\mathrm{M_{\star}}$, SFR and $\mathrm{L_{AGN}}$. Subtle departures are most likely attributed to differences in methodologies as elucidated below.

The SMFs from \citet{songEvolutionGalaxyStellar2016} were determined by convolving the UV luminosity function from \citet{finkelsteinEvolutionGalaxyRestFrame2015} with the $\mathrm{M_{UV}-M_{\star}}$. On the other hand, \citet{stefanonGalaxyStellarMass2021} compute the SMF, but the calculation of $\mathrm{M_{\star}}$ is not always from full SED fitting, especially for the galaxies without a significant Spitzer/IRAC detection (e.g., their Fig. 4). The SMFs from \citet{navarro-carreraConstraintsFaintEnd2023,harveyEPOCHSIVSED2024,shuntovCOSMOSWebStellarMass2025} are most appropriate comparison data sets as they use JWST photometry; although, they used different SED fitting codes. 

All of the SFRFs against which we compare \citep{bouwensUVLuminosityFunctions2015a,bouwensEvolutionUVLF2023a,harikaneComprehensiveStudyGalaxies2023,donnanEvolutionGalaxyUV2023, perez-gonzalezLife30Probing2023a, robertsonEarliestGalaxiesJADES2024, adamsEPOCHSIIUltraviolet2024} were converted from computed UV luminosity functions as $\mathrm{SFR = 0.63 \times 1.28 \times 10^{-28} \times L_{UV}}$, where $\mathrm{L_{UV}}$ is the UV luminosity at \SI{1500}{\angstrom} in units of $\mathrm{erg \, s^{-1} \, Hz^{-1}}$ and the factor of $0.63$ is the relevant conversion of a Salpeter to Chabrier IMF \citep{madaudickinson2014}. This $\mathrm{L_{UV}}$ to SFR conversion is a function of at least metallicity and the shape of the SFH, and is therefore a potential source of systematic difference. 

A likely source of systematic uncertainty driving the difference between our AGNLFs and the literature \citep{akinsCOSMOSWebOverabundancePhysical2024,kokorevCensusPhotometricallySelected2024,greeneUNCOVERSpectroscopyConfirms2024} is that the literature AGNLFs are calculated for photometrically selected compact and red sources, many of which are LRDs. We did not apply any such selection to our sample. The largest difference between our results and the literature is observed in the $7.5 \leq z < 9.5$ bin where our results tend to be $\approx 1$~dex below the literature. Indeed, this difference is similar to the fraction of sources in that redshift bin that were better fit with $\mathrm{Pro_{Stellar+AGN}}$ compared to $\mathrm{Pro_{Stellar}}$, $\log_{10}(65/484) \approx -0.9$, as presented in Figure~\ref{fig:z_distribution}. 

The SMF and SFRF show general agreement with both hydrodynamical \citep[e.g., \textsc{simba}, \textsc{Flares}, ][]{daveSIMBACosmologicalSimulations2019,lovellFirstLightReionisation2020,vijayanFirstLightReionization2021a} and semi-analytic \citep[e.g., \textsc{Shark},][]{lagosQuenchingMassiveGalaxies2024b} models\footnote{Every simulation shown is the `fiducial' model as reported in their introductory papers.} up to $z\approx 9.5$. With the exception of \textsc{Shark}, at $z>9.5$ the simulations show a dearth of resolved galaxies, $\mathrm{M_{\star} \lesssim 10^{8.5}}$, making it difficult to compare against these observations. 

\textsc{Shark}, however, employs the P-Millenium dark-matter only simulation \citep{baugh19pmill} as the basis for the semi-analytic model to achieve better resolution of low mass galaxies and smaller time increments between snapshots compared to the first version presented in \citet{lagosSharkIntroducingOpen2018}. The SFRFs at all redshifts are generally in agreement with \textsc{Shark}. The SMFs mostly agree in terms of their shape at all redshifts, except that the normalization is $\gtrsim 0.5$~dex for our results in the two highest redshift bins. The AGNLF in \textsc{Shark} is in general $\approx 1$~dex higher in normalization with a steeper bright-end slope compared to our results in all redshift bins. 

We remark that the simulations themselves show differences that are manifested from their respective implementation of the physical model, despite converging upon the $z\approx 0$ SMF. Recently, \citet{Lagos24b} showed that the different implementation of AGN feedback in these simulations (plus a few others analysed in that paper) led to galaxies quenching very differently at $z\gtrsim 5$, leading to different formation histories and connections with SMBH properties. While beyond the scope of this work, JWST is well suited to start distinguishing between models in the high redshift frontier \citep[e.g.,][]{shenHighredshiftPredictionsIllustrisTNG2022,dsilvaUnveilingMainSequence2023b}. 

\subsection{Cosmic SMH/SFRH/AGNH}
\label{subsec:csfhcagnh}
The cosmic evolution of $\mathrm{M_{\star}}$, SFR, $\mathrm{L_{AGN}}$ summarizes the history of baryon conversion and astrophysical photon production from the present day to within the first few hundreds of millions of years after the Big Bang. The cosmic evolution is found by taking the distributions in each redshift bin presented in Section~\ref{sec:distributions} and integrating them as follows:
\begin{equation}
    \label{eq:rho_integral}
    \rho_{X} = \int^{X_{hi}}_{X_{lo}} X \times \Phi(X) dX,
\end{equation}
where $\Phi(X)$ is the number density distribution per unit comoving volume of the astrophysical quantity $X$. 

\begin{figure}
    \centering
    \includegraphics[width=\columnwidth]{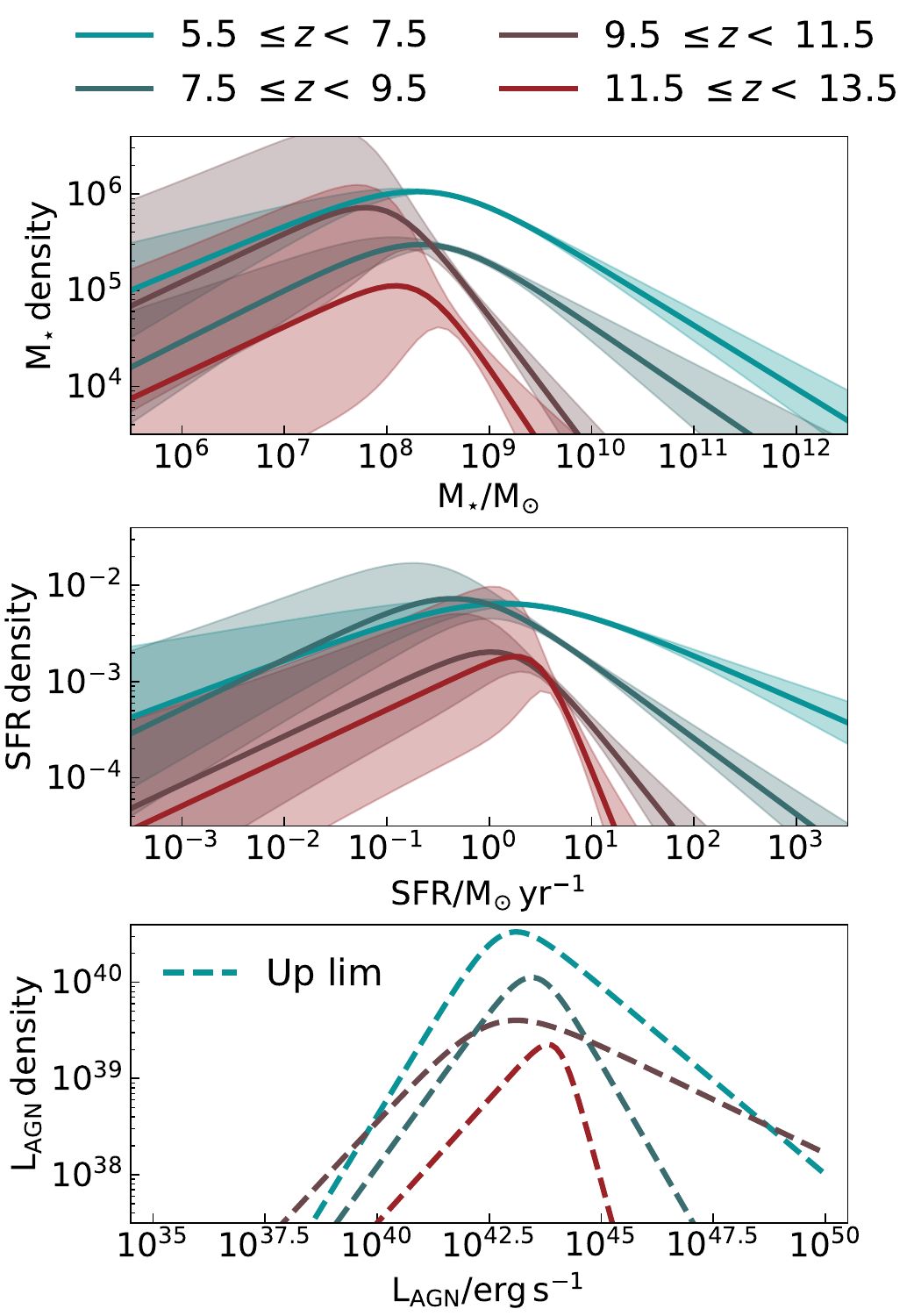}
    \caption{Density distributions of $\mathrm{M_{\star}}$ (\textit{top}), $\mathrm{SFR}$ (\textit{middle}) and $\mathrm{L_{AGN}}$ (\textit{bottom}), obtained from Equation~\ref{eq:rho_integral} and using the fitted double power law functions from Figure~\ref{fig:all_distributions}. A similar colour scheme is used here as in Figure~\ref{fig:all_distributions}.
    }
    \label{fig:all_densities}
\end{figure}

Numerically, the integration limits should subtend to infinity for a complete census of $\mathrm{M_{\star}}$, SFR, $\mathrm{L_{AGN}}$. Physically, however, these quantities must vanish at some finite lower and upper limit. We integrated over the domain shown in Figure~\ref{fig:all_densities}. The integration limits for $\mathrm{M_{\star}}$ SFR were chosen to encompass $\approx 7$ decades, while the integration limits for $\mathrm{L_{AGN}}$ were chosen to cover the bounds of $\mathrm{L_{AGN}}$ imposed by \textsc{ProSpect} from our SED fitting \citep[see Tab. 1 in][]{thorneDeepExtragalacticVIsible2022}. As can be seen in Figure~\ref{fig:all_densities}, the density distributions for $\mathrm{M_{\star}}$, SFR and $\mathrm{L_{AGN}}$, the integrands of Equation~\ref{eq:rho_integral}, are unimodal and concave-downward meaning that the integrals over the entire domain are convergent. Contributions to the integral are ever diminishing for values of the domain far beyond the peak. 

In the highest redshift bin, we also calculated the direct sum of each quantity per unit comoving volume since the integrals rely on the most significant extrapolations of the fitted distributions. We did this in every redshift bin for the $\mathrm{L_{AGN}}$ as our calculation for the CAGNH lower bound, using the $\mathrm{L_{AGN}^{lolim}}$ for the $\mathrm{Pro_{Stellar}}$ preferred fits. We omitted $\mathrm{L_{AGN} > 10^{46} \, erg \, s^{-1}}$, as done for the distribution function fitting, because it is not yet clear if the light from these galaxies is truly dominated by AGN. The SEDs of these galaxies showed the characteristic `v-shape' of little red dots (LRDs), with dominant AGN emission from our \textsc{ProSpect} fit. Without redder wavelengths, we may be overestimating the AGN contribution in these objects. Including $\mathrm{L_{AGN} > 10^{46} \, erg \, s^{-1}}$ produces $\approx 2$~dex more $\mathrm{L_{AGN}}$ density from this direct sum.

\begin{figure*}
    \centering
    \includegraphics[width=\linewidth]{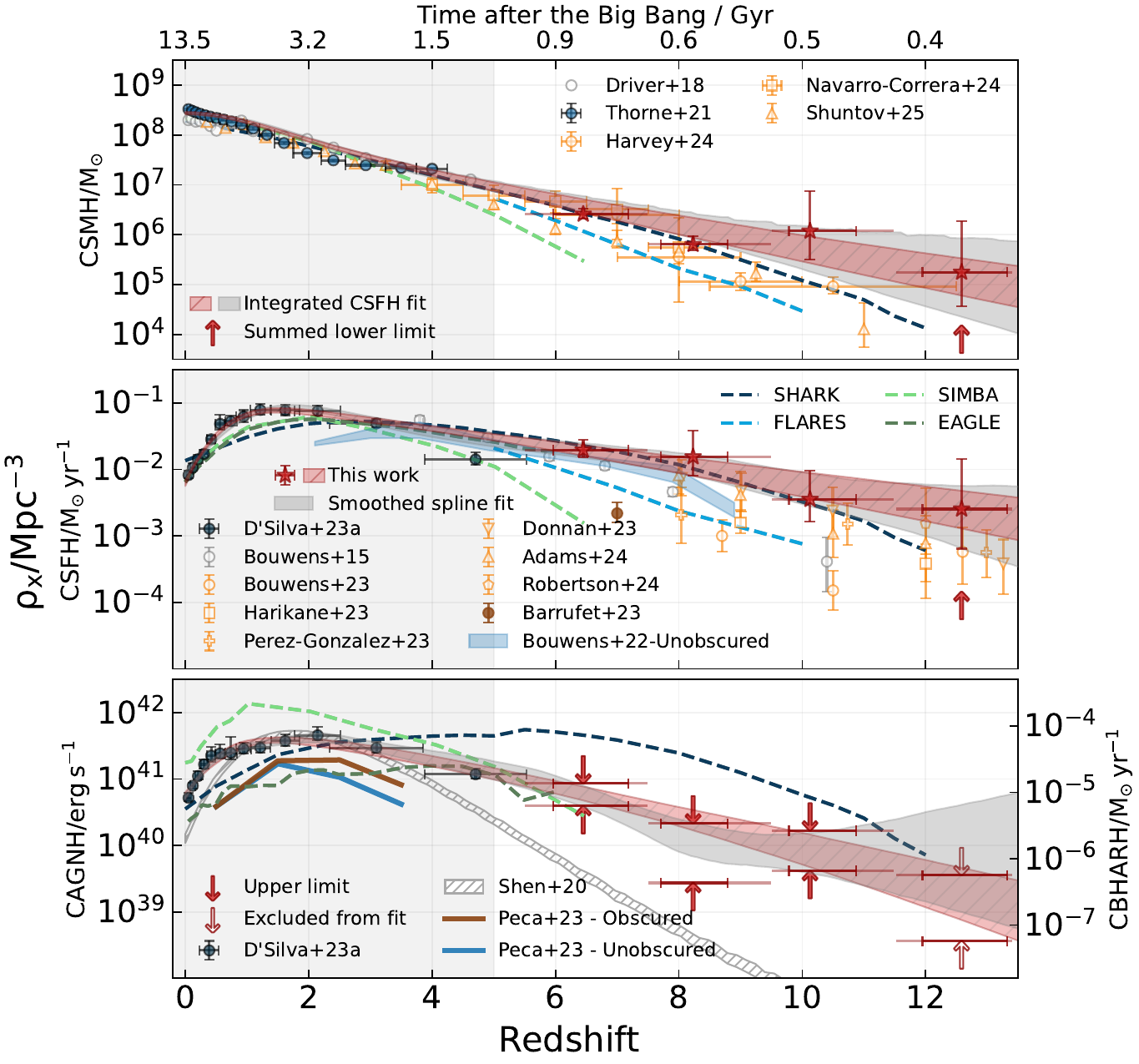}
    \caption{\textit{Top:} Cosmic stellar mass history. The novel results of this work and their $1\sigma$ uncertainties are shown as red stars with error bars. The horizontal error bars show both the $1\sigma$ of the redshift distribution in the bin and the edges of the redshift bin. Both the red and grey filled lines show the integral of the CSFH fit from the middle panel. We show results from \citet{thorneDeepExtragalacticVIsible2021, driverGAMAG10COSMOS3DHST2018, harveyEPOCHSIVSED2024,navarro-carreraConstraintsFaintEnd2023,shuntovCOSMOSWebStellarMass2025}, as per the figure legend. The grey shaded region marks the state-of-the art SED fitting results before JWST ($z\lesssim5$). \textit{Middle:} Cosmic star formation history. The red, filled lines show the \nth{16}-\nth{84} percentile ranges of the fitted model. The grey filled lines show the \nth{16}-\nth{84} percentile ranges of the smoothed spline fit. The colour scheme is the same as the top panel. We show results from \citet{dsilvaGAMADEVILSCosmic2023,bouwensUVLuminosityFunctions2015a, bouwens29GalaxiesMagnified2022, bouwensEvolutionUVLF2023a,harikaneComprehensiveStudyGalaxies2023,donnanEvolutionGalaxyUV2023, perez-gonzalezLife30Probing2023a,adamsEPOCHSIIUltraviolet2024,robertsonEarliestGalaxiesJADES2024,barrufetALMAREBELSSurvey2023}. \textit{Bottom:} Cosmic AGN luminosity history. The colour scheme is the same as the top panel. The upward facing arrows show the results for the direct sum quantities, summing only $\mathrm{L_{AGN} \leq 10^{46} \, erg \, s^{-1}}$ and the downward facing arrows show the results for the $\mathrm{L_{AGN}}$ upper limits. The mid point of this lower and upper bound was used for fitting. The highest redshift point was excluded from the model fit. We show results from \citet{dsilvaGAMADEVILSCosmic2023,shenBolometricQuasarLuminosity2020,pecaCosmicEvolutionAGN2023}. The cosmic black hole accretion rate history was computed as in Equation~\ref{eq:cbharh}. In all panels, results from both hydrodynamical \citep[e.g., \textsc{simba}, \textsc{Eagle}, \textsc{Flares}, ][]{daveSIMBACosmologicalSimulations2019,schayeEAGLEProjectSimulating2015a, crainEAGLESimulationsGalaxy2015a, furlongEvolutionGalaxyStellar2015, lovellFirstLightReionisation2020,vijayanFirstLightReionization2021a} and semi-analytic \citep[e.g., \textsc{Shark},][]{lagosSharkIntroducingOpen2018,lagosQuenchingMassiveGalaxies2024b} models are shown as dashed lines.}
    \label{fig:cosmic_X_history}
\end{figure*}
Figure~\ref{fig:cosmic_X_history} shows the redshift evolution of $\mathrm{\rho_{M_{\star}}}$, $\mathrm{\rho_{SFR}}$ and $\mathrm{\rho_{L_{AGN}}}$, all obtained from integrating the density distributions directly. The lower limit, summed quantities, are shown as upward facing arrows. For the CSFH and the CAGNH we fitted \texttt{massfunc\_snorm\_trunc} functions \citep[Eqs. 1-5 in ][]{dsilvaGAMADEVILSCosmic2023}\footnote{We fitted with respect to $z$, using $\mathtt{magemax}=30$ and $\mathtt{mtrunc}=2$} to the combined results from \citet{dsilvaGAMADEVILSCosmic2023} at $z \lesssim 3$ and this work. This is the same functional form used to parametrize galaxy SFHs in \textsc{ProSpect} and is essentially a skewed, unimodal, normal distribution. For the CAGNH we fitted the mid points of the lower and upper limits. The highest redshift point from \citet{dsilvaGAMADEVILSCosmic2023} was excluded in the CSFH/CAGNH fitting because their data suffered the greatest selection effects in that bin. As further discussed in Section~\ref{sec:caveats}, it is not clear whether all of the AGN luminosity density has been accounted for, especially, at $z\approx11.5-13.5$ because of the limited rest-frame wavelengths that we probe with the F444W filter. As such, we omitted the highest redshift CAGNH point from the fit. For comparison we fitted the CSFH with a $6$ parameter smoothed spline function that was found to produce the best reduced $\chi^{2}$ value, while for the CAGNH we fitted a $7$ parameter smoothed spline function. Using these more flexible smoothed spline functions instead of the fixed \texttt{massfunc\_snorm\_trunc} functions nevertheless gives us similar results for the CSFH and CAGNH. 

Instead of fitting a separate function to the CSMH data points, we obtained a model CSMF by integrating the fitted CSFH. In detail, this will depend on both the IMF and the set of isochrones within the stellar population synthesis library to properly account for the mass returned to the ISM from stellar evolution \citep[e.g.,][]{robothamProGenyNewSimple2024}. We used \textsc{ProSpect} itself to calculate the total mass retained in stars as per the CSFH and with the appropriate \citet{bruzualStellarPopulationSynthesis2003} stellar population synthesis library. However, we confirmed that using a fixed return fraction of $41\%$, which is calculated under the assumption of instantaneous recycling and the closed box model for the \citet{chabrierGalacticStellarSubstellar2003} IMF \citep[e.g.,][]{madaudickinson2014}, does not significantly change the results within the uncertainties. We find that the integrated CSFH produces $\approx 0.25$~dex more stellar mass density than compared to the direct SMF integrals from \citet{thorneDeepExtragalacticVIsible2021} at $z\approx 3$, consistent with the discrepancy previously reported in the literature \citep[e.g.,][]{wilkinsRecalibratingCosmicStar2019,shuntovCOSMOSWebStellarMass2025}.

That our points for the CSFH are consistent with previous literature results is interesting as \citet{dsilvaStarFormationAGN2023} found a tentative $\approx 0.4$~dex reduction in the CSFH compared to the same literature values at $z\gtrsim 9.5$ when also including an AGN component in the SED fitting. This inclusion was hence inferred as a pathway to alleviate the tension of excessive star formation compared to models of typical baryon conversion efficiencies inside of dark matter haloes \citep[e.g.,][]{harikaneGOLDRUSHIVLuminosity2022}. This analysis, however, improves on \citet{dsilvaStarFormationAGN2023} because we have far greater number statistics at $z\gtrsim 9.5$ and also used the DIC to decide between $\mathrm{Pro_{Stellar}}$ and $\mathrm{Pro_{Stellar + AGN}}$ models, suggesting that we cannot entirely explain the difference of the observational CSFH to the typical models as AGN contamination. Despite this, we found that the $\mathrm{M_{\star}}$ between \textsc{ProSpect} runs can vary significantly, with far lower $\mathrm{M_{\star}}$ when including an AGN component \citep[consistent with][]{dsilvaStarFormationAGN2023}, meaning that methods to account for potential AGN components and appropriate model selection are critical when modelling SEDs at these redshifts. Concerning the efficiency of star formation, we note that \textsc{Shark} shows $\approx 0.5$~dex higher CSFH compared to our results at $z\approx 5.5-10.5$, due especially to the presence of faint galaxies, demonstrating that the observed abundance of CSFH at these redshifts is somewhat reconcilable with physical models of star formation in dark matter halos. 

The CSMH, CSFH and CAGNH exhibit steady increases over $z=13.5\to 5.5$. The CSMH rises by $\approx 1.9$~dex and the CSFH rises by $\approx 1$~dex, demonstrating the rapid assembly of stellar mass shortly after the Big Bang and up to the peak of CSFH at $z\approx 1-2$. The CAGNH rises by $\approx 1.1$~dex over $z=10.5 \to 5.5$. The rise is more extreme over $z = 13.5 \to 5.5$, tending to rise by $\approx 2$~dex, but, as mentioned earlier, measurements of the AGN luminosity density at $z\approx 11.5-13.5$ are uncertain. The CAGNH could also be flat between $z\approx 13.5-11.5$ as per the shape and width of the smoothed spline fit instead of the more rigid \texttt{massfunc\_snorm\_trunc} function. Both star formation and accretion on the SMBH are fuelled by the gas supply, so if star formation is efficient we might further anticipate AGN power to behave similarly. Our fitted CAGNH is $\gtrsim 1$~dex greater at $z\gtrsim 5.5$ throughout compared to the measurements from \citet{shenBolometricQuasarLuminosity2020}, who compute AGN bolometric luminosity functions from $ z \approx 0-7$. The reason for this is likely due to the ability for JWST to uncover more AGN than was previously possible \citep[e.g.,][]{kokorevCensusPhotometricallySelected2024,akinsCOSMOSWebOverabundancePhysical2024}.

\subsection{Quantifying baryon conversion}
\label{sec:baryon}
The phase of the baryon budget usable for fuelling star formation and SMBH accretion (e.g., largely molecular instead of atomic or ionised) was in a surplus during these earliest epochs \citep{lagosCosmicEvolutionAtomic2011}, especially at $z \gtrsim 5.5$. As galaxies were only ramping up their star formation and SMBH accretion, they were yet to deplete the supply of the usable reservoir. Additionally, we can expect that feedback from star formation and SMBH accretion was yet to adequately heat the gas and prevent it from entering the necessary cool phase for fuelling, and/or eject it from the galaxy \citep[e.g.,][]{schayePhysicsDrivingCosmic2010}. 

To quantify baryon conversion, we computed the cosmic black hole accretion rate history (CBHARH) as
\begin{equation}
    \label{eq:cbharh}
    \mathrm{CBHARH = CAGNH / \epsilon c^{2}},
\end{equation} where $\epsilon = 0.1$ is the radiative efficiency and $c$ is the speed of light, and divide that by the CSFH for a dimensionless baryon conversion ratio. Comparing this ratio of the fitted CSFH and CBHARH, we see that the CBHARH experiences a greater increase from $z\approx 10.5 (13.5) \to 5.5$ than the CSFH by a factor of $\approx 3 (6)$, suggesting that the rate of change of SMBH growth tends to outpace the equivalent rate of change of the stellar mass assembly in this redshift range, similar in effect to $z\gtrsim 5$ SMBHs tending to be over-massive compared to the stellar mass of their host galaxies \citep[e.g.,][]{maiolinoJADESDiversePopulation2023a,pacucciJWSTCEERSJADES2023,kormendyCoevolutionNotSupermassive2013}. This increase is insensitive to the choice of constant radiative efficiency, $\epsilon$, that can vary between $0.06-0.4$ depending on the spin of the SMBH \citep[e.g.,][]{shakuraBlackHolesBinary1973}. In a follow up paper (D'Silva et al. in prep.), we will investigate the affect of the CSFH and CBHARH on reionisation. 

\section{Caveats}
\label{sec:caveats}
While we have endeavoured to be as consistent and conservative as possible in preparing this work with the data and tools on hand, systematic uncertainties are likely to be under-represented. We here discuss sources of systematic uncertainties and prospects for the future.

\subsection{Photometric redshifts}
In the absence of spectroscopic confirmation, we can only compute $z_{\mathrm{phot}}$ up to a certain confidence level. This is especially pertinent in the high redshift frontier because the limited photometric bands from even the combination of HST and JWST, means spectral features are sparsely sampled. The results of Sections~\ref{sec:eazy} and \ref{sec:prospect} have demonstrated that template incompleteness may be a significant source of uncertainty in template fitting codes. Fortunately, as the NIRSpec spectroscopic database continues to grow, this issue can be alleviated. 

\subsection{SED modelling} 
\label{sect:sed_model}
Although \textsc{ProSpect} was designed to be highly flexible, we operate in a constrained mode to inhibit parameter degeneracies. A key assumption is a universal Chabrier IMF. The metal-poor terrain of the early Universe seems to afford the formation of massive, $\mathrm{M_{\star} \approx 100 M_{\odot}}$, stars due to the lack of metal-line cooling to effectively fragment the primordial gas clouds at scales smaller than the Jean's length of unpolluted clouds \citep[e.g.,][]{brommFirstStars2004}. Additionally, this may be a pathway to producing massive seed black holes either as remnants of population III stellar evolution or via direct collapse \citep[e.g.,][]{latifFormationSupermassiveBlack2016}. \citet{harveyEPOCHSIVSED2024} showed that adopting a top-heavy IMF when SED fitting JWST galaxies can reduce the $\mathrm{M_{\star}}$ of $z\approx 12$ galaxies by $\approx 0.5$~dex. Although, the goodness-of-fit between top-heavy and more standard IMFs did not significantly change, indicating that photometry alone cannot discern between alternative IMFs. On the other hand, the implementation of the Chabrier IMF in \textsc{Shark} was shown to reproduce the observed UV-NIR-FIR properties of galaxies from $0<z<10$ \citep{lagosFarultravioletFarinfraredGalaxy2019}. 

For $z\gtrsim 5.5$ galaxies, we rely on rest-frame UV to optical wavelengths with this data to separate star formation from AGN when interpreting the SEDs. To explore any biases subsumed into the bolometric $\mathrm{L_{AGN}}$ from being limited to these rest-frame wavelengths, we did a simple test using a low redshift sample. \citet{thorneDeepExtragalacticVIsible2022} presented $0.09-30\mu$m \textsc{ProSpect} SED fits of a sample of 41 bright AGN with minimal host galaxy contamination at $z<0.68$ from \citet{brownSpectralEnergyDistributions2019a}. We refitted these 41 AGN but iteratively truncated the wavelength range as $\lambda_{i}/\mu\mathrm{m} \leq 4.4 (1 + z_{i})$, where $z_{i}=6.5,8.5,10.5,12.5$. This allowed us to test the performance of \textsc{ProSpect} when we only have access to rest-frame wavelengths covered by the longest F444W filter in each of our four redshift bins.
\begin{figure*}
    \centering
    \includegraphics[width=\linewidth]{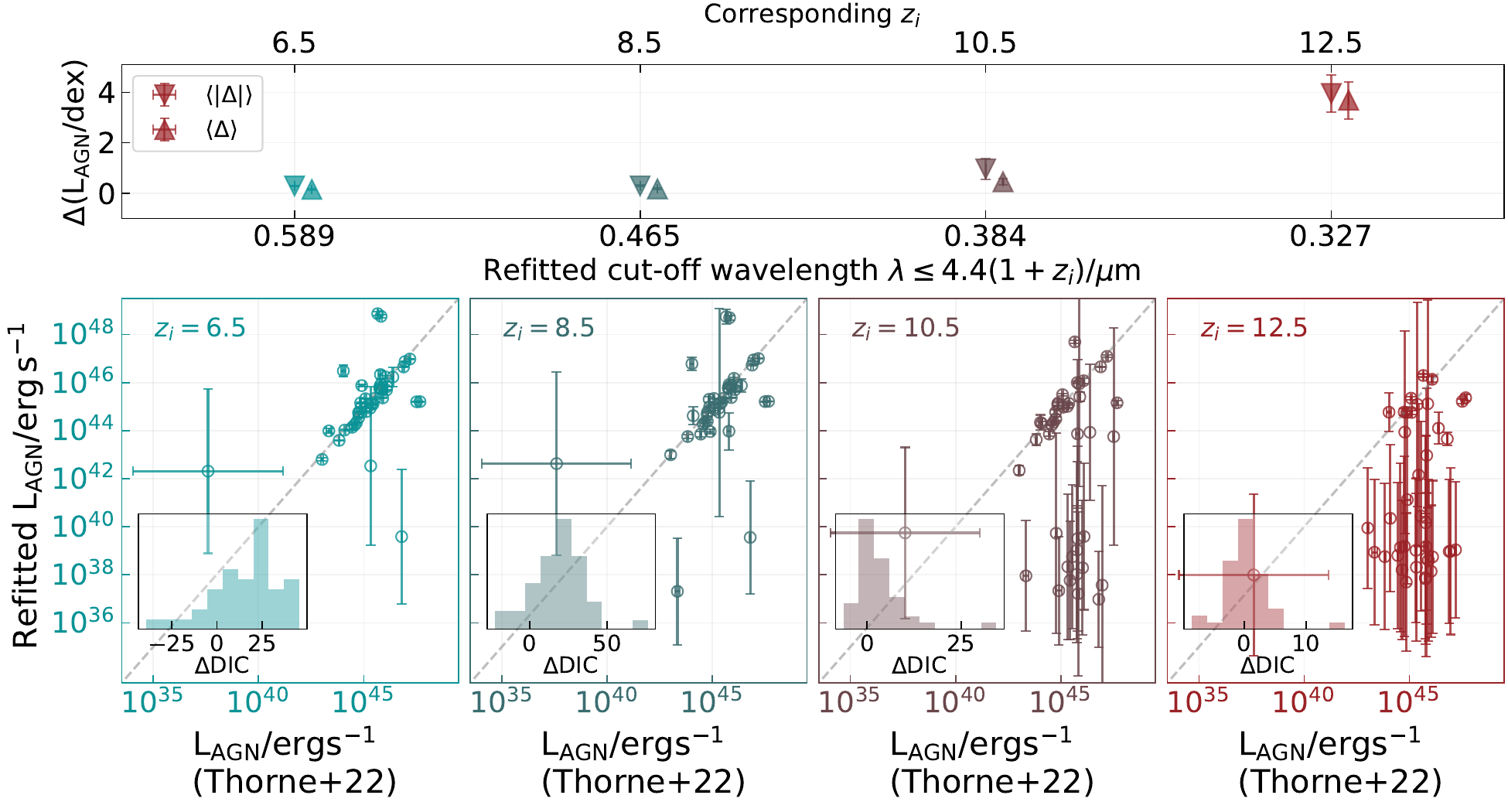}
    \caption{\textit{Bottom:} Refitted $\mathrm{L_{AGN}}$ against $\mathrm{L_{AGN}}$ reported by \citet{thorneDeepExtragalacticVIsible2022} of $0.09-30\mu$m \textsc{ProSpect} fits of 41 bright AGN from \citet{brownSpectralEnergyDistributions2019a}. From left to right, we successively truncated the fitted wavelengths as $\lambda_{i}/\mu\mathrm{m} \leq 4.4 (1 + z_{i})$, where $z_{i}=6.5,8.5,10.5,12.5$ to demonstrate how the rest-frame wavelength coverage in redshift bins biases the $\mathrm{L_{AGN}}$ extraction from \textsc{ProSpect} SED fitting. We also show in every panel the histogram of the $\Delta \mathrm{DIC}$ as in Equation~\ref{eq:dic}. In all panels, the grey dashed line is the 1:1 relation. \textit{Top:} The median absolute differences, $\left< | \Delta | \right> = \left< | \mathrm{Thorne+22 - Refitted} | \right>$, are shown with down facing triangles. The median differences, $\left< \Delta \right> = \left< \mathrm{Thorne+22 - Refitted} \right>$, are shown with up facing triangles and offset for clarity. We computed an uncertainty range on these quantities by recomputing them 1000 times after sampling the $\mathrm{L_{AGN}}$ about their $1\sigma$ uncertainties. The uncertainties are smaller than the size of the point for the first three data points.}
    \label{fig:brown19}
\end{figure*}
Figure~\ref{fig:brown19} shows the results of this experiment. We calculated the median absolute difference between the two values, $\left< | \Delta | \right> = \left< | \mathrm{Thorne+22 - Refitted} | \right>$, to show the absolute performance of recovering the $\mathrm{L_{AGN}}$. We also calculated just the median difference, $\left< \Delta \right> = \left< \mathrm{Thorne+22 - Refitted} \right>$, to show the relative performance, since with fewer filters we expected to recover systematically less of the $\mathrm{L_{AGN}}$. An uncertainty range on these quantities was calculated via a Monte-Carlo method, by recomputing these values 1000 times after sampling the $\mathrm{L_{AGN}}$ about their $1\sigma$ errors.

Up to $z\approx 8.5$, we can generally recover the $\mathrm{L_{AGN}}$ within $\approx 0.2$~dex and at $z\approx 10.5$ we can recover it within $\approx 0.4$~dex. At $z\gtrsim 10.5$, \textsc{ProSpect} however tends to systematically underestimate the $\mathrm{L_{AGN}}$ by as much as $\approx 4.0$~dex. For comparison, we also show the $\Delta \mathrm{DIC}$, as in Equation~\ref{eq:dic}, where most of the values up to $z\approx 10.5$ prefer $\mathrm{Pro_{Stellar+AGN}}$. While the 41 AGN from \citet{brownSpectralEnergyDistributions2019a} may not necessarily reflect the population of AGN that may exist at $z\gtrsim 10.5$, this test demonstrates that precise quantification of the potential AGN density at $z\gtrsim 10.5$ remains elusive without strong observational constraints at redder wavelengths.

On a related note, it is not yet clear if AGN or stellar emission dominates the SED of LRDs without redder wavelengths \citep[e.g.,][]{perez-gonzalezWhatNatureLittle2024,kocevskiRiseFaintRed2024a}. The abundance of $\mathrm{L_{AGN} \gtrsim 10^{46} \, erg \, s^{-1}}$ AGN, many of which appear as LRDs, may be overestimated as stellar emission may have been misattributed as AGN, indicating that SED fitting with only HST and NIRCam photometry struggles to accurately constrain the AGN power. With the addition of the Mid-Infrared Instrument (MIRI) on JWST to probe into the rest-frame IR at these redshifts, the SEDs of some LRDs have indeed been shown to be adequately described by stellar emission, with potentially sub-dominant AGN contributions \citep{perez-gonzalezWhatNatureLittle2024,williamsGalaxiesMissedHubble2024}. Better understanding the effect of the modelling assumptions remains an exciting challenge for the community. 

\subsection{Obscured star formation} 
The consensus from the literature is that $\approx 10-30\%$ of the CSFH at $z\approx 7$ is enshrouded in dust as inferred from the rest-frame IR unveiled by the \textit{Atacama Large Millimeter Array} (ALMA) that would remain undetected in rest-frame UV samples \citep[e.g.,][]{caseyBrightestGalaxiesDark2018,zavalaEvolutionIRLuminosity2021,algeraALMAREBELSSurvey2023,chiangCosmicInfraredBackground2025}. With JWST, the situation has improved at $z\approx 7$ because now we can probe redder wavelengths than HST that are less affected by dust-obscuration and more sensitive to evolved stellar populations that may not have significant rest-frame UV emission \citep[e.g.,][]{barrufetUnveilingNatureInfrared2023,weibelGalaxyBuildupFirst2024a}.

This improvement is demonstrated in Figure~\ref{fig:cosmic_X_history} where it can be seen that we recover $\approx 0.1$~dex more CSFH than earlier unobscured measurements from \citet{bouwens29GalaxiesMagnified2022} who only used HST observations. We also show in Figure~\ref{fig:cosmic_X_history} the measurement from \citet{barrufetALMAREBELSSurvey2023} who integrated the IR luminosity function that was itself inferred from ALMA $88\mu$m and $158\mu$m observations. We computed the ratio of this IR measurement to the total sum of both obscured and unobscured CSFH. Using the measurements of \citet{bouwens29GalaxiesMagnified2022} as the unobscured component, we find that the obscured fraction is $\approx 16\%$. Now, using our CSFH unobscured component, we find that the obscured fraction is $\approx 12\%$, reflecting the $\approx 0.1$~dex higher CSFH we measured compared to \citet{bouwens29GalaxiesMagnified2022}. At $z \gtrsim 7$, censuses of star formation will once again become restricted to the rest-frame UV. Since the obscured fraction is $\sim 10\%$ at $z\approx 7$, we expect that a potentially missing contribution from dust obscured star formation at even higher redshifts will reside in the ranges of the uncertainties presented in Figure~\ref{fig:cosmic_X_history}.

\subsection{Obscured AGN}
Similarly, the proportion of heavily obscured AGN may be $\gtrsim 50\%$ even at $z\approx 1-4$ \citep[e.g.,][]{tasnimanannaAccretionHistoryAGNs2019,pecaCosmicEvolutionAGN2023}, as inferred from their X-ray properties. For comparison, shown in Figure~\ref{fig:cosmic_X_history} are the results of the CAGNH from \citet{pecaCosmicEvolutionAGN2023} measured from both obscured and unobscured AGN. We used their fitted X-ray luminosity functions, converted to bolometric luminosities with the $2-10$~keV bolometric corrections from \citet{netzerBolometricCorrectionFactors2019} and integrated them over the same range used for our own calculation. As expected the obscured CAGNH, as inferred from the X-ray measurements, is dominant toward $z\approx 4$.

It is worth highlighting that X-rays are mostly obscured due to photoelectric absorption and scattering from gas in contrast to UV photons (with wavelengths larger than the Ly$\alpha$ rest wavelength) that are mostly obscured by dust grains. The obscured fraction from X-rays should, in principle, converge to the obscuration fractions from rest-frame UV-optical samples provided that the hydrogen column density threshold for obscuration is $\mathrm{N_{H} \gtrsim 10^{22} \, cm^{-2}}$ \citep[e.g.,][and references therein]{hickoxObscuredActiveGalactic2018}. However, at $z>4$ there is relatively more gas than dust, and so the expectation is that the X-rays are more susceptible to be obscured than the UV and optical photons.

MIRI has discovered populations of MIR bright AGN at $z\approx 0-5$ \citep[e.g.,]{yangCEERSKeyPaper2023,maiolinoJWSTMeetsChandra2025a} without significant X-ray detections. Analyses of LRDs report that the weak X-ray emission cannot be entirely explained by optically thick tori because the required absorber column densities ($\mathrm{N_{H} \gtrsim 10^{24} - 10^{25} \, cm^{-2}}$) are far higher than expected for typical type-1 AGN \citep[$\mathrm{N_{H} \gtrsim 10^{20} - 10^{23} \, cm^{-2}}$,][]{anannaXRayViewLittle2024,yueStackingXrayObservations2024,kocevskiRiseFaintRed2024a}. \citet{lyuActiveGalacticNuclei2024} report that $\approx 80$\% of MIRI identified AGN are lacking X-ray detections that, if not due to optically thick torii, may be a result of intrinsic X-ray faintness due to the absence of strong corona at low metallicity. Hence, MIRI observations will be critical for uncovering UV-optical obscured AGN by probing redder rest-frame wavelengths that will be both far less affected by obscuration and able to better probe the emission of the torus than the wavelengths studied here.

Additionally, we have treated all of our galaxies as singular spatial units, employing the same attenuation law via \textsc{ProSpect} over the entire geography of each galaxy. AGN activity is however confined to the innermost regions of galaxies that, at least in nearby galaxies, exhibits the greatest obscuration \citep{kimAnalysisSpatiallyResolved2019}. Without higher spatial resolution, any significantly obscured AGN component may be washed out by the light of the surrounding host galaxy. As such, we may also be missing a proportion of the AGN activity because of the inability to observe spatially resolved trends in these $z\gtrsim 5.5$ galaxies. 

Whether NIRCam has missed a significant fraction of obscured AGN activity, beyond the ranges of the uncertainties presented in Figure~\ref{fig:cosmic_X_history}, remains to be confirmed. This situation can improve by turning to spectroscopic observations of broad components in the $\mathrm{H\alpha}$ line \citep[e.g.,][]{larsonCEERSDiscoveryAccreting2023a} and/or metallicity-independent emission line ratio diagnostics \citep[e.g.,][]{ublerGANIFSMassiveBlack2023a}, along with further deep, high resolution, multiwavelength constraints from X-rays, MIRI, Sub-mm and beyond. A future work will analyse the X-ray properties of these sources also detected in the NuSTAR NEP survey, thus providing insights into the AGN obscuration.

\subsection{Survey variance}
According to the survey variance calculator\footnote{\url{https://cosmocalc.icrar.org/}} of \citet{driverQuantifyingCosmicVariance2010}, survey variance in our fields varies from $\approx 57$\% in the smallest PEARLS parallel fields in the highest redshift bin to $\approx 20$\% in the largest PRIMER COSMOS in the lowest redshift bin. The situation improves by combining the fields. Employing the volume-weighted sum of squares method of \citet{mosterCosmicVarianceCookbook2011}, suggests $\approx 2$\% survey variance in the highest redshift bin and $\lesssim 1$\% in the lowest. Survey variance may however increase toward lower $\mathrm{M_{\star}}$, SFR and $\mathrm{L_{AGN}}$ where the constraining power comes from deep, narrow surveys like JADES and NGDEEP; but, as can be seen from Figure~\ref{fig:all_densities}, contributions to the density from lower values diminish. Hence, we expect that survey variance is a sub-dominant source of uncertainty in the $\mathrm{M_{star}}$, SFR and $\mathrm{L_{AGN}}$ space densities compared to the others discussed in this section. As the JWST database continues to expand cosmic variance will be further reduced. 

\section{Summary} \label{sec:summary}
JWST continues to enlighten our grasp of star formation and AGN from $z\approx 13.5\to5.5$. For the first time, the interface of star formation and the growth of SMBHs has been consistently connected from $z=0$ to $z=13.5$, encompassing $\gtrsim 13$ Gyr of cosmic time. The key results are:

\begin{itemize}
    \item The novel, open-source software suite \textsc{JumProPe} for processing JWST images was described. \textsc{JumProPe} is an all-in-one package that can query the MAST archive, calibrate images using the JWST Calibration Pipeline, remove $1/f$ noise and wisp artefacts, produce mosaics with \textsc{ProPane} and finally perform multi-band measurements with \textsc{ProFound}. 
    
    \item We searched for $z\gtrsim 5.5$ galaxies from 400 arcmin$^{2}$ of JWST fields processed with \textsc{JumProPe} (see Table~\ref{tab:survey_limits}). A sample of 3751 $z \approx 5.5-13.5$ galaxies was obtained with a two-phased photometric redshift approach, striving for $95$\% completeness. Redshifts for the high redshift sample were initially obtained using EAZY and then refined by further SED fitting with \textsc{ProSpect}. With the \textsc{ProSpect} SED fits we obtained the $\mathrm{M_{\star}}$, SFR and $\mathrm{L_{AGN}}$.

    \item We performed two sets of \textsc{ProSpect} SED fits: (i). $\mathrm{Pro_{Stellar}}$ where only stellar components were used to characterise the SEDs (ii). $\mathrm{Pro_{Stellar+AGN}}$ that also included an AGN component. The DIC was used to determine which model was preferred. For the galaxies with preferred $\mathrm{Pro_{Stellar}}$ fits, we considered two extremes of AGN contribution to the SED: $\mathrm{L_{AGN}^{lolim}} = 0$ and $\mathrm{L_{AGN}^{uplim}}$ that was found from an unobscured AGN template that contributed at most $20\%$ of the flux in the combined F277W$+$F356W$+$F444W filters. This allowed us to compute a lower and upper bound of possible AGN contribution to these galaxy SEDs. 

    \item The inferred $\mathrm{M_{\star}}$, SFR and $\mathrm{L_{AGN}}$ were used to compute the SMF, SFRF and AGNLF in four redshift bins at $z \approx 5.5-13.5$. We computed both binned quantities and double power-law fits (Equation~\ref{eq:dpl}) with \textsc{dftools}. However, to fit the AGNLF we could only use the $\mathrm{L_{AGN}^{uplim}}$ for the $\mathrm{Pro_{Stellar}}$ preferred SED fits. 

    \item The SMF, SFRF and AGNLF were integrated to then compute the CSMH, CSFH and CAGNH. Our results for the CSMH, CSFH and CAGNH are consistent with previous results in the literature. The CSMH rises by $\approx 2$~dex over $z=13.5\to5.5$, while the CSFH rises by $\approx 1.1$~dex, reflecting the rapid assembly of stellar mass shortly after the Big Bang. The CAGNH rises by $\approx 1$~dex over $z=10.5\to5.5$, and in this redshift range the dimensionless baryon conversion ratio of the CBHARH (converted from the CAGNH) to the CSFH increases by a factor of $\approx 3$.

    \item Systematic uncertainties remain an obstruction to further deciphering the mysteries of the early Universe. Fortunately, this situation can only get better as JWST continues to build up its imaging and spectroscopic database, to be used in harmony with multiwavelength facilities covering the breadth of the electromagnetic spectrum. 
\end{itemize}

\section*{Data availability}
The JWST data presented in this article were obtained from the Mikulski Archive for Space Telescopes (MAST) at the Space Telescope Science Institute. The specific observations analyzed can be accessed via \dataset[doi: 10.17909/1c6w-dy61]{https://doi.org/10.17909/1c6w-dy61}. The processing pipeline \textsc{JumProPe} can be found on \href{https://github.com/JordanDSilva/JUMPROPE.git}{GitHub}. Catalogues will be made available upon reasonable request to the corresponding author.

\section*{Acknowledgements}
We thank the anonymous referee for providing constructive feedback that improved the quality of this work. J.C.J.D is supported by the Australian Government Research Training Program (RTP) Scholarship. CL is a recipient of the ARC Discovery Project DP210101945. ASGR acknowledges funding by the Australian Research Council (ARC) Future Fellowship scheme (FT200100375). NA, TH and C.J.C acknowledge support from the ERC Advanced Investigator Grant EPOCHS (788113). RAW, SHC, and RAJ acknowledge support from NASA JWST Interdisciplinary Scientist grants NAG5-12460, NNX14AN10G and 80NSSC18K0200 from GSFC. CNAW acknowledges funding from the JWST/NIRCam contract NASS-0215 to the University of Arizona.

This research was supported by the Australian Research Council Centre of Excellence for All Sky Astrophysics in 3 Dimensions (ASTRO 3D), through project number CE170100013.  This work was supported by resources provided by The Pawsey Supercomputing Centre with funding from the Australian Government and the Government of Western Australia.

This work is based on observations made with the NASA/ESA Hubble Space Telescope (HST) and NASA/ESA/CSA James Webb Space Telescope (JWST) obtained from the Mikulski Archive for Space Telescopes (MAST) at the Space Telescope Science Institute (STScI), which is operated by the Association of Universities for Research in Astronomy, Inc., under NASA contract NAS 5-03127 for JWST, and NAS 5–26555 for HST. The observations used in this work are associated with JWST programs 1176, 2738, 1345, 2079, 1180, 1210, 3250 and 1837. We acknowledge all those involved in the development of the JWST and the execution of these large observing programs. 

\bibliography{ref}{}

\begin{thebibliography}{}
\expandafter\ifx\csname natexlab\endcsname\relax\def\natexlab#1{#1}\fi
\providecommand{\url}[1]{\href{#1}{#1}}
\providecommand{\dodoi}[1]{doi:~\href{http://doi.org/#1}{\nolinkurl{#1}}}
\providecommand{\doeprint}[1]{\href{http://ascl.net/#1}{\nolinkurl{http://ascl.net/#1}}}
\providecommand{\doarXiv}[1]{\href{https://arxiv.org/abs/#1}{\nolinkurl{https://arxiv.org/abs/#1}}}

\bibitem[{Acquaviva {et~al.}(2015)Acquaviva, Raichoor, \&
  Gawiser}]{acquavivaSimultaneousEstimationPhotometric2015}
Acquaviva, V., Raichoor, A., \& Gawiser, E. 2015, \apj, 804, 8,
  \dodoi{10.1088/0004-637X/804/1/8}

\bibitem[{Adams {et~al.}(2024)Adams, Conselice, Austin, Harvey, Ferreira,
  Trussler, Juod{\v z}balis, Li, Windhorst, Cohen, Jansen, Summers, Tompkins,
  Driver, Robotham, D'Silva, Yan, Coe, Frye, Grogin, Koekemoer, Marshall,
  Pirzkal, Ryan, Maksym, Rutkowski, Willmer, Hammel, Nonino, Bhatawdekar,
  Wilkins, Bradley, Broadhurst, Cheng, Dole, Hathi, \&
  Zitrin}]{adamsEPOCHSIIUltraviolet2024}
Adams, N.~J., Conselice, C.~J., Austin, D., {et~al.} 2024, \apj, 965, 169,
  \dodoi{10.3847/1538-4357/ad2a7b}

\bibitem[{Akins {et~al.}(2024)Akins, Casey, Lambrides, Allen, Andika, Brinch,
  Champagne, Cooper, Ding, Drakos, Faisst, Finkelstein, Franco, Fujimoto,
  Gentile, Gillman, Gozaliasl, Harish, Hayward, Hirschmann, Ilbert, Kartaltepe,
  Kocevski, Koekemoer, Kokorev, Liu, Long, McCracken, McKinney, Onoue,
  Paquereau, Renzini, Rhodes, Robertson, Shuntov, Silverman, Tanaka, Toft,
  Trakhtenbrot, Valentino, \& Zavala}]{akinsCOSMOSWebOverabundancePhysical2024}
Akins, H.~B., Casey, C.~M., Lambrides, E., {et~al.} 2024, {{COSMOS-Web}}:
  {{The}} over-Abundance and Physical Nature of "Little Red
  Dots"--{{Implications}} for Early Galaxy and {{SMBH}} Assembly,  arXiv.
\newblock \doarXiv{2406.10341}

\bibitem[{Algera {et~al.}(2023)Algera, Inami, Oesch, Sommovigo, Bouwens,
  Topping, Schouws, Stefanon, Stark, Aravena, Barrufet, {da~Cunha}, Dayal,
  Endsley, Ferrara, Fudamoto, Gonzalez, Graziani, Hodge, Hygate, {de~Looze},
  Nanayakkara, Schneider, \& {van~der~Werf}}]{algeraALMAREBELSSurvey2023}
Algera, H. S.~B., Inami, H., Oesch, P.~A., {et~al.} 2023, Monthly Notices of
  the Royal Astronomical Society, 518, 6142, \dodoi{10.1093/mnras/stac3195}

\bibitem[{{Ananna} {et~al.}(2024){Ananna}, {Bogd{\'a}n}, {Kov{\'a}cs},
  {Natarajan}, \& {Hickox}}]{anannaXRayViewLittle2024}
{Ananna}, T.~T., {Bogd{\'a}n}, {\'A}., {Kov{\'a}cs}, O.~E., {Natarajan}, P., \&
  {Hickox}, R.~C. 2024, \apjl, 969, L18, \dodoi{10.3847/2041-8213/ad5669}

\bibitem[{Arrabal~Haro {et~al.}(2023)Arrabal~Haro, Dickinson, Finkelstein,
  Fujimoto, Fern{\'a}ndez, Kartaltepe, Jung, Cole, Burgarella, Chworowsky,
  Hutchison, Morales, Papovich, Simons, Amor{\'i}n, Backhaus, Bagley,
  Bisigello, Calabr{\`o}, Castellano, Cleri, Dav{\'e}, Dekel, Ferguson,
  Fontana, Gawiser, Giavalisco, Harish, Hathi, Hirschmann, Holwerda,
  {Huertas-Company}, Koekemoer, Larson, Lucas, Mobasher,
  {P{\'e}rez-Gonz{\'a}lez}, Pirzkal, Rose, Santini, Trump, {de la Vega}, Wang,
  Weiner, Wilkins, Yang, Yung, \&
  Zavala}]{arrabalharoSpectroscopicConfirmationCEERS2023a}
Arrabal~Haro, P., Dickinson, M., Finkelstein, S.~L., {et~al.} 2023, \apj, 951,
  L22, \dodoi{10.3847/2041-8213/acdd54}

\bibitem[{{Arrabal Haro} {et~al.}(2023){Arrabal Haro}, {Dickinson},
  {Finkelstein}, {Kartaltepe}, {Donnan}, {Burgarella}, {Carnall}, {Cullen},
  {Dunlop}, {Fern{\'a}ndez}, {Fujimoto}, {Jung}, {Krips}, {Larson}, {Papovich},
  {P{\'e}rez-Gonz{\'a}lez}, {Amor{\'\i}n}, {Bagley}, {Buat}, {Casey},
  {Chworowsky}, {Cohen}, {Ferguson}, {Giavalisco}, {Huertas-Company},
  {Hutchison}, {Kocevski}, {Koekemoer}, {Lucas}, {McLeod}, {McLure}, {Pirzkal},
  {Seill{\'e}}, {Trump}, {Weiner}, {Wilkins}, \&
  {Zavala}}]{arrabalharoConfirmationRefutationVery2023}
{Arrabal Haro}, P., {Dickinson}, M., {Finkelstein}, S.~L., {et~al.} 2023, \nat,
  622, 707, \dodoi{10.1038/s41586-023-06521-7}

\bibitem[{Atek {et~al.}(2018)Atek, Richard, Kneib, \&
  Schaerer}]{atekExtremeFaintEnd2018}
Atek, H., Richard, J., Kneib, J.-P., \& Schaerer, D. 2018, \mnras, 479, 5184,
  \dodoi{10.1093/mnras/sty1820}

\bibitem[{{Bagley} {et~al.}(2023){Bagley}, {Finkelstein}, {Koekemoer},
  {Ferguson}, {Arrabal Haro}, {Dickinson}, {Kartaltepe}, {Papovich},
  {P{\'e}rez-Gonz{\'a}lez}, {Pirzkal}, {Somerville}, {Willmer}, {Yang}, {Yung},
  {Fontana}, {Grazian}, {Grogin}, {Hirschmann}, {Kewley}, {Kirkpatrick},
  {Kocevski}, {Lotz}, {Medrano}, {Morales}, {Pentericci}, {Ravindranath},
  {Trump}, {Wilkins}, {Calabr{\`o}}, {Cooper}, {Costantin}, {de la Vega},
  {Hilbert}, {Hutchison}, {Larson}, {Lucas}, {McGrath}, {Ryan}, {Wang}, \&
  {Wuyts}}]{bagleyCEERSEpochNIRCam2023a}
{Bagley}, M.~B., {Finkelstein}, S.~L., {Koekemoer}, A.~M., {et~al.} 2023,
  \apjl, 946, L12, \dodoi{10.3847/2041-8213/acbb08}

\bibitem[{{Bagley} {et~al.}(2024){Bagley}, {Pirzkal}, {Finkelstein},
  {Papovich}, {Berg}, {Lotz}, {Leung}, {Ferguson}, {Koekemoer}, {Dickinson},
  {Kartaltepe}, {Kocevski}, {Somerville}, {Yung}, {Backhaus}, {Casey},
  {Castellano}, {Ch{\'a}vez Ortiz}, {Chworowsky}, {Cox}, {Dav{\'e}}, {Davis},
  {Estrada-Carpenter}, {Fontana}, {Fujimoto}, {Gardner}, {Giavalisco},
  {Grazian}, {Grogin}, {Hathi}, {Hutchison}, {Jaskot}, {Jung}, {Kewley},
  {Kirkpatrick}, {Larson}, {Matharu}, {Natarajan}, {Pentericci},
  {P{\'e}rez-Gonz{\'a}lez}, {Ravindranath}, {Rothberg}, {Ryan}, {Shen},
  {Simons}, {Snyder}, {Trump}, \& {Wilkins}}]{bagleyNextGenerationDeep2024}
{Bagley}, M.~B., {Pirzkal}, N., {Finkelstein}, S.~L., {et~al.} 2024, \apjl,
  965, L6, \dodoi{10.3847/2041-8213/ad2f31}

\bibitem[{{Baldwin} {et~al.}(1981){Baldwin}, {Phillips}, \&
  {Terlevich}}]{baldwinClassificationParametersEmissionline1981b}
{Baldwin}, J.~A., {Phillips}, M.~M., \& {Terlevich}, R. 1981, \pasp, 93, 5,
  \dodoi{10.1086/130766}

\bibitem[{Barrufet {et~al.}(2023{\natexlab{a}})Barrufet, Oesch, Bouwens, Inami,
  Sommovigo, Algera, {da~Cunha}, Aravena, Dayal, Ferrara, Fudamoto, Gonzalez,
  Graziani, Hygate, {de~Looze}, Nanayakkara, Pallottini, Schneider, Stefanon,
  Topping, \& {van~der~Werf}}]{barrufetALMAREBELSSurvey2023}
Barrufet, L., Oesch, P.~A., Bouwens, R., {et~al.} 2023{\natexlab{a}}, Monthly
  Notices of the Royal Astronomical Society, 522, 3926,
  \dodoi{10.1093/mnras/stad1259}

\bibitem[{Barrufet {et~al.}(2023{\natexlab{b}})Barrufet, Oesch, Weibel,
  Brammer, Bezanson, Bouwens, Fudamoto, Gonzalez, Gottumukkala, Illingworth,
  Heintz, Holden, Labbe, Magee, Naidu, Nelson, Stefanon, Smit, {van~Dokkum},
  Weaver, \& Williams}]{barrufetUnveilingNatureInfrared2023}
Barrufet, L., Oesch, P.~A., Weibel, A., {et~al.} 2023{\natexlab{b}}, Monthly
  Notices of the Royal Astronomical Society, 522, 449,
  \dodoi{10.1093/mnras/stad947}

\bibitem[{{Baugh} {et~al.}(2019){Baugh}, {Gonzalez-Perez}, {Lagos}, {Lacey},
  {Helly}, {Jenkins}, {Frenk}, {Benson}, {Bower}, \& {Cole}}]{baugh19pmill}
{Baugh}, C.~M., {Gonzalez-Perez}, V., {Lagos}, C. D.~P., {et~al.} 2019, \mnras,
  483, 4922, \dodoi{10.1093/mnras/sty3427}

\bibitem[{Behroozi {et~al.}(2013)Behroozi, Wechsler, \&
  Conroy}]{behrooziAverageStarFormation2013a}
Behroozi, P.~S., Wechsler, R.~H., \& Conroy, C. 2013, \apj, 770, 57,
  \dodoi{10.1088/0004-637X/770/1/57}

\bibitem[{Bellstedt \& Robotham(2024)}]{bellstedtProGenyIIImpact2024}
Bellstedt, S., \& Robotham, A. S.~G. 2024, arXiv e-prints, arXiv:2410.17698,
  \dodoi{10.48550/arXiv.2410.17698}

\bibitem[{Bellstedt {et~al.}(2020{\natexlab{a}})Bellstedt, Driver, Robotham,
  Davies, Bogue, Cook, Hashemizadeh, Koushan, Taylor, Thorne, Turner, \&
  Wright}]{bellstedtGalaxyMassAssembly2020a}
Bellstedt, S., Driver, S.~P., Robotham, A. S.~G., {et~al.} 2020{\natexlab{a}},
  \mnras, 496, 3235, \dodoi{10.1093/mnras/staa1466}

\bibitem[{Bellstedt {et~al.}(2020{\natexlab{b}})Bellstedt, Robotham, Driver,
  Thorne, Davies, Lagos, Stevens, Taylor, Baldry, Moffett, Hopkins, \&
  Phillipps}]{bellstedtGalaxyMassAssembly2020}
Bellstedt, S., Robotham, A. S.~G., Driver, S.~P., {et~al.} 2020{\natexlab{b}},
  \mnras, 498, 5581, \dodoi{10.1093/mnras/staa2620}

\bibitem[{Bouwens {et~al.}(2022)Bouwens, Illingworth, Ellis, Oesch, \&
  Stefanon}]{bouwens29GalaxiesMagnified2022}
Bouwens, R.~J., Illingworth, G.~D., Ellis, R.~S., Oesch, P.~A., \& Stefanon, M.
  2022, \apj, 940, 55, \dodoi{10.3847/1538-4357/ac86d1}

\bibitem[{Bouwens {et~al.}(2015)Bouwens, Illingworth, Oesch, Trenti, Labb{\'e},
  Bradley, Carollo, {van Dokkum}, Gonzalez, Holwerda, Franx, Spitler, Smit, \&
  Magee}]{bouwensUVLuminosityFunctions2015a}
Bouwens, R.~J., Illingworth, G.~D., Oesch, P.~A., {et~al.} 2015, \apj, 803, 34,
  \dodoi{10.1088/0004-637X/803/1/34}

\bibitem[{Bouwens {et~al.}(2021)Bouwens, Oesch, Stefanon, Illingworth, Labbe,
  Reddy, Atek, Montes, Naidu, Nanayakkara, Nelson, \&
  Wilkins}]{bouwensNewDeterminationsUV2021}
Bouwens, R.~J., Oesch, P.~A., Stefanon, M., {et~al.} 2021, \aj, 162, 47,
  \dodoi{10.3847/1538-3881/abf83e}

\bibitem[{Bouwens {et~al.}(2023)Bouwens, Stefanon, Brammer, Oesch,
  {Herard-Demanche}, Illingworth, Matthee, Naidu, {van Dokkum}, \& {van
  Leeuwen}}]{bouwensEvolutionUVLF2023a}
Bouwens, R.~J., Stefanon, M., Brammer, G., {et~al.} 2023, \mnras, 523, 1036,
  \dodoi{10.1093/mnras/stad1145}

\bibitem[{Bowler {et~al.}(2020)Bowler, Jarvis, Dunlop, McLure, McLeod, Adams,
  {Milvang-Jensen}, \& McCracken}]{bowlerLackEvolutionVery2020a}
Bowler, R. A.~A., Jarvis, M.~J., Dunlop, J.~S., {et~al.} 2020, Monthly Notices
  of the Royal Astronomical Society, 493, 2059, \dodoi{10.1093/mnras/staa313}

\bibitem[{Brammer {et~al.}(2008)Brammer, {van Dokkum}, \&
  Coppi}]{brammerEAZYFastPublic2008}
Brammer, G.~B., {van Dokkum}, P.~G., \& Coppi, P. 2008, \apj, 686, 1503,
  \dodoi{10.1086/591786}

\bibitem[{{Brandt} \& {Alexander}(2015)}]{brandtCosmicXraySurveys2015}
{Brandt}, W.~N., \& {Alexander}, D.~M. 2015, \aapr, 23, 1,
  \dodoi{10.1007/s00159-014-0081-z}

\bibitem[{Bromm \& Larson(2004)}]{brommFirstStars2004}
Bromm, V., \& Larson, R.~B. 2004, \araa, 42, 79,
  \dodoi{10.1146/annurev.astro.42.053102.134034}

\bibitem[{Brown {et~al.}(2019)Brown, Duncan, Landt, Kirk, Ricci, Kamraj,
  Salvato, \& Ananna}]{brownSpectralEnergyDistributions2019a}
Brown, M. J.~I., Duncan, K.~J., Landt, H., {et~al.} 2019, Monthly Notices of
  the Royal Astronomical Society, 489, 3351, \dodoi{10.1093/mnras/stz2324}

\bibitem[{Bruzual \& Charlot(2003)}]{bruzualStellarPopulationSynthesis2003}
Bruzual, G., \& Charlot, S. 2003, \mnras, 344, 1000,
  \dodoi{10.1046/j.1365-8711.2003.06897.x}

\bibitem[{{Bunker} {et~al.}(2024){Bunker}, {Cameron}, {Curtis-Lake},
  {Jakobsen}, {Carniani}, {Curti}, {Witstok}, {Maiolino}, {D'Eugenio},
  {Looser}, {Willott}, {Bonaventura}, {Hainline}, {{\"U}bler}, {Willmer},
  {Saxena}, {Smit}, {Alberts}, {Arribas}, {Baker}, {Baum}, {Bhatawdekar},
  {Bowler}, {Boyett}, {Charlot}, {Chen}, {Chevallard}, {Circosta}, {DeCoursey},
  {de Graaff}, {Egami}, {Eisenstein}, {Endsley}, {Ferruit}, {Giardino},
  {Hausen}, {Helton}, {Hviding}, {Ji}, {Johnson}, {Jones}, {Kumari}, {Laseter},
  {L{\"u}tzgendorf}, {Maseda}, {Nelson}, {Parlanti}, {Perna}, {Rauscher},
  {Rawle}, {Rix}, {Rieke}, {Robertson}, {Rodr{\'\i}guez Del Pino}, {Sandles},
  {Scholtz}, {Sharpe}, {Skarbinski}, {Stark}, {Sun}, {Tacchella}, {Topping},
  {Villanueva}, {Wallace}, {Williams}, \&
  {Woodrum}}]{bunkerJADESNIRSpecInitial2023}
{Bunker}, A.~J., {Cameron}, A.~J., {Curtis-Lake}, E., {et~al.} 2024, \aap, 690,
  A288, \dodoi{10.1051/0004-6361/202347094}

\bibitem[{Casey {et~al.}(2018)Casey, A.~Zavala, Spilker, {da Cunha}, Hodge,
  Hung, Staguhn, Finkelstein, \& Drew}]{caseyBrightestGalaxiesDark2018}
Casey, C.~M., A.~Zavala, J., Spilker, J., {et~al.} 2018, The Astrophysical
  Journal, 862, 77, \dodoi{10.3847/1538-4357/aac82d}

\bibitem[{Chabrier(2003)}]{chabrierGalacticStellarSubstellar2003}
Chabrier, G. 2003, \pasp, 115, 763, \dodoi{10.1086/376392}

\bibitem[{Charlot \& Fall(2000)}]{charlotSimpleModelAbsorption2000}
Charlot, S., \& Fall, S.~M. 2000, \apj, 539, 718, \dodoi{10.1086/309250}

\bibitem[{Chiang {et~al.}(2025)Chiang, Makiya, \&
  M{\'e}nard}]{chiangCosmicInfraredBackground2025}
Chiang, Y.-K., Makiya, R., \& M{\'e}nard, B. 2025, arXiv e-prints,
  arXiv:2504.05384, \dodoi{10.48550/arXiv.2504.05384}

\bibitem[{{Conroy}(2013)}]{conroyModelingPanchromaticSpectral2013a}
{Conroy}, C. 2013, \araa, 51, 393, \dodoi{10.1146/annurev-astro-082812-141017}

\bibitem[{Conroy \& Gunn(2010)}]{conroyPropagationUncertaintiesStellar2010}
Conroy, C., \& Gunn, J.~E. 2010, \apj, 712, 833,
  \dodoi{10.1088/0004-637X/712/2/833}

\bibitem[{{Conselice} {et~al.}(2024){Conselice}, {Adams}, {Harvey}, {Austin},
  {Ferreira}, {Ormerod}, {Duan}, {Trussler}, {Li}, {Juodzbalis}, {Westcott},
  {Harris}, {Seeyave}, {Bluck}, {Windhorst}, {Bhatawdekar}, {Coe}, {Cohen},
  {Cheng}, {Driver}, {Frye}, {Furtak}, {Grogin}, {Hathi}, {Holwerda}, {Jansen},
  {Koekemoer}, {Marshall}, {Nonino}, {Robotham}, {Summers}, {Wilkins},
  {Willmer}, {Yan}, \& {Zitrin}}]{conseliceEPOCHSDiscoveryStar2024}
{Conselice}, C.~J., {Adams}, N., {Harvey}, T., {et~al.} 2024, arXiv e-prints,
  arXiv:2407.14973, \dodoi{10.48550/arXiv.2407.14973}

\bibitem[{{Crain} {et~al.}(2015){Crain}, {Schaye}, {Bower}, {Furlong},
  {Schaller}, {Theuns}, {Dalla Vecchia}, {Frenk}, {McCarthy}, {Helly},
  {Jenkins}, {Rosas-Guevara}, {White}, \&
  {Trayford}}]{crainEAGLESimulationsGalaxy2015a}
{Crain}, R.~A., {Schaye}, J., {Bower}, R.~G., {et~al.} 2015, \mnras, 450, 1937,
  \dodoi{10.1093/mnras/stv725}

\bibitem[{Dale {et~al.}(2014)Dale, Helou, Magdis, Armus, {D{\'i}az-Santos}, \&
  Shi}]{daleTwoparameterModelInfrared2014}
Dale, D.~A., Helou, G., Magdis, G.~E., {et~al.} 2014, \apj, 784, 83,
  \dodoi{10.1088/0004-637X/784/1/83}

\bibitem[{{Dav{\'e}} {et~al.}(2019){Dav{\'e}}, {Angl{\'e}s-Alc{\'a}zar},
  {Narayanan}, {Li}, {Rafieferantsoa}, \&
  {Appleby}}]{daveSIMBACosmologicalSimulations2019}
{Dav{\'e}}, R., {Angl{\'e}s-Alc{\'a}zar}, D., {Narayanan}, D., {et~al.} 2019,
  \mnras, 486, 2827, \dodoi{10.1093/mnras/stz937}

\bibitem[{{Davies} {et~al.}(2018){Davies}, {Robotham}, {Driver}, {Lagos},
  {Cortese}, {Mannering}, {Foster}, {Lidman}, {Hashemizadeh}, {Koushan},
  {O'Toole}, {Baldry}, {Bilicki}, {Bland-Hawthorn}, {Bremer}, {Brown},
  {Bryant}, {Catinella}, {Croom}, {Grootes}, {Holwerda}, {Jarvis}, {Maddox},
  {Meyer}, {Moffett}, {Phillipps}, {Taylor}, {Windhorst}, \&
  {Wolf}}]{daviesDeepExtragalacticVIsible2018}
{Davies}, L.~J.~M., {Robotham}, A.~S.~G., {Driver}, S.~P., {et~al.} 2018,
  \mnras, 480, 768, \dodoi{10.1093/mnras/sty1553}

\bibitem[{{Davies} {et~al.}(2021){Davies}, {Thorne}, {Robotham}, {Bellstedt},
  {Driver}, {Adams}, {Bilicki}, {Bowler}, {Bravo}, {Cortese}, {Foster},
  {Grootes}, {H{\"a}u{\ss}ler}, {Hashemizadeh}, {Holwerda}, {Hurley}, {Jarvis},
  {Lidman}, {Maddox}, {Meyer}, {Paolillo}, {Phillipps}, {Radovich}, {Siudek},
  {Vaccari}, \& {Windhorst}}]{daviesDeepExtragalacticVIsible2021a}
{Davies}, L.~J.~M., {Thorne}, J.~E., {Robotham}, A.~S.~G., {et~al.} 2021,
  \mnras, 506, 256, \dodoi{10.1093/mnras/stab1601}

\bibitem[{Davies {et~al.}(2022)Davies, Thorne, Bellstedt, Bravo, Robotham,
  Driver, Cook, Cortese, D'Silva, Grootes, Holwerda, Hopkins, Jarvis, Lidman,
  Phillipps, \& Siudek}]{daviesDeepExtragalacticVIsible2022}
Davies, L. J.~M., Thorne, J.~E., Bellstedt, S., {et~al.} 2022, \mnras, 509,
  4392, \dodoi{10.1093/mnras/stab3145}

\bibitem[{Donnan {et~al.}(2023)Donnan, McLeod, Dunlop, McLure, Carnall, Begley,
  Cullen, Hamadouche, Bowler, Magee, McCracken, {Milvang-Jensen}, Moneti, \&
  Targett}]{donnanEvolutionGalaxyUV2023}
Donnan, C.~T., McLeod, D.~J., Dunlop, J.~S., {et~al.} 2023, Monthly Notices of
  the Royal Astronomical Society, 518, 6011, \dodoi{10.1093/mnras/stac3472}

\bibitem[{{Donnan} {et~al.}(2024){Donnan}, {McLure}, {Dunlop}, {McLeod},
  {Magee}, {Arellano-C{\'o}rdova}, {Barrufet}, {Begley}, {Bowler}, {Carnall},
  {Cullen}, {Ellis}, {Fontana}, {Illingworth}, {Grogin}, {Hamadouche},
  {Koekemoer}, {Liu}, {Mason}, {Santini}, \&
  {Stanton}}]{donnanJWSTPRIMERNew2024}
{Donnan}, C.~T., {McLure}, R.~J., {Dunlop}, J.~S., {et~al.} 2024, \mnras, 533,
  3222, \dodoi{10.1093/mnras/stae2037}

\bibitem[{{Driver} \& {Robotham}(2010)}]{driverQuantifyingCosmicVariance2010}
{Driver}, S.~P., \& {Robotham}, A. S.~G. 2010, \mnras, 407, 2131,
  \dodoi{10.1111/j.1365-2966.2010.17028.x}

\bibitem[{{Driver} {et~al.}(2011){Driver}, {Hill}, {Kelvin}, {Robotham},
  {Liske}, {Norberg}, {Baldry}, {Bamford}, {Hopkins}, {Loveday}, {Peacock},
  {Andrae}, {Bland-Hawthorn}, {Brough}, {Brown}, {Cameron}, {Ching}, {Colless},
  {Conselice}, {Croom}, {Cross}, {de Propris}, {Dye}, {Drinkwater}, {Ellis},
  {Graham}, {Grootes}, {Gunawardhana}, {Jones}, {van Kampen}, {Maraston},
  {Nichol}, {Parkinson}, {Phillipps}, {Pimbblet}, {Popescu}, {Prescott},
  {Roseboom}, {Sadler}, {Sansom}, {Sharp}, {Smith}, {Taylor}, {Thomas},
  {Tuffs}, {Wijesinghe}, {Dunne}, {Frenk}, {Jarvis}, {Madore}, {Meyer},
  {Seibert}, {Staveley-Smith}, {Sutherland}, \&
  {Warren}}]{driverGalaxyMassAssembly2011a}
{Driver}, S.~P., {Hill}, D.~T., {Kelvin}, L.~S., {et~al.} 2011, \mnras, 413,
  971, \dodoi{10.1111/j.1365-2966.2010.18188.x}

\bibitem[{Driver {et~al.}(2018)Driver, Andrews, {da Cunha}, Davies, Lagos,
  Robotham, Vinsen, Wright, Alpaslan, {Bland-Hawthorn}, Bourne, Brough, Bremer,
  Cluver, Colless, Conselice, Dunne, Eales, Gomez, Holwerda, Hopkins, Kafle,
  Kelvin, Loveday, Liske, Maddox, Phillipps, Pimbblet, Rowlands, Sansom,
  Taylor, Wang, \& Wilkins}]{driverGAMAG10COSMOS3DHST2018}
Driver, S.~P., Andrews, S.~K., {da Cunha}, E., {et~al.} 2018, \mnras, 475,
  2891, \dodoi{10.1093/mnras/stx2728}

\bibitem[{{Driver} {et~al.}(2022){Driver}, {Bellstedt}, {Robotham}, {Baldry},
  {Davies}, {Liske}, {Obreschkow}, {Taylor}, {Wright}, {Alpaslan}, {Bamford},
  {Bauer}, {Bland-Hawthorn}, {Bilicki}, {Bravo}, {Brough}, {Casura}, {Cluver},
  {Colless}, {Conselice}, {Croom}, {de Jong}, {D'Eugenio}, {De Propris},
  {Dogruel}, {Drinkwater}, {Dvornik}, {Farrow}, {Frenk}, {Giblin}, {Graham},
  {Grootes}, {Gunawardhana}, {Hashemizadeh}, {H{\"a}u{\ss}ler}, {Heymans},
  {Hildebrandt}, {Holwerda}, {Hopkins}, {Jarrett}, {Heath Jones}, {Kelvin},
  {Koushan}, {Kuijken}, {Lara-L{\'o}pez}, {Lange}, {L{\'o}pez-S{\'a}nchez},
  {Loveday}, {Mahajan}, {Meyer}, {Moffett}, {Napolitano}, {Norberg}, {Owers},
  {Radovich}, {Raouf}, {Peacock}, {Phillipps}, {Pimbblet}, {Popescu}, {Said},
  {Sansom}, {Seibert}, {Sutherland}, {Thorne}, {Tuffs}, {Turner}, {van der
  Wel}, {van Kampen}, \& {Wilkins}}]{driverGalaxyMassAssembly2022a}
{Driver}, S.~P., {Bellstedt}, S., {Robotham}, A. S.~G., {et~al.} 2022, \mnras,
  513, 439, \dodoi{10.1093/mnras/stac472}

\bibitem[{{D'Silva} {et~al.}(2023{\natexlab{a}}){D'Silva}, {Driver}, {Lagos},
  {Robotham}, {Summers}, \& {Windhorst}}]{dsilvaStarFormationAGN2023}
{D'Silva}, J. C.~J., {Driver}, S.~P., {Lagos}, C. D.~P., {et~al.}
  2023{\natexlab{a}}, \apjl, 959, L18, \dodoi{10.3847/2041-8213/ad103e}

\bibitem[{{D'Silva} {et~al.}(2023{\natexlab{b}}){D'Silva}, {Lagos}, {Davies},
  {Lovell}, \& {Vijayan}}]{dsilvaUnveilingMainSequence2023b}
{D'Silva}, J. C.~J., {Lagos}, C. D.~P., {Davies}, L. J.~M., {Lovell}, C.~C., \&
  {Vijayan}, A.~P. 2023{\natexlab{b}}, \mnras, 518, 456,
  \dodoi{10.1093/mnras/stac2878}

\bibitem[{{D'Silva} {et~al.}(2023{\natexlab{c}}){D'Silva}, {Driver}, {Lagos},
  {Robotham}, {Bellstedt}, {Davies}, {Thorne}, {Bland-Hawthorn}, {Bravo},
  {Holwerda}, {Phillipps}, {Seymour}, {Siudek}, \&
  {Windhorst}}]{dsilvaGAMADEVILSCosmic2023}
{D'Silva}, J. C.~J., {Driver}, S.~P., {Lagos}, C. D.~P., {et~al.}
  2023{\natexlab{c}}, \mnras, 524, 1448, \dodoi{10.1093/mnras/stad1974}

\bibitem[{D'Silva {et~al.}(2025)}]{jordandsilva_2025_15086450}
D'Silva, J. C.~J., {et~al.} 2025, JordanDSilva/JUMPROPE: JUMPROPE MIRI and
  CHUNK, v1.3.2,  Zenodo, \dodoi{10.5281/zenodo.15086450}

\bibitem[{{Eisenstein} {et~al.}(2023){Eisenstein}, {Willott}, {Alberts},
  {Arribas}, {Bonaventura}, {Bunker}, {Cameron}, {Carniani}, {Charlot},
  {Curtis-Lake}, {D'Eugenio}, {Endsley}, {Ferruit}, {Giardino}, {Hainline},
  {Hausen}, {Jakobsen}, {Johnson}, {Maiolino}, {Rieke}, {Rieke}, {Rix},
  {Robertson}, {Stark}, {Tacchella}, {Williams}, {Willmer}, {Baker}, {Baum},
  {Bhatawdekar}, {Boyett}, {Chen}, {Chevallard}, {Circosta}, {Curti},
  {Danhaive}, {DeCoursey}, {de Graaff}, {Dressler}, {Egami}, {Helton},
  {Hviding}, {Ji}, {Jones}, {Kumari}, {L{\"u}tzgendorf}, {Laseter}, {Looser},
  {Lyu}, {Maseda}, {Nelson}, {Parlanti}, {Perna}, {Pusk{\'a}s}, {Rawle},
  {Rodr{\'\i}guez Del Pino}, {Sandles}, {Saxena}, {Scholtz}, {Sharpe},
  {Shivaei}, {Silcock}, {Simmonds}, {Skarbinski}, {Smit}, {Stone}, {Suess},
  {Sun}, {Tang}, {Topping}, {{\"U}bler}, {Villanueva}, {Wallace}, {Whitler},
  {Witstok}, \& {Woodrum}}]{eisensteinOverviewJWSTAdvanced2023a}
{Eisenstein}, D.~J., {Willott}, C., {Alberts}, S., {et~al.} 2023, arXiv
  e-prints, arXiv:2306.02465, \dodoi{10.48550/arXiv.2306.02465}

\bibitem[{Feltre {et~al.}(2012)Feltre, Hatziminaoglou, Fritz, \&
  Franceschini}]{feltreSmoothClumpyDust2012a}
Feltre, A., Hatziminaoglou, E., Fritz, J., \& Franceschini, A. 2012, Monthly
  Notices of the Royal Astronomical Society, 426, 120,
  \dodoi{10.1111/j.1365-2966.2012.21695.x}

\bibitem[{{Ferreira} {et~al.}(2023){Ferreira}, {Conselice}, {Sazonova},
  {Ferrari}, {Caruana}, {Tohill}, {Lucatelli}, {Adams}, {Irodotou}, {Marshall},
  {Roper}, {Lovell}, {Verma}, {Austin}, {Trussler}, \&
  {Wilkins}}]{ferreiraJWSTHubbleSequence2023}
{Ferreira}, L., {Conselice}, C.~J., {Sazonova}, E., {et~al.} 2023, \apj, 955,
  94, \dodoi{10.3847/1538-4357/acec76}

\bibitem[{Finkelstein {et~al.}(2015)Finkelstein, Ryan~Jr., Papovich, Dickinson,
  Song, Somerville, Ferguson, Salmon, Giavalisco, Koekemoer, Ashby, Behroozi,
  Castellano, Dunlop, Faber, Fazio, Fontana, Grogin, Hathi, Jaacks, Kocevski,
  Livermore, McLure, Merlin, Mobasher, Newman, Rafelski, Tilvi, \&
  Willner}]{finkelsteinEvolutionGalaxyRestFrame2015}
Finkelstein, S.~L., Ryan~Jr., R.~E., Papovich, C., {et~al.} 2015, \apj, 810,
  71, \dodoi{10.1088/0004-637X/810/1/71}

\bibitem[{Finkelstein {et~al.}(2022{\natexlab{a}})Finkelstein, Bagley, Song,
  Larson, Papovich, Dickinson, Finkelstein, Koekemoer, Pirzkal, Somerville,
  Yung, Behroozi, Ferguson, Giavalisco, Grogin, Hathi, Hutchison, Jung,
  Kocevski, Kawinwanichakij, {Rojas-Ruiz}, Ryan, Snyder, \&
  Tacchella}]{finkelsteinCensusBright85112022}
Finkelstein, S.~L., Bagley, M., Song, M., {et~al.} 2022{\natexlab{a}}, The
  Astrophysical Journal, 928, 52, \dodoi{10.3847/1538-4357/ac3aed}

\bibitem[{Finkelstein {et~al.}(2022{\natexlab{b}})Finkelstein, Bagley, Haro,
  Dickinson, Ferguson, Kartaltepe, Papovich, Burgarella, Kocevski,
  {Huertas-Company}, Iyer, Larson, {P{\'e}rez-Gonz{\'a}lez}, Rose, Tacchella,
  Wilkins, Chworowsky, Medrano, Morales, Somerville, Yung, Fontana, Giavalisco,
  Grazian, Grogin, Kewley, Koekemoer, Kirkpatrick, Kurczynski, Lotz,
  Pentericci, Pirzkal, Ravindranath, Ryan~Jr., Trump, Yang, Almaini,
  Amor{\'i}n, Annunziatella, Backhaus, Barro, Behroozi, Bell, Bhatawdekar,
  Bisigello, Bromm, Buat, Buitrago, Calabr{\'o}, Casey, Castellano, Ortiz,
  Ciesla, Cleri, Cohen, Cole, Cooke, Cooper, Cooray, Costantin, Cox, Croton,
  Daddi, Dav{\'e}, {de la Vega}, Dekel, Elbaz, {Estrada-Carpenter}, Faber,
  Fern{\'a}ndez, Finkelstein, Freundlich, Fujimoto, {Garc{\'i}a-Argum{\'a}nez},
  Gardner, Gawiser, {G{\'o}mez-Guijarro}, Guo, Hamilton, Hathi, Holwerda,
  Hirschmann, Hutchison, Jaskot, Jha, Jogee, Juneau, Jung, Kassin, Bail, Leung,
  Lucas, Magnelli, Mantha, Matharu, McGrath, McIntosh, Merlin, Mobasher,
  Newman, Nicholls, Pandya, Rafelski, Ronayne, Santini, Seill{\'e}, Shah, Shen,
  Simons, Snyder, Stanway, Straughn, Teplitz, Vanderhoof, {Vega-Ferrero}, Wang,
  Weiner, Willmer, Wuyts, \& Zavala}]{finkelsteinLongTimeAgo2022}
Finkelstein, S.~L., Bagley, M.~B., Haro, P.~A., {et~al.} 2022{\natexlab{b}},
  \apjl, 940, L55, \dodoi{10.3847/2041-8213/ac966e}

\bibitem[{{Finkelstein} {et~al.}(2023){Finkelstein}, {Bagley}, {Ferguson},
  {Wilkins}, {Kartaltepe}, {Papovich}, {Yung}, {Arrabal Haro}, {Behroozi},
  {Dickinson}, {Kocevski}, {Koekemoer}, {Larson}, {Le Bail}, {Morales},
  {P{\'e}rez-Gonz{\'a}lez}, {Burgarella}, {Dav{\'e}}, {Hirschmann},
  {Somerville}, {Wuyts}, {Bromm}, {Casey}, {Fontana}, {Fujimoto}, {Gardner},
  {Giavalisco}, {Grazian}, {Grogin}, {Hathi}, {Hutchison}, {Jha}, {Jogee},
  {Kewley}, {Kirkpatrick}, {Long}, {Lotz}, {Pentericci}, {Pierel}, {Pirzkal},
  {Ravindranath}, {Ryan}, {Trump}, {Yang}, {Bhatawdekar}, {Bisigello}, {Buat},
  {Calabr{\`o}}, {Castellano}, {Cleri}, {Cooper}, {Croton}, {Daddi}, {Dekel},
  {Elbaz}, {Franco}, {Gawiser}, {Holwerda}, {Huertas-Company}, {Jaskot},
  {Leung}, {Lucas}, {Mobasher}, {Pandya}, {Tacchella}, {Weiner}, \&
  {Zavala}}]{finkelsteinCEERSKeyPaper2023a}
{Finkelstein}, S.~L., {Bagley}, M.~B., {Ferguson}, H.~C., {et~al.} 2023, \apjl,
  946, L13, \dodoi{10.3847/2041-8213/acade4}

\bibitem[{Fitzpatrick \&
  Massa(2007)}]{fitzpatrickAnalysisShapesInterstellar2007a}
Fitzpatrick, E.~L., \& Massa, D. 2007, \apj, 663, 320, \dodoi{10.1086/518158}

\bibitem[{Fritz {et~al.}(2006)Fritz, Franceschini, \&
  Hatziminaoglou}]{fritz06agnmodel}
Fritz, J., Franceschini, A., \& Hatziminaoglou, E. 2006, \mnras, 366, 767,
  \dodoi{10.1111/j.1365-2966.2006.09866.x}

\bibitem[{Frye {et~al.}(2023)Frye, Pascale, Foo, Leimbach, Garuda, Robles,
  Summers, Diaz, Kamieneski, Furtak, Cohen, Diego, Beauchesne, Windhorst,
  Willner, Koekemoer, Zitrin, Caminha, Caputi, Coe, Conselice, Dai, Dole,
  Driver, Grogin, Harrington, Jansen, Kneib, Lehnert, Lowenthal, Marshall,
  Menanteau, Pampliega, Pirzkal, Polletta, Richard, Robotham, Ryan, Rutkowski,
  Sif{\'o}n, Tompkins, Wang, Yan, \& Yun}]{fryeJWSTPEARLSView2023}
Frye, B.~L., Pascale, M., Foo, N., {et~al.} 2023, \apj, 952, 81,
  \dodoi{10.3847/1538-4357/acd929}

\bibitem[{{Furlong} {et~al.}(2015){Furlong}, {Bower}, {Theuns}, {Schaye},
  {Crain}, {Schaller}, {Dalla Vecchia}, {Frenk}, {McCarthy}, {Helly},
  {Jenkins}, \& {Rosas-Guevara}}]{furlongEvolutionGalaxyStellar2015}
{Furlong}, M., {Bower}, R.~G., {Theuns}, T., {et~al.} 2015, \mnras, 450, 4486,
  \dodoi{10.1093/mnras/stv852}

\bibitem[{Gardner {et~al.}(2006)Gardner, Mather, Clampin, Doyon, Greenhouse,
  Hammel, Hutchings, Jakobsen, Lilly, Long, Lunine, McCaughrean, Mountain,
  Nella, Rieke, Rieke, Rix, Smith, Sonneborn, Stiavelli, Stockman, Windhorst,
  \& Wright}]{gardnerJamesWebbSpace2006a}
Gardner, J.~P., Mather, J.~C., Clampin, M., {et~al.} 2006, \ssr, 123, 485,
  \dodoi{10.1007/s11214-006-8315-7}

\bibitem[{Gebhardt {et~al.}(2000)Gebhardt, Bender, Bower, Dressler, Faber,
  Filippenko, Green, Grillmair, Ho, Kormendy, Lauer, Magorrian, Pinkney,
  Richstone, \& Tremaine}]{gebhardtRelationshipNuclearBlack2000a}
Gebhardt, K., Bender, R., Bower, G., {et~al.} 2000, \apj, 539, L13,
  \dodoi{10.1086/312840}

\bibitem[{Glazebrook {et~al.}(2024)Glazebrook, Nanayakkara, Schreiber, Lagos,
  Kawinwanichakij, Jacobs, Chittenden, Brammer, Kacprzak, Labbe, Marchesini,
  Marsan, Oesch, Papovich, Remus, Tran, Esdaile, \&
  {Chandro-Gomez}}]{glazebrookMassiveGalaxyThat2024}
Glazebrook, K., Nanayakkara, T., Schreiber, C., {et~al.} 2024, Nature, 628,
  277, \dodoi{10.1038/s41586-024-07191-9}

\bibitem[{Greene {et~al.}(2024)Greene, Labbe, Goulding, Furtak, Chemerynska,
  Kokorev, Dayal, Volonteri, Williams, Wang, Setton, Burgasser, Bezanson, Atek,
  Brammer, Cutler, Feldmann, Fujimoto, Glazebrook, {de Graaff}, Khullar, Leja,
  Marchesini, Maseda, Matthee, Miller, Naidu, Nanayakkara, Oesch, Pan,
  Papovich, Price, {van Dokkum}, Weaver, Whitaker, \&
  Zitrin}]{greeneUNCOVERSpectroscopyConfirms2024}
Greene, J.~E., Labbe, I., Goulding, A.~D., {et~al.} 2024, \apj, 964, 39,
  \dodoi{10.3847/1538-4357/ad1e5f}

\bibitem[{{Grogin} {et~al.}(2011){Grogin}, {Kocevski}, {Faber}, {Ferguson},
  {Koekemoer}, {Riess}, {Acquaviva}, {Alexander}, {Almaini}, {Ashby}, {Barden},
  {Bell}, {Bournaud}, {Brown}, {Caputi}, {Casertano}, {Cassata}, {Castellano},
  {Challis}, {Chary}, {Cheung}, {Cirasuolo}, {Conselice}, {Roshan Cooray},
  {Croton}, {Daddi}, {Dahlen}, {Dav{\'e}}, {de Mello}, {Dekel}, {Dickinson},
  {Dolch}, {Donley}, {Dunlop}, {Dutton}, {Elbaz}, {Fazio}, {Filippenko},
  {Finkelstein}, {Fontana}, {Gardner}, {Garnavich}, {Gawiser}, {Giavalisco},
  {Grazian}, {Guo}, {Hathi}, {H{\"a}ussler}, {Hopkins}, {Huang}, {Huang},
  {Jha}, {Kartaltepe}, {Kirshner}, {Koo}, {Lai}, {Lee}, {Li}, {Lotz}, {Lucas},
  {Madau}, {McCarthy}, {McGrath}, {McIntosh}, {McLure}, {Mobasher},
  {Moustakas}, {Mozena}, {Nandra}, {Newman}, {Niemi}, {Noeske}, {Papovich},
  {Pentericci}, {Pope}, {Primack}, {Rajan}, {Ravindranath}, {Reddy}, {Renzini},
  {Rix}, {Robaina}, {Rodney}, {Rosario}, {Rosati}, {Salimbeni}, {Scarlata},
  {Siana}, {Simard}, {Smidt}, {Somerville}, {Spinrad}, {Straughn}, {Strolger},
  {Telford}, {Teplitz}, {Trump}, {van der Wel}, {Villforth}, {Wechsler},
  {Weiner}, {Wiklind}, {Wild}, {Wilson}, {Wuyts}, {Yan}, \&
  {Yun}}]{groginCANDELS2011}
{Grogin}, N.~A., {Kocevski}, D.~D., {Faber}, S.~M., {et~al.} 2011, \apjs, 197,
  35, \dodoi{10.1088/0067-0049/197/2/35}

\bibitem[{Harikane {et~al.}(2024)Harikane, Nakajima, Ouchi, Umeda, Isobe, Ono,
  Xu, \& Zhang}]{harikanePureSpectroscopicConstraints2024}
Harikane, Y., Nakajima, K., Ouchi, M., {et~al.} 2024, \apj, 960, 56,
  \dodoi{10.3847/1538-4357/ad0b7e}

\bibitem[{Harikane {et~al.}(2022)Harikane, Ono, Ouchi, Liu, Sawicki, Shibuya,
  Behroozi, He, Shimasaku, Arnouts, Coupon, Fujimoto, Gwyn, Huang, Inoue,
  Kashikawa, Komiyama, Matsuoka, \& Willott}]{harikaneGOLDRUSHIVLuminosity2022}
Harikane, Y., Ono, Y., Ouchi, M., {et~al.} 2022, \apjs, 259, 20,
  \dodoi{10.3847/1538-4365/ac3dfc}

\bibitem[{{Harikane} {et~al.}(2023){Harikane}, {Zhang}, {Nakajima}, {Ouchi},
  {Isobe}, {Ono}, {Hatano}, {Xu}, \& {Umeda}}]{harikaneJWSTNIRSpecFirst2023b}
{Harikane}, Y., {Zhang}, Y., {Nakajima}, K., {et~al.} 2023, \apj, 959, 39,
  \dodoi{10.3847/1538-4357/ad029e}

\bibitem[{Harikane {et~al.}(2023)Harikane, Ouchi, Oguri, Ono, Nakajima, Isobe,
  Umeda, Mawatari, \& Zhang}]{harikaneComprehensiveStudyGalaxies2023}
Harikane, Y., Ouchi, M., Oguri, M., {et~al.} 2023, \apjs, 265, 5,
  \dodoi{10.3847/1538-4365/acaaa9}

\bibitem[{Harvey {et~al.}(2024)Harvey, Conselice, Adams, Austin, Juod{\v
  z}balis, Trussler, Li, Ormerod, Ferreira, Lovell, Duan, Westcott, Harris,
  Bhatawdekar, Coe, Cohen, Caruana, Cheng, Driver, Frye, Furtak, Grogin, Hathi,
  Holwerda, Jansen, Koekemoer, Marshall, Nonino, Vijayan, Wilkins, Windhorst,
  Willmer, Yan, \& Zitrin}]{harveyEPOCHSIVSED2024}
Harvey, T., Conselice, C.~J., Adams, N.~J., {et~al.} 2024, The Astrophysical
  Journal, 978, 89, \dodoi{10.3847/1538-4357/ad8c29}

\bibitem[{Hickox \& Alexander(2018)}]{hickoxObscuredActiveGalactic2018}
Hickox, R.~C., \& Alexander, D.~M. 2018, Annual Review of Astronomy and
  Astrophysics, 56, 625, \dodoi{10.1146/annurev-astro-081817-051803}

\bibitem[{Juod{\v z}balis {et~al.}(2023)Juod{\v z}balis, Conselice, Singh,
  Adams, Ormerod, Harvey, Austin, Volonteri, Cohen, Jansen, Summers, Windhorst,
  D'Silva, Koekemoer, Coe, Driver, Frye, Grogin, Marshall, Nonino, Pirzkal,
  Robotham, Ryan, Ortiz~III, Tompkins, Willmer, \&
  Yan}]{juodzbalisEPOCHSVIIDiscovery2023}
Juod{\v z}balis, I., Conselice, C.~J., Singh, M., {et~al.} 2023, \mnras, 525,
  1353, \dodoi{10.1093/mnras/stad2396}

\bibitem[{Kartaltepe {et~al.}(2023)Kartaltepe, Rose, Vanderhoof, McGrath,
  Costantin, Cox, Yung, Kocevski, Wuyts, Andrews, Bagley, Finkelstein, Amorin,
  Haro, Backhaus, Behroozi, Bisigello, Calabro, Casey, Coogan, Croton, {de la
  Vega}, Dickinson, Cooper, Fontana, Franco, Grazian, Grogin, Hathi, Holwerda,
  {Huertas-Company}, Iyer, Jogee, Jung, Kewley, Kirkpatrick, Koekemoer, Liu,
  Lotz, Lucas, Newman, Pacifici, Pandya, Papovich, Pentericci,
  {Perez-Gonzalez}, Petersen, Pirzkal, Rafelski, Ravindranath, Simons, Snyder,
  Somerville, Stanway, Straughn, Tacchella, Trump, {Vega-Ferrero}, Wilkins,
  Yang, \& Zavala}]{kartaltepeCEERSKeyPaper2023a}
Kartaltepe, J.~S., Rose, C., Vanderhoof, B.~N., {et~al.} 2023, \apjl, 946, L15,
  \dodoi{10.3847/2041-8213/acad01}

\bibitem[{Katsianis {et~al.}(2019)Katsianis, Zheng, Gonzalez, Blanc, Lagos,
  Davies, Camps, Tr{\v c}ka, Baes, Schaye, Trayford, Theuns, \&
  Stalevski}]{katsianisEvolvingMassdependentSsSFRM2019}
Katsianis, A., Zheng, X., Gonzalez, V., {et~al.} 2019, \apj, 879, 11,
  \dodoi{10.3847/1538-4357/ab1f8d}

\bibitem[{Kauffmann \& Haehnelt(2000)}]{kauffmannUnifiedModelEvolution2000}
Kauffmann, G., \& Haehnelt, M. 2000, \mnras, 311, 576,
  \dodoi{10.1046/j.1365-8711.2000.03077.x}

\bibitem[{{Keel} {et~al.}(2023){Keel}, {Windhorst}, {Jansen}, {Cohen},
  {Summers}, {Holwerda}, {Bradford}, {Robertson}, {Ferrami}, {Wyithe}, {Yan},
  {Conselice}, {Driver}, {Robotham}, {Grogin}, {Willmer}, {Koekemoer}, {Frye},
  {Hathi}, {Ryan}, {Pirzkal}, {Marshall}, {Coe}, {Diego}, {Broadhurst},
  {Rutkowski}, {Wang}, {Willner}, {Petric}, {Cheng}, \&
  {Zitrin}}]{keelJWSTPEARLSDust2023a}
{Keel}, W.~C., {Windhorst}, R.~A., {Jansen}, R.~A., {et~al.} 2023, \aj, 165,
  166, \dodoi{10.3847/1538-3881/acbdff}

\bibitem[{{Kim} {et~al.}(2019){Kim}, {Jansen}, {Windhorst}, {Cohen}, \&
  {McCabe}}]{kimAnalysisSpatiallyResolved2019}
{Kim}, D., {Jansen}, R.~A., {Windhorst}, R.~A., {Cohen}, S.~H., \& {McCabe},
  T.~J. 2019, \apj, 884, 21, \dodoi{10.3847/1538-4357/ab385c}

\bibitem[{Kocevski {et~al.}(2023)Kocevski, Onoue, Inayoshi, Trump,
  Arrabal~Haro, Grazian, Dickinson, Finkelstein, Kartaltepe, Hirschmann, Aird,
  Holwerda, Fujimoto, Juneau, Amor{\'i}n, Backhaus, Bagley, Barro, Bell,
  Bisigello, Calabr{\`o}, Cleri, Cooper, Ding, Grogin, Ho, Hutchison, Inoue,
  Jiang, Jones, Koekemoer, Li, Li, McGrath, Molina, Papovich,
  {P{\'e}rez-Gonz{\'a}lez}, Pirzkal, Wilkins, Yang, \&
  Yung}]{kocevskiHiddenLittleMonsters2023a}
Kocevski, D.~D., Onoue, M., Inayoshi, K., {et~al.} 2023, \apj, 954, L4,
  \dodoi{10.3847/2041-8213/ace5a0}

\bibitem[{Kocevski {et~al.}(2024)Kocevski, Finkelstein, Barro, Taylor,
  Calabr{\`o}, Laloux, Buchner, Trump, Leung, Yang, Dickinson,
  {P{\'e}rez-Gonz{\'a}lez}, Pacucci, Inayoshi, Somerville, McGrath, Akins,
  Bagley, Bisigello, Bowler, Carnall, Casey, Cheng, Cleri, Costantin, Cullen,
  Davis, Donnan, Dunlop, Ellis, Ferguson, Fujimoto, Fontana, Giavalisco,
  Grazian, Grogin, Hathi, Hirschmann, {Huertas-Company}, Holwerda, Illingworth,
  Juneau, Kartaltepe, Koekemoer, Li, Lucas, Magee, Mason, McLeod, McLure,
  Napolitano, Papovich, Pirzkal, Rodighiero, Santini, Wilkins, \&
  Yung}]{kocevskiRiseFaintRed2024a}
Kocevski, D.~D., Finkelstein, S.~L., Barro, G., {et~al.} 2024, The {{Rise}} of
  {{Faint}}, {{Red AGN}} at \$z{$>$}4\$: {{A Sample}} of {{Little Red Dots}} in
  the {{JWST Extragalactic Legacy Fields}}, \dodoi{10.48550/arXiv.2404.03576}

\bibitem[{{Koekemoer} {et~al.}(2011){Koekemoer}, {Faber}, {Ferguson}, {Grogin},
  {Kocevski}, {Koo}, {Lai}, {Lotz}, {Lucas}, {McGrath}, {Ogaz}, {Rajan},
  {Riess}, {Rodney}, {Strolger}, {Casertano}, {Castellano}, {Dahlen},
  {Dickinson}, {Dolch}, {Fontana}, {Giavalisco}, {Grazian}, {Guo}, {Hathi},
  {Huang}, {van der Wel}, {Yan}, {Acquaviva}, {Alexander}, {Almaini}, {Ashby},
  {Barden}, {Bell}, {Bournaud}, {Brown}, {Caputi}, {Cassata}, {Challis},
  {Chary}, {Cheung}, {Cirasuolo}, {Conselice}, {Roshan Cooray}, {Croton},
  {Daddi}, {Dav{\'e}}, {de Mello}, {de Ravel}, {Dekel}, {Donley}, {Dunlop},
  {Dutton}, {Elbaz}, {Fazio}, {Filippenko}, {Finkelstein}, {Frazer}, {Gardner},
  {Garnavich}, {Gawiser}, {Gruetzbauch}, {Hartley}, {H{\"a}ussler},
  {Herrington}, {Hopkins}, {Huang}, {Jha}, {Johnson}, {Kartaltepe},
  {Khostovan}, {Kirshner}, {Lani}, {Lee}, {Li}, {Madau}, {McCarthy},
  {McIntosh}, {McLure}, {McPartland}, {Mobasher}, {Moreira}, {Mortlock},
  {Moustakas}, {Mozena}, {Nandra}, {Newman}, {Nielsen}, {Niemi}, {Noeske},
  {Papovich}, {Pentericci}, {Pope}, {Primack}, {Ravindranath}, {Reddy},
  {Renzini}, {Rix}, {Robaina}, {Rosario}, {Rosati}, {Salimbeni}, {Scarlata},
  {Siana}, {Simard}, {Smidt}, {Snyder}, {Somerville}, {Spinrad}, {Straughn},
  {Telford}, {Teplitz}, {Trump}, {Vargas}, {Villforth}, {Wagner}, {Wandro},
  {Wechsler}, {Weiner}, {Wiklind}, {Wild}, {Wilson}, {Wuyts}, \&
  {Yun}}]{koekemoerCANDELS2011}
{Koekemoer}, A.~M., {Faber}, S.~M., {Ferguson}, H.~C., {et~al.} 2011, \apjs,
  197, 36, \dodoi{10.1088/0067-0049/197/2/36}

\bibitem[{Kokorev {et~al.}(2024)Kokorev, Caputi, Greene, Dayal, Trebitsch,
  Cutler, Fujimoto, Labb{\'e}, Miller, Iani, {Navarro-Carrera}, \&
  Rinaldi}]{kokorevCensusPhotometricallySelected2024}
Kokorev, V., Caputi, K.~I., Greene, J.~E., {et~al.} 2024, \apj, 968, 38,
  \dodoi{10.3847/1538-4357/ad4265}

\bibitem[{Kormendy \& Ho(2013)}]{kormendyCoevolutionNotSupermassive2013}
Kormendy, J., \& Ho, L.~C. 2013, \araa, 51, 511,
  \dodoi{10.1146/annurev-astro-082708-101811}

\bibitem[{Kroupa(2002)}]{kroupaInitialMassFunction2002}
Kroupa, P. 2002, Science, 295, 82, \dodoi{10.1126/science.1067524}

\bibitem[{Labb{\'e} {et~al.}(2023)Labb{\'e}, {van Dokkum}, Nelson, Bezanson,
  Suess, Leja, Brammer, Whitaker, Mathews, Stefanon, \&
  Wang}]{labbePopulationRedCandidate2023}
Labb{\'e}, I., {van Dokkum}, P., Nelson, E., {et~al.} 2023, Nature, 616, 266,
  \dodoi{10.1038/s41586-023-05786-2}

\bibitem[{Lacey {et~al.}(2016)Lacey, Baugh, Frenk, Benson, Bower, Cole,
  {Gonzalez-Perez}, Helly, Lagos, \&
  Mitchell}]{laceyUnifiedMultiwavelengthModel2016}
Lacey, C.~G., Baugh, C.~M., Frenk, C.~S., {et~al.} 2016, Monthly Notices of the
  Royal Astronomical Society, 462, 3854, \dodoi{10.1093/mnras/stw1888}

\bibitem[{Lagos {et~al.}(2011)Lagos, Baugh, Lacey, Benson, Kim, \&
  Power}]{lagosCosmicEvolutionAtomic2011}
Lagos, C. D.~P., Baugh, C.~M., Lacey, C.~G., {et~al.} 2011, \mnras, 418, 1649,
  \dodoi{10.1111/j.1365-2966.2011.19583.x}

\bibitem[{{Lagos} {et~al.}(2018){Lagos}, {Tobar}, {Robotham}, {Obreschkow},
  {Mitchell}, {Power}, \& {Elahi}}]{lagosSharkIntroducingOpen2018}
{Lagos}, C. d.~P., {Tobar}, R.~J., {Robotham}, A. S.~G., {et~al.} 2018, \mnras,
  481, 3573, \dodoi{10.1093/mnras/sty2440}

\bibitem[{{Lagos} {et~al.}(2019){Lagos}, {Robotham}, {Trayford}, {Tobar},
  {Bravo}, {Bellstedt}, {Davies}, {Driver}, {Elahi}, {Obreschkow}, \&
  {Power}}]{lagosFarultravioletFarinfraredGalaxy2019}
{Lagos}, C. d.~P., {Robotham}, A. S.~G., {Trayford}, J.~W., {et~al.} 2019,
  \mnras, 489, 4196, \dodoi{10.1093/mnras/stz2427}

\bibitem[{{Lagos} {et~al.}(2024){Lagos}, {Bravo}, {Tobar}, {Obreschkow},
  {Power}, {Robotham}, {Proctor}, {Hansen}, {Chandro-G{\'o}mez}, \&
  {Carrivick}}]{lagosQuenchingMassiveGalaxies2024b}
{Lagos}, C. d.~P., {Bravo}, M., {Tobar}, R., {et~al.} 2024, \mnras, 531, 3551,
  \dodoi{10.1093/mnras/stae1024}

\bibitem[{Lagos {et~al.}(2025)Lagos, Valentino, Wright, {de~Graaff},
  Glazebrook, De~Lucia, Robotham, Nanayakkara, {Chandro-Gomez}, Bravo, Baugh,
  Harborne, Hirschmann, Fontanot, Xie, \& Chittenden}]{Lagos24b}
Lagos, C. d.~P., Valentino, F., Wright, R.~J., {et~al.} 2025, Monthly Notices
  of the Royal Astronomical Society, 536, 2324, \dodoi{10.1093/mnras/stae2626}

\bibitem[{{Langeroodi} \& {Hjorth}(2023)}]{langeroodiLittleRedDots2023}
{Langeroodi}, D., \& {Hjorth}, J. 2023, \apjl, 957, L27,
  \dodoi{10.3847/2041-8213/acfeec}

\bibitem[{{Larson} {et~al.}(2023){Larson}, {Finkelstein}, {Kocevski},
  {Hutchison}, {Trump}, {Arrabal Haro}, {Bromm}, {Cleri}, {Dickinson},
  {Fujimoto}, {Kartaltepe}, {Koekemoer}, {Papovich}, {Pirzkal}, {Tacchella},
  {Zavala}, {Bagley}, {Behroozi}, {Champagne}, {Cole}, {Jung}, {Morales},
  {Yang}, {Zhang}, {Zitrin}, {Amor{\'\i}n}, {Burgarella}, {Casey}, {Ch{\'a}vez
  Ortiz}, {Cox}, {Chworowsky}, {Fontana}, {Gawiser}, {Grazian}, {Grogin},
  {Harish}, {Hathi}, {Hirschmann}, {Holwerda}, {Juneau}, {Leung}, {Lucas},
  {McGrath}, {P{\'e}rez-Gonz{\'a}lez}, {Rigby}, {Seill{\'e}}, {Simons}, {de La
  Vega}, {Weiner}, {Wilkins}, {Yung}, \& {Ceers
  Team}}]{larsonCEERSDiscoveryAccreting2023a}
{Larson}, R.~L., {Finkelstein}, S.~L., {Kocevski}, D.~D., {et~al.} 2023, \apjl,
  953, L29, \dodoi{10.3847/2041-8213/ace619}

\bibitem[{Larson {et~al.}(2023)Larson, Hutchison, Bagley, Finkelstein, Yung,
  Somerville, Hirschmann, Brammer, Holwerda, Papovich, Morales, \&
  Wilkins}]{larsonSpectralTemplatesOptimal2023}
Larson, R.~L., Hutchison, T.~A., Bagley, M., {et~al.} 2023, \apj, 958, 141,
  \dodoi{10.3847/1538-4357/acfed4}

\bibitem[{Latif \& Ferrara(2016)}]{latifFormationSupermassiveBlack2016}
Latif, M.~A., \& Ferrara, A. 2016, \pasa, 33, e051,
  \dodoi{10.1017/pasa.2016.41}

\bibitem[{Leethochawalit {et~al.}(2022)Leethochawalit, Trenti, Morishita,
  {Roberts-Borsani}, \&
  Treu}]{leethochawalitQuantitativeAssessmentCompleteness2022a}
Leethochawalit, N., Trenti, M., Morishita, T., {Roberts-Borsani}, G., \& Treu,
  T. 2022, \mnras, 509, 5836, \dodoi{10.1093/mnras/stab3265}

\bibitem[{{Lotz} {et~al.}(2017){Lotz}, {Koekemoer}, {Coe}, {Grogin}, {Capak},
  {Mack}, {Anderson}, {Avila}, {Barker}, {Borncamp}, {Brammer}, {Durbin},
  {Gunning}, {Hilbert}, {Jenkner}, {Khandrika}, {Levay}, {Lucas}, {MacKenty},
  {Ogaz}, {Porterfield}, {Reid}, {Robberto}, {Royle}, {Smith},
  {Storrie-Lombardi}, {Sunnquist}, {Surace}, {Taylor}, {Williams}, {Bullock},
  {Dickinson}, {Finkelstein}, {Natarajan}, {Richard}, {Robertson}, {Tumlinson},
  {Zitrin}, {Flanagan}, {Sembach}, {Soifer}, \& {Mountain}}]{lotzHFF2017}
{Lotz}, J.~M., {Koekemoer}, A., {Coe}, D., {et~al.} 2017, \apj, 837, 97,
  \dodoi{10.3847/1538-4357/837/1/97}

\bibitem[{Lovell {et~al.}(2023)Lovell, Harrison, Harikane, Tacchella, \&
  Wilkins}]{lovellExtremeValueStatistics2023}
Lovell, C.~C., Harrison, I., Harikane, Y., Tacchella, S., \& Wilkins, S.~M.
  2023, \mnras, 518, 2511, \dodoi{10.1093/mnras/stac3224}

\bibitem[{{Lovell} {et~al.}(2021){Lovell}, {Vijayan}, {Thomas}, {Wilkins},
  {Barnes}, {Irodotou}, \& {Roper}}]{lovellFirstLightReionisation2020}
{Lovell}, C.~C., {Vijayan}, A.~P., {Thomas}, P.~A., {et~al.} 2021, \mnras, 500,
  2127, \dodoi{10.1093/mnras/staa3360}

\bibitem[{{Lyu} {et~al.}(2024){Lyu}, {Alberts}, {Rieke}, {Shivaei},
  {P{\'e}rez-Gonz{\'a}lez}, {Sun}, {Hainline}, {Baum}, {Bonaventura}, {Bunker},
  {Egami}, {Eisenstein}, {Florian}, {Ji}, {Johnson}, {Morrison}, {Rieke},
  {Robertson}, {Rujopakarn}, {Tacchella}, {Scholtz}, \&
  {Willmer}}]{lyuActiveGalacticNuclei2024}
{Lyu}, J., {Alberts}, S., {Rieke}, G.~H., {et~al.} 2024, \apj, 966, 229,
  \dodoi{10.3847/1538-4357/ad3643}

\bibitem[{{Madau} \& {Dickinson}(2014)}]{madaudickinson2014}
{Madau}, P., \& {Dickinson}, M. 2014, \araa, 52, 415,
  \dodoi{10.1146/annurev-astro-081811-125615}

\bibitem[{Magorrian {et~al.}(1998)Magorrian, Tremaine, Richstone, Bender,
  Bower, Dressler, Faber, Gebhardt, Green, Grillmair, Kormendy, \&
  Lauer}]{magorrianDemographyMassiveDark1998}
Magorrian, J., Tremaine, S., Richstone, D., {et~al.} 1998, \aj, 115, 2285,
  \dodoi{10.1086/300353}

\bibitem[{Maiolino {et~al.}(2023)Maiolino, Scholtz, {Curtis-Lake}, Carniani,
  Baker, {de Graaff}, Tacchella, {\"U}bler, D'Eugenio, Witstok, Curti, Arribas,
  Bunker, Charlot, Chevallard, Eisenstein, Egami, Ji, Jones, Lyu, Rawle,
  Robertson, Rujopakarn, Perna, Sun, Venturi, Williams, \&
  Willott}]{maiolinoJADESDiversePopulation2023a}
Maiolino, R., Scholtz, J., {Curtis-Lake}, E., {et~al.} 2023, {{JADES}}. {{The}}
  Diverse Population of Infant {{Black Holes}} at 4,
  \dodoi{10.48550/arXiv.2308.01230}

\bibitem[{Maiolino {et~al.}(2025)Maiolino, Risaliti, Signorini, Trefoloni,
  Juod{\v z}balis, Scholtz, {\"U}bler, D'Eugenio, Carniani, Fabian, Ji,
  Mazzolari, Bertola, Brusa, Bunker, Charlot, Comastri, Cresci, DeCoursey,
  Egami, Fiore, Gilli, Perna, Tacchella, \&
  Venturi}]{maiolinoJWSTMeetsChandra2025a}
Maiolino, R., Risaliti, G., Signorini, M., {et~al.} 2025, Monthly Notices of
  the Royal Astronomical Society, 538, 1921, \dodoi{10.1093/mnras/staf359}

\bibitem[{Marley {et~al.}(2021)Marley, Saumon, Visscher, Lupu, Freedman,
  Morley, Fortney, Seay, Smith, Teal, \& Wang}]{marleySonoraBrownDwarf2021a}
Marley, M.~S., Saumon, D., Visscher, C., {et~al.} 2021, \apj, 920, 85,
  \dodoi{10.3847/1538-4357/ac141d}

\bibitem[{{Matsuoka} {et~al.}(2018){Matsuoka}, {Strauss}, {Kashikawa}, {Onoue},
  {Iwasawa}, {Tang}, {Lee}, {Imanishi}, {Nagao}, {Akiyama}, {Asami}, {Bosch},
  {Furusawa}, {Goto}, {Gunn}, {Harikane}, {Ikeda}, {Izumi}, {Kawaguchi},
  {Kato}, {Kikuta}, {Kohno}, {Komiyama}, {Lupton}, {Minezaki}, {Miyazaki},
  {Murayama}, {Niida}, {Nishizawa}, {Noboriguchi}, {Oguri}, {Ono}, {Ouchi},
  {Price}, {Sameshima}, {Schulze}, {Shirakata}, {Silverman}, {Sugiyama},
  {Tait}, {Takada}, {Takata}, {Tanaka}, {Toba}, {Utsumi}, {Wang}, \&
  {Yamashita}}]{matsuokaSubaruHighzExploration2018a}
{Matsuoka}, Y., {Strauss}, M.~A., {Kashikawa}, N., {et~al.} 2018, \apj, 869,
  150, \dodoi{10.3847/1538-4357/aaee7a}

\bibitem[{Matthee {et~al.}(2024)Matthee, Naidu, Brammer, Chisholm, Eilers,
  Goulding, Greene, Kashino, Labbe, Lilly, Mackenzie, Oesch, Weibel, Wuyts,
  Xiao, Bordoloi, Bouwens, {van Dokkum}, Illingworth, Kramarenko, Maseda,
  Mason, Meyer, Nelson, Reddy, Shivaei, Simcoe, \&
  Yue}]{mattheeLittleRedDots2024}
Matthee, J., Naidu, R.~P., Brammer, G., {et~al.} 2024, \apj, 963, 129,
  \dodoi{10.3847/1538-4357/ad2345}

\bibitem[{Merloni \& Heinz(2008)}]{merloniSynthesisModelAGN2008}
Merloni, A., \& Heinz, S. 2008, \mnras, 388, 1011,
  \dodoi{10.1111/j.1365-2966.2008.13472.x}

\bibitem[{{Moster} {et~al.}(2011){Moster}, {Somerville}, {Newman}, \&
  {Rix}}]{mosterCosmicVarianceCookbook2011}
{Moster}, B.~P., {Somerville}, R.~S., {Newman}, J.~A., \& {Rix}, H.-W. 2011,
  \apj, 731, 113, \dodoi{10.1088/0004-637X/731/2/113}

\bibitem[{Nakajima {et~al.}(2023)Nakajima, Ouchi, Isobe, Harikane, Zhang, Ono,
  Umeda, \& Oguri}]{nakajimaJWSTCensusMassMetallicity2023a}
Nakajima, K., Ouchi, M., Isobe, Y., {et~al.} 2023, \apjs, 269, 33,
  \dodoi{10.3847/1538-4365/acd556}

\bibitem[{{Navarro-Carrera} {et~al.}(2024){Navarro-Carrera}, {Rinaldi},
  {Caputi}, {Iani}, {Kokorev}, \& {van
  Mierlo}}]{navarro-carreraConstraintsFaintEnd2023}
{Navarro-Carrera}, R., {Rinaldi}, P., {Caputi}, K.~I., {et~al.} 2024, \apj,
  961, 207, \dodoi{10.3847/1538-4357/ad0df6}

\bibitem[{Netzer(2019)}]{netzerBolometricCorrectionFactors2019}
Netzer, H. 2019, Monthly Notices of the Royal Astronomical Society, 488, 5185,
  \dodoi{10.1093/mnras/stz2016}

\bibitem[{Obreschkow {et~al.}(2018)Obreschkow, Murray, Robotham, \&
  Westmeier}]{obreschkowEddingtonDemonInferring2018}
Obreschkow, D., Murray, S.~G., Robotham, A. S.~G., \& Westmeier, T. 2018,
  \mnras, 474, 5500, \dodoi{10.1093/mnras/stx3155}

\bibitem[{{O'Brien} {et~al.}(2024){O'Brien}, {Jansen}, {Grogin}, {Cohen},
  {Smith}, {Silver}, {Maksym}, {Windhorst}, {Carleton}, {Koekemoer}, {Hathi},
  {Willmer}, {Frye}, {Alpaslan}, {Ashby}, {Ashcraft}, {Bonoli}, {Brisken},
  {Cappelluti}, {Civano}, {Conselice}, {Dhillon}, {Driver}, {Duncan}, {Dupke},
  {Elvis}, {Fazio}, {Finkelstein}, {Gim}, {Griffiths}, {Hammel}, {Hyun}, {Im},
  {Jones}, {Kim}, {Ladjelate}, {Larson}, {Malhotra}, {Marshall}, {Milam},
  {Pierel}, {Rhoads}, {Rodney}, {R{\"o}ttgering}, {Rutkowski}, {Ryan}, {Ward},
  {White}, {van Weeren}, {Zhao}, {Summers}, {D'Silva}, {Ortiz}, {Robotham},
  {Coe}, {Nonino}, {Pirzkal}, {Yan}, \&
  {Acharya}}]{obrienTREASUREHUNTTransientsVariability2024}
{O'Brien}, R., {Jansen}, R.~A., {Grogin}, N.~A., {et~al.} 2024, \apjs, 272, 19,
  \dodoi{10.3847/1538-4365/ad3948}

\bibitem[{Oesch {et~al.}(2018)Oesch, Bouwens, Illingworth, Labb{\'e}, \&
  Stefanon}]{oeschDearth10Galaxies2018}
Oesch, P.~A., Bouwens, R.~J., Illingworth, G.~D., Labb{\'e}, I., \& Stefanon,
  M. 2018, \apj, 855, 105, \dodoi{10.3847/1538-4357/aab03f}

\bibitem[{{Oke} \& {Gunn}(1983)}]{okeSecondaryStandardStars1983}
{Oke}, J.~B., \& {Gunn}, J.~E. 1983, \apj, 266, 713, \dodoi{10.1086/160817}

\bibitem[{Pacucci {et~al.}(2023)Pacucci, Nguyen, Carniani, Maiolino, \&
  Fan}]{pacucciJWSTCEERSJADES2023}
Pacucci, F., Nguyen, B., Carniani, S., Maiolino, R., \& Fan, X. 2023, \apjl,
  957, L3, \dodoi{10.3847/2041-8213/ad0158}

\bibitem[{{Padovani} {et~al.}(2017){Padovani}, {Alexander}, {Assef}, {De
  Marco}, {Giommi}, {Hickox}, {Richards}, {Smol{\v{c}}i{\'c}},
  {Hatziminaoglou}, {Mainieri}, \&
  {Salvato}}]{padovaniActiveGalacticNuclei2017a}
{Padovani}, P., {Alexander}, D.~M., {Assef}, R.~J., {et~al.} 2017, \aapr, 25,
  2, \dodoi{10.1007/s00159-017-0102-9}

\bibitem[{Peca {et~al.}(2023)Peca, Cappelluti, Urry, LaMassa, Marchesi, Ananna,
  Balokovi{\'c}, Sanders, Auge, Treister, Powell, Turner, Kirkpatrick, \&
  Tian}]{pecaCosmicEvolutionAGN2023}
Peca, A., Cappelluti, N., Urry, C.~M., {et~al.} 2023, The Astrophysical
  Journal, 943, 162, \dodoi{10.3847/1538-4357/acac28}

\bibitem[{{P{\'e}rez-Gonz{\'a}lez} {et~al.}(2023){P{\'e}rez-Gonz{\'a}lez},
  Costantin, Langeroodi, Rinaldi, Annunziatella, Ilbert, Colina,
  {N{\o}rgaard-Nielsen}, Greve, {\"O}stlin, Wright, {Alonso-Herrero},
  {\'A}lvarez-M{\'a}rquez, Caputi, Eckart, Le~F{\`e}vre, Labiano,
  {Garc{\'i}a-Mar{\'i}n}, Hjorth, Kendrew, Pye, Tikkanen, {van der Werf},
  Walter, Ward, Bik, Boogaard, Bosman, G{\'o}mez, Gillman, Iani, Jermann,
  Melinder, Meyer, Moutard, {van Dishoek}, Henning, Lagage, Guedel, Peissker,
  Ray, Vandenbussche, {Garc{\'i}a-Argum{\'a}nez}, \&
  Mar{\'i}a~M{\'e}rida}]{perez-gonzalezLife30Probing2023a}
{P{\'e}rez-Gonz{\'a}lez}, P.~G., Costantin, L., Langeroodi, D., {et~al.} 2023,
  The Astrophysical Journal Letters, 951, L1, \dodoi{10.3847/2041-8213/acd9d0}

\bibitem[{{P{\'e}rez-Gonz{\'a}lez} {et~al.}(2024){P{\'e}rez-Gonz{\'a}lez},
  {Barro}, {Rieke}, {Lyu}, {Rieke}, {Alberts}, {Williams}, {Hainline}, {Sun},
  {Pusk{\'a}s}, {Annunziatella}, {Baker}, {Bunker}, {Egami}, {Ji}, {Johnson},
  {Robertson}, {Rodr{\'\i}guez Del Pino}, {Rujopakarn}, {Shivaei}, {Tacchella},
  {Willmer}, \& {Willott}}]{perez-gonzalezWhatNatureLittle2024}
{P{\'e}rez-Gonz{\'a}lez}, P.~G., {Barro}, G., {Rieke}, G.~H., {et~al.} 2024,
  \apj, 968, 4, \dodoi{10.3847/1538-4357/ad38bb}

\bibitem[{{Planck Collaboration} {et~al.}(2013){Planck Collaboration}, Ade,
  Aghanim, Arnaud, Ashdown, {Atrio-Barandela}, Aumont, Baccigalupi, Balbi,
  Banday, Barreiro, Bartlett, Battaner, Benabed, Beno{\^i}t, Bernard,
  Bersanelli, Bonaldi, Bond, Borrill, Bouchet, Burigana, Cabella, Cardoso,
  Catalano, Cay{\'o}n, Chary, Chiang, Christensen, Clements, Colombo, Coulais,
  Crill, Cuttaia, Danese, D'Arcangelo, Davis, de~Bernardis, de~Gasperis,
  de~Rosa, de~Zotti, Delabrouille, Dickinson, Diego, Dobler, Dole, Donzelli,
  Dor{\'e}, D{\"o}rl, Douspis, Dupac, Efstathiou, En{\ss}lin, Eriksen, Finelli,
  Forni, Frailis, Franceschi, Galeotta, Ganga, Giard, Giardino,
  {Gonz{\'a}lez-Nuevo}, G{\'o}rski, Gratton, Gregorio, Gruppuso, Hansen,
  Harrison, Helou, {Henrot-Versill{\'e}}, {Hern{\'a}ndez-Monteagudo},
  Hildebrandt, Hivon, Hobson, Holmes, Hornstrup, Hovest, Huffenberger, Jaffe,
  Jagemann, Jewell, Jones, Juvela, Keih{\"a}nen, Knoche, Knox, Kunz,
  {Kurki-Suonio}, Lagache, L{\"a}hteenm{\"a}ki, Lamarre, Lasenby, Lawrence,
  Leach, Leonardi, Lilje, {Linden-V{\o}rnle}, {L{\'o}pez-Caniego}, Lubin,
  {Mac{\'i}as-P{\'e}rez}, Maffei, Maino, Mandolesi, Maris, Marshall, Martin,
  {Mart{\'i}nez-Gonz{\'a}lez}, Masi, Massardi, Matarrese, Matthai, Mazzotta,
  Meinhold, Melchiorri, Mendes, Mennella, Mitra, Moneti, Montier, Morgante,
  Munshi, Murphy, Naselsky, Natoli, {N{\o}rgaard-Nielsen}, Noviello, Novikov,
  Novikov, Osborne, Pajot, Paladini, Paoletti, Partridge, Pearson, Perdereau,
  Perrotta, Piacentini, Piat, Pierpaoli, Pietrobon, Plaszczynski,
  Pointecouteau, Polenta, Ponthieu, Popa, Poutanen, Pratt, Prunet, Puget,
  Rachen, Rebolo, Reinecke, Renault, Ricciardi, Riller, Ristorcelli, Rocha,
  Rosset, {Rubi{\~n}o-Mart{\'i}n}, Rusholme, Sandri, Savini, Schaefer, Scott,
  Smoot, Spencer, Stivoli, Sudiwala, {Suur-Uski}, Sygnet, Tauber, Terenzi,
  Toffolatti, Tomasi, Tristram, T{\"u}rler, Umana, Valenziano, Tent, Vielva,
  Villa, Vittorio, Wade, Wandelt, White, Yvon, Zacchei, \&
  Zonca}]{adePlanckIntermediateResults2013}
{Planck Collaboration}, Ade, P. a.~R., Aghanim, N., {et~al.} 2013, \aap, 554,
  A139, \dodoi{10.1051/0004-6361/201220271}

\bibitem[{{Rieke} {et~al.}(2023){Rieke}, {Robertson}, {Tacchella}, {Hainline},
  {Johnson}, {Hausen}, {Ji}, {Willmer}, {Eisenstein}, {Pusk{\'a}s}, {Alberts},
  {Arribas}, {Baker}, {Baum}, {Bhatawdekar}, {Bonaventura}, {Boyett}, {Bunker},
  {Cameron}, {Carniani}, {Charlot}, {Chevallard}, {Chen}, {Curti},
  {Curtis-Lake}, {Danhaive}, {DeCoursey}, {Dressler}, {Egami}, {Endsley},
  {Helton}, {Hviding}, {Kumari}, {Looser}, {Lyu}, {Maiolino}, {Maseda},
  {Nelson}, {Rieke}, {Rix}, {Sandles}, {Saxena}, {Sharpe}, {Shivaei},
  {Skarbinski}, {Smit}, {Stark}, {Stone}, {Suess}, {Sun}, {Topping},
  {{\"U}bler}, {Villanueva}, {Wallace}, {Williams}, {Willott}, {Whitler},
  {Witstok}, \& {Woodrum}}]{riekeJADESInitialData2023}
{Rieke}, M.~J., {Robertson}, B., {Tacchella}, S., {et~al.} 2023, \apjs, 269,
  16, \dodoi{10.3847/1538-4365/acf44d}

\bibitem[{Robertson {et~al.}(2024)Robertson, Johnson, Tacchella, Eisenstein,
  Hainline, Arribas, Baker, Bunker, Carniani, Cargile, Carreira, Charlot,
  Chevallard, Curti, {Curtis-Lake}, D'Eugenio, Egami, Hausen, Helton, Jakobsen,
  Ji, Jones, Maiolino, Maseda, Nelson, {P{\'e}rez-Gonz{\'a}lez}, Pusk{\'a}s,
  Rieke, Smit, Sun, {\"U}bler, Whitler, Williams, Willmer, Willott, \&
  Witstok}]{robertsonEarliestGalaxiesJADES2024}
Robertson, B., Johnson, B.~D., Tacchella, S., {et~al.} 2024, The Astrophysical
  Journal, 970, 31, \dodoi{10.3847/1538-4357/ad463d}

\bibitem[{Robotham \& Bellstedt(2024)}]{robothamProGenyNewSimple2024}
Robotham, A. S.~G., \& Bellstedt, S. 2024, arXiv e-prints, arXiv:2410.17697,
  \dodoi{10.48550/arXiv.2410.17697}

\bibitem[{Robotham {et~al.}(2020)Robotham, Bellstedt, Lagos, Thorne, Davies,
  Driver, \& Bravo}]{robothamProSpectGeneratingSpectral2020}
Robotham, A. S.~G., Bellstedt, S., Lagos, C. d.~P., {et~al.} 2020, \mnras, 495,
  905, \dodoi{10.1093/mnras/staa1116}

\bibitem[{Robotham {et~al.}(2018)Robotham, Davies, Driver, Koushan, Taranu,
  Casura, \& Liske}]{robothamProFoundSourceExtraction2018}
Robotham, A. S.~G., Davies, L. J.~M., Driver, S.~P., {et~al.} 2018, \mnras,
  476, 3137, \dodoi{10.1093/mnras/sty440}

\bibitem[{Robotham {et~al.}(2023)Robotham, D'Silva, Windhorst, Jansen, Summers,
  Driver, Wilmer, \& Bellstedt}]{robothamDynamicWispRemoval2023}
Robotham, A. S.~G., D'Silva, J. C.~J., Windhorst, R.~A., {et~al.} 2023, \pasp,
  135, 085003, \dodoi{10.1088/1538-3873/acea42}

\bibitem[{Robotham {et~al.}(2017)Robotham, Taranu, Tobar, Moffett, \&
  Driver}]{robothamProFitBayesianProfile2017}
Robotham, A. S.~G., Taranu, D.~S., Tobar, R., Moffett, A., \& Driver, S.~P.
  2017, \mnras, 466, 1513, \dodoi{10.1093/mnras/stw3039}

\bibitem[{Robotham {et~al.}(2024)Robotham, Tobar, Bellstedt, Casura, Cook,
  D'Silva, Davies, Driver, Li, \& Garate-Nuñez}]{robotham_propane_2024}
Robotham, A. S.~G., Tobar, R., Bellstedt, S., {et~al.} 2024, \mnras, 528, 5046,
  \dodoi{10.1093/mnras/stae349}

\bibitem[{Schaye {et~al.}(2010)Schaye, Dalla~Vecchia, Booth, Wiersma, Theuns,
  Haas, Bertone, Duffy, McCarthy, \& {van de
  Voort}}]{schayePhysicsDrivingCosmic2010}
Schaye, J., Dalla~Vecchia, C., Booth, C.~M., {et~al.} 2010, \mnras, 402, 1536,
  \dodoi{10.1111/j.1365-2966.2009.16029.x10.48550/arXiv.0909.5196}

\bibitem[{{Schaye} {et~al.}(2015){Schaye}, {Crain}, {Bower}, {Furlong},
  {Schaller}, {Theuns}, {Dalla Vecchia}, {Frenk}, {McCarthy}, {Helly},
  {Jenkins}, {Rosas-Guevara}, {White}, {Baes}, {Booth}, {Camps}, {Navarro},
  {Qu}, {Rahmati}, {Sawala}, {Thomas}, \&
  {Trayford}}]{schayeEAGLEProjectSimulating2015a}
{Schaye}, J., {Crain}, R.~A., {Bower}, R.~G., {et~al.} 2015, \mnras, 446, 521,
  \dodoi{10.1093/mnras/stu2058}

\bibitem[{Schechter(1976)}]{schechterAnalyticExpressionLuminosity1976a}
Schechter, P. 1976, \apj, 203, 297, \dodoi{10.1086/154079}

\bibitem[{Schneider {et~al.}(2023)Schneider, Valiante, Trinca, Graziani,
  Volonteri, \& Maiolino}]{schneiderAreWeSurprised2023b}
Schneider, R., Valiante, R., Trinca, A., {et~al.} 2023, \mnras, 526, 3250,
  \dodoi{10.1093/mnras/stad2503}

\bibitem[{Shakura \& Sunyaev(1973)}]{shakuraBlackHolesBinary1973}
Shakura, N.~I., \& Sunyaev, R.~A. 1973, \aap, 24, 337

\bibitem[{Shen {et~al.}(2020)Shen, Hopkins, {Faucher-Gigu{\`e}re}, Alexander,
  Richards, Ross, \& Hickox}]{shenBolometricQuasarLuminosity2020}
Shen, X., Hopkins, P.~F., {Faucher-Gigu{\`e}re}, C.-A., {et~al.} 2020, \mnras,
  495, 3252, \dodoi{10.1093/mnras/staa1381}

\bibitem[{{Shen} {et~al.}(2022){Shen}, {Vogelsberger}, {Nelson}, {Tacchella},
  {Hernquist}, {Springel}, {Marinacci}, \&
  {Torrey}}]{shenHighredshiftPredictionsIllustrisTNG2022}
{Shen}, X., {Vogelsberger}, M., {Nelson}, D., {et~al.} 2022, \mnras, 510, 5560,
  \dodoi{10.1093/mnras/stab3794}

\bibitem[{Shuntov {et~al.}(2025)Shuntov, Ilbert, Toft, {Arango-Toro}, Akins,
  Casey, Franco, Harish, Kartaltepe, Koekemoer, McCracken, Paquereau, Laigle,
  Bethermin, Dubois, Drakos, Faisst, Gozaliasl, Gillman, Hayward, Hirschmann,
  {Huertas-Company}, Jespersen, Jin, Kokorev, Lambrides, Borgne, Liu, Magdis,
  Massey, McPartland, Mercier, McCleary, McKinney, Oesch, Renzini, Rhodes,
  Rich, Robertson, Sanders, Trebitsch, Tresse, Valentino, Vijayan, Weaver,
  Weibel, Wilkins, \& Yang}]{shuntovCOSMOSWebStellarMass2025}
Shuntov, M., Ilbert, O., Toft, S., {et~al.} 2025, Astronomy \& Astrophysics,
  695, A20, \dodoi{10.1051/0004-6361/202452570}

\bibitem[{Song {et~al.}(2016)Song, Finkelstein, Ashby, Grazian, Lu, Papovich,
  Salmon, Somerville, Dickinson, Duncan, Faber, Fazio, Ferguson, Fontana, Guo,
  Hathi, Lee, Merlin, \& Willner}]{songEvolutionGalaxyStellar2016}
Song, M., Finkelstein, S.~L., Ashby, M. L.~N., {et~al.} 2016, \apj, 825, 5,
  \dodoi{10.3847/0004-637X/825/1/5}

\bibitem[{Stefanon {et~al.}(2021)Stefanon, Bouwens, Labb{\'e}, Illingworth,
  Gonzalez, \& Oesch}]{stefanonGalaxyStellarMass2021}
Stefanon, M., Bouwens, R.~J., Labb{\'e}, I., {et~al.} 2021, \apj, 922, 29,
  \dodoi{10.3847/1538-4357/ac1bb6}

\bibitem[{Tasnim~Ananna {et~al.}(2019)Tasnim~Ananna, Treister, Megan~Urry,
  Ricci, Kirkpatrick, LaMassa, Buchner, Civano, Tremmel, \&
  Marchesi}]{tasnimanannaAccretionHistoryAGNs2019}
Tasnim~Ananna, T., Treister, E., Megan~Urry, C., {et~al.} 2019, The
  Astrophysical Journal, 871, 240, \dodoi{10.3847/1538-4357/aafb77}

\bibitem[{Thorne {et~al.}(2021)Thorne, Robotham, Davies, Bellstedt, Driver,
  Bravo, Bremer, Holwerda, Hopkins, Lagos, Phillipps, Siudek, Taylor, \&
  Wright}]{thorneDeepExtragalacticVIsible2021}
Thorne, J.~E., Robotham, A. S.~G., Davies, L. J.~M., {et~al.} 2021, \mnras,
  505, 540, \dodoi{10.1093/mnras/stab1294}

\bibitem[{Thorne {et~al.}(2022{\natexlab{a}})Thorne, Robotham, Bellstedt,
  Davies, Cook, Cortese, Holwerda, Phillipps, \&
  Siudek}]{thorneDEVILSCosmicEvolution2022}
Thorne, J.~E., Robotham, A. S.~G., Bellstedt, S., {et~al.} 2022{\natexlab{a}},
  \mnras, 517, 6035, \dodoi{10.1093/mnras/stac3082}

\bibitem[{Thorne {et~al.}(2022{\natexlab{b}})Thorne, Robotham, Davies,
  Bellstedt, Brown, Croom, Delvecchio, Groves, Jarvis, Shabala, Seymour,
  Whittam, Bravo, Cook, Driver, Holwerda, Phillipps, \&
  Siudek}]{thorneDeepExtragalacticVIsible2022}
Thorne, J.~E., Robotham, A. S.~G., Davies, L. J.~M., {et~al.}
  2022{\natexlab{b}}, \mnras, 509, 4940, \dodoi{10.1093/mnras/stab3208}

\bibitem[{Trump {et~al.}(2015)Trump, Sun, Zeimann, Luck, Bridge, Grier, Hagen,
  Juneau, {Montero-Dorta}, Rosario, Brandt, Ciardullo, \&
  Schneider}]{trumpBIASESOPTICALLINERATIO2015}
Trump, J.~R., Sun, M., Zeimann, G.~R., {et~al.} 2015, The Astrophysical
  Journal, 811, 26, \dodoi{10.1088/0004-637X/811/1/26}

\bibitem[{{{\"U}bler} {et~al.}(2023){{\"U}bler}, {Maiolino}, {Curtis-Lake},
  {P{\'e}rez-Gonz{\'a}lez}, {Curti}, {Perna}, {Arribas}, {Charlot}, {Marshall},
  {D'Eugenio}, {Scholtz}, {Bunker}, {Carniani}, {Ferruit}, {Jakobsen}, {Rix},
  {Rodr{\'\i}guez Del Pino}, {Willott}, {Boeker}, {Cresci}, {Jones}, {Kumari},
  \& {Rawle}}]{ublerGANIFSMassiveBlack2023a}
{{\"U}bler}, H., {Maiolino}, R., {Curtis-Lake}, E., {et~al.} 2023, \aap, 677,
  A145, \dodoi{10.1051/0004-6361/202346137}

\bibitem[{{Vijayan} {et~al.}(2021){Vijayan}, {Lovell}, {Wilkins}, {Thomas},
  {Barnes}, {Irodotou}, {Kuusisto}, \&
  {Roper}}]{vijayanFirstLightReionization2021a}
{Vijayan}, A.~P., {Lovell}, C.~C., {Wilkins}, S.~M., {et~al.} 2021, \mnras,
  501, 3289, \dodoi{10.1093/mnras/staa3715}

\bibitem[{Weibel {et~al.}(2024)Weibel, Oesch, Barrufet, Gottumukkala, Ellis,
  Santini, Weaver, Allen, Bouwens, Bowler, Brammer, Carnall, Cullen, Dayal,
  Dickinson, Donnan, Dunlop, Giavalisco, Grogin, Illingworth, Koekemoer, Labbe,
  Marchesini, McLeod, McLure, Naidu, {P{\'e}rez-Gonz{\'a}lez}, Shuntov,
  Stefanon, Toft, \& Xiao}]{weibelGalaxyBuildupFirst2024a}
Weibel, A., Oesch, P.~A., Barrufet, L., {et~al.} 2024, Monthly Notices of the
  Royal Astronomical Society, 533, 1808, \dodoi{10.1093/mnras/stae1891}

\bibitem[{Weigel {et~al.}(2016)Weigel, Schawinski, \&
  Bruderer}]{weigelStellarMassFunctions2016}
Weigel, A.~K., Schawinski, K., \& Bruderer, C. 2016, \mnras, 459, 2150,
  \dodoi{10.1093/mnras/stw756}

\bibitem[{Wilkins {et~al.}(2019)Wilkins, Lovell, \&
  Stanway}]{wilkinsRecalibratingCosmicStar2019}
Wilkins, S.~M., Lovell, C.~C., \& Stanway, E.~R. 2019, Monthly Notices of the
  Royal Astronomical Society, 490, 5359, \dodoi{10.1093/mnras/stz2894}

\bibitem[{{Williams} {et~al.}(2024){Williams}, {Alberts}, {Ji}, {Hainline},
  {Lyu}, {Rieke}, {Endsley}, {Suess}, {Sun}, {Johnson}, {Florian}, {Shivaei},
  {Rujopakarn}, {Baker}, {Bhatawdekar}, {Boyett}, {Bunker}, {Cameron},
  {Carniani}, {Charlot}, {Curtis-Lake}, {DeCoursey}, {de Graaff}, {Egami},
  {Eisenstein}, {Gibson}, {Hausen}, {Helton}, {Maiolino}, {Maseda}, {Nelson},
  {P{\'e}rez-Gonz{\'a}lez}, {Rieke}, {Robertson}, {Saxena}, {Tacchella},
  {Willmer}, \& {Willott}}]{williamsGalaxiesMissedHubble2024}
{Williams}, C.~C., {Alberts}, S., {Ji}, Z., {et~al.} 2024, \apj, 968, 34,
  \dodoi{10.3847/1538-4357/ad3f17}

\bibitem[{Windhorst {et~al.}(2011)Windhorst, Cohen, Hathi, McCarthy, Ryan, Yan,
  Baldry, Driver, Frogel, Hill, Kelvin, Koekemoer, Mechtley, O'Connell,
  Robotham, Rutkowski, Seibert, Tuffs, Balick, Bond, Bushouse, Calzetti,
  Crockett, Disney, Dopita, Hall, Holtzman, Kaviraj, Kimble, MacKenty,
  Mutchler, Paresce, Saha, Silk, Trauger, Walker, Whitmore, \&
  Young}]{windhorstHubbleSpaceTelescope2011a}
Windhorst, R.~A., Cohen, S.~H., Hathi, N.~P., {et~al.} 2011, \apjs, 193, 27,
  \dodoi{10.1088/0067-0049/193/2/27}

\bibitem[{Windhorst {et~al.}(2023)Windhorst, Cohen, Jansen, Summers, Tompkins,
  Conselice, Driver, Yan, Coe, Frye, Grogin, Koekemoer, Marshall, O'Brien,
  Pirzkal, Robotham, Ryan, Willmer, Carleton, Diego, Keel, Porto, Redshaw,
  Scheller, Wilkins, Willner, Zitrin, Adams, Austin, Arendt, Beacom,
  Bhatawdekar, Bradley, Broadhurst, Cheng, Civano, Dai, Dole, D'Silva, Duncan,
  Fazio, Ferrami, Ferreira, Finkelstein, Furtak, Gim, Griffiths, Hammel,
  Harrington, Hathi, Holwerda, Honor, Huang, Hyun, Im, Joshi, Kamieneski,
  Kelly, Larson, Li, Lim, Ma, Maksym, Manzoni, Meena, Milam, Nonino, Pascale,
  Petric, Pierel, {del Carmen Polletta}, R{\"o}ttgering, Rutkowski, Smail,
  Straughn, Strolger, Swirbul, Trussler, Wang, Welch, B.~Wyithe, Yun,
  Zackrisson, Zhang, \& Zhao}]{windhorstJWSTPEARLSPrime2023}
Windhorst, R.~A., Cohen, S.~H., Jansen, R.~A., {et~al.} 2023, \aj, 165, 13,
  \dodoi{10.3847/1538-3881/aca163}

\bibitem[{{Wright} {et~al.}(2024){Wright}, {Somerville}, {Lagos}, {Schaller},
  {Dav{\'e}}, {Angl{\'e}s-Alc{\'a}zar}, \&
  {Genel}}]{wrightBaryonCycleModern2024}
{Wright}, R.~J., {Somerville}, R.~S., {Lagos}, C. d.~P., {et~al.} 2024, \mnras,
  532, 3417, \dodoi{10.1093/mnras/stae1688}

\bibitem[{Yang {et~al.}(2023)Yang, Caputi, Papovich, Arrabal~Haro, Bagley,
  Behroozi, Bell, Bisigello, Buat, Burgarella, Cheng, Cleri, Dav{\'e},
  Dickinson, Elbaz, Ferguson, Finkelstein, Grogin, Hathi, Hirschmann, Holwerda,
  {Huertas-Company}, Hutchison, Iani, Kartaltepe, Kirkpatrick, Kocevski,
  Koekemoer, Kokorev, Larson, Lucas, {P{\'e}rez-Gonz{\'a}lez}, Rinaldi, Shen,
  Trump, {de la Vega}, Yung, \& Zavala}]{yangCEERSKeyPaper2023}
Yang, G., Caputi, K.~I., Papovich, C., {et~al.} 2023, The Astrophysical Journal
  Letters, 950, L5, \dodoi{10.3847/2041-8213/acd639}

\bibitem[{{Yue} {et~al.}(2024){Yue}, {Eilers}, {Ananna}, {Panagiotou}, {Kara},
  \& {Miyaji}}]{yueStackingXrayObservations2024}
{Yue}, M., {Eilers}, A.-C., {Ananna}, T.~T., {et~al.} 2024, \apjl, 974, L26,
  \dodoi{10.3847/2041-8213/ad7eba}

\bibitem[{Zavala {et~al.}(2021)Zavala, Casey, Manning, Aravena, Bethermin,
  Caputi, Clements, da~Cunha, Drew, Finkelstein, Fujimoto, Hayward, Hodge,
  Kartaltepe, Knudsen, Koekemoer, Long, Magdis, Man, Popping, Sanders,
  Scoville, Sheth, Staguhn, Toft, Treister, Vieira, \&
  Yun}]{zavalaEvolutionIRLuminosity2021}
Zavala, J.~A., Casey, C.~M., Manning, S.~M., {et~al.} 2021, The Astrophysical
  Journal, 909, 165, \dodoi{10.3847/1538-4357/abdb27}

\bibitem[{{Zavala} {et~al.}(2023){Zavala}, {Buat}, {Casey}, {Finkelstein},
  {Burgarella}, {Bagley}, {Ciesla}, {Daddi}, {Dickinson}, {Ferguson}, {Franco},
  {Jim{\'e}nez-Andrade}, {Kartaltepe}, {Koekemoer}, {Le Bail}, {Murphy},
  {Papovich}, {Tacchella}, {Wilkins}, {Aretxaga}, {Behroozi}, {Champagne},
  {Fontana}, {Giavalisco}, {Grazian}, {Grogin}, {Kewley}, {Kocevski},
  {Kirkpatrick}, {Lotz}, {Pentericci}, {P{\'e}rez-Gonz{\'a}lez}, {Pirzkal},
  {Ravindranath}, {Somerville}, {Trump}, {Yang}, {Yung}, {Almaini},
  {Amor{\'\i}n}, {Annunziatella}, {Arrabal Haro}, {Backhaus}, {Barro}, {Bell},
  {Bhatawdekar}, {Bisigello}, {Buitrago}, {Calabr{\`o}}, {Castellano},
  {Ch{\'a}vez Ortiz}, {Chworowsky}, {Cleri}, {Cohen}, {Cole}, {Cooke},
  {Cooper}, {Cooray}, {Costantin}, {Cox}, {Croton}, {Dav{\'e}}, {de La Vega},
  {Dekel}, {Elbaz}, {Estrada-Carpenter}, {Fern{\'a}ndez}, {Finkelstein},
  {Freundlich}, {Fujimoto}, {Garc{\'\i}a-Argum{\'a}nez}, {Gardner}, {Gawiser},
  {G{\'o}mez-Guijarro}, {Guo}, {Hamilton}, {Hathi}, {Holwerda}, {Hirschmann},
  {Huertas-Company}, {Hutchison}, {Iyer}, {Jaskot}, {Jha}, {Jogee}, {Juneau},
  {Jung}, {Kassin}, {Kurczynski}, {Larson}, {Leung}, {Long}, {Lucas},
  {Magnelli}, {Mantha}, {Matharu}, {McGrath}, {McIntosh}, {Medrano}, {Merlin},
  {Mobasher}, {Morales}, {Newman}, {Nicholls}, {Pandya}, {Rafelski}, {Ronayne},
  {Rose}, {Ryan}, {Santini}, {Seill{\'e}}, {Shah}, {Shen}, {Simons}, {Snyder},
  {Stanway}, {Straughn}, {Teplitz}, {Vanderhoof}, {Vega-Ferrero}, {Wang},
  {Weiner}, {Willmer}, {Wuyts}, \& {Ceers
  Team}}]{zavalaDustyStarburstsMasquerading2023a}
{Zavala}, J.~A., {Buat}, V., {Casey}, C.~M., {et~al.} 2023, \apjl, 943, L9,
  \dodoi{10.3847/2041-8213/acacfe}

\end{thebibliography}
\bibliographystyle{aasjournal}

\appendix

\section{Tabulated double power law fits to the SMF, SFRF, AGNLF}
The parameters of the double power law function (Equation~\ref{eq:dpl}) are reported in Table~\ref{tab:dpl_fits}.
\begin{table}[h!]
    \centering
    \begin{tabular}{c c c c c c}
       Quantity & $\bar{z}$ & $\log_{10}(\phi^{*})$ & $\alpha$ & $\beta$ & $\log_{10}(x^{*})$ \\
       \hline
       \hline
        $\mathrm{M_{\star}}$ 
      & $ 6.450$ & $-2.126 \pm 0.239$ & $-0.551 \pm 0.191$ & $-1.659 \pm 0.103$ & $8.445 \pm 0.211$ \\
      & $ 8.231$ & $-2.666 \pm 0.328$ & $-0.464 \pm 0.197$ & $-1.730 \pm 0.189$ & $8.436 \pm 0.280$ \\
      & $10.126$ & $-1.920 \pm 1.234$ & $-0.496 \pm 0.200$ & $-2.418 \pm 0.280$ & $8.029 \pm 0.560$ \\
      & $12.583$ & $-3.044 \pm 1.585$ & $-0.500 \pm 0.200$ & $-2.609 \pm 0.683$ & $8.328 \pm 0.747$ \\

        \hline 

        SFR
      & $ 6.450$ & $-2.118 \pm 0.434$ & $-0.602 \pm 0.225$ & $-1.470 \pm 0.113$ & $0.228  \pm 0.405$ \\
      & $ 8.231$ & $-1.636 \pm 0.747$ & $-0.485 \pm 0.201$ & $-1.789 \pm 0.114$ & $-0.209 \pm 0.466$ \\
      & $10.126$ & $-2.690 \pm 0.732$ & $-0.498 \pm 0.200$ & $-2.342 \pm 0.359$ & $0.253  \pm 0.415$ \\
      & $12.583$ & $-3.038 \pm 1.018$ & $-0.500 \pm 0.200$ & $-3.673 \pm 0.985$ & $0.490  \pm 0.384$ \\

        \hline

        $\mathrm{L_{AGN} \, upper \, limit}$
      & $ 6.450$ & $-2.046$ & $-0.232$ & $-1.391$ & $42.848$ \\
      & $ 8.231$ & $-3.169$ & $-0.356$ & $-1.807$ & $43.518$ \\
      & $10.126$ & $-2.754$ & $-0.498$ & $-1.224$ & $42.627$ \\
      & $12.583$ & $-4.463$ & $-0.485$ & $-2.744$ & $44.050$ \\
       
        \hline
    \end{tabular}
    \caption{double power law fits (see Equation~\ref{eq:dpl}) to the SMF, SFRF and AGNLF. Values are the best fit parameters and the $1\sigma$ uncertainties.}
    \label{tab:dpl_fits}
\end{table}

\section{Tabulated CSMH, CSFH, CAGNH}
We fitted $\mathtt{massfunc\_snorm\_trunc}$ functions \citep[Eqs. 1-5 in ][]{dsilvaGAMADEVILSCosmic2023} to the CSFH and CAGNH from $z \approx 0 \to 13.5$, combining the results of this work and \citet{dsilvaGAMADEVILSCosmic2023}. We fitted with respect to $z$ and use $\mathtt{magemax}=30$ and $\mathtt{mtrunc}=2$. The values for the CSMH, CSFH, CAGNH and the fit parameters are reported in Table~\ref{tab:cosmic_densities}.
\begin{table}[h!]
    \centering
    \begin{tabular}{c c c c c c}
       Values&$\bar{z}$ & $\log_{10}$(CSMH) & $\log_{10}$(CSFH) & $\log_{10}$(CAGNH) summed lower limit & $\log_{10}(\mathrm{CAGNH}$) upper limit \\
       \hline
       \hline

        -- & $6.450 $ & $6.413 \pm 0.098$ & $-1.707 \pm 0.132$ & $40.593$ & $40.931$ \\
        -- & $8.231 $ & $5.807 \pm 0.145$ & $-1.809 \pm 0.335$ & $39.435$ & $40.330$ \\
        -- & $10.126$ & $6.074 \pm 0.687$ & $-2.451 \pm 0.387$ & $39.615$ & $40.223$ \\
        -- & $12.583$ & $5.249 \pm 0.859$ & $-2.596 \pm 0.674$ & $38.559$ & $39.551$ \\
       
        \hline
        \hline

        Fit parameters & $\log_{10}$(\texttt{mX}) & \texttt{mpeak} & \texttt{mperiod} & \texttt{mskew} & --\\
       \hline
       \hline
        CSFH & $-1.111 \pm 0.035$ & $1.497 \pm 0.103$ & $1.105 \pm 0.062$ & $-0.447 \pm 0.038$ & --\\
        CAGNH & $41.595 \pm 0.042$ & $1.493 \pm 0.27$ & $1.136 \pm 0.188$ & $-0.368 \pm 0.055$ & -- \\

        \hline
\end{tabular}
\caption{Values and $1\sigma$ uncertainties for the CSMH, CSFH and CAGNH (see Equation~\ref{eq:rho_integral}). Values and $1\sigma$ uncertainties of the $\mathtt{massfunc\_snorm\_trunc}$ parameters.}
\label{tab:cosmic_densities}
\end{table}

\end{document}